\documentclass[a4paper,11pt]{article}

\usepackage{jcappub}
\allowdisplaybreaks
\usepackage{graphicx}
\usepackage{epstopdf}

\newcommand{\be}{\begin{equation}}
\newcommand{\ee}{\end{equation}}
\newcommand{\bea}{\begin{eqnarray}}
\newcommand{\eea}{\end{eqnarray}}
\newcommand{\nn}{\nonumber}
\def\nl{\nonumber \\ &}

\title{Next-to-next-to-leading order gravitational spin-squared potential via 
	the effective field theory for spinning objects in the post-Newtonian scheme}

\author[a,b]{Mich\`ele Levi}
\author[c,d]{and Jan Steinhoff}

\affiliation[a]{Universit\'e Pierre et Marie Curie, CNRS-UMR 7095, 
Institut d'Astrophysique de Paris,\\ 
98 bis Boulevard Arago, 75014 Paris, France} 
\affiliation[b]{Sorbonne Universit\'es, Institut Lagrange de Paris,\\ 
98 bis Boulevard Arago, 75014 Paris, France} 
\affiliation[c]{Max-Planck-Institute for Gravitational Physics 
(Albert-Einstein-Institute),\\ 
Am M{\"u}hlenberg 1, 14476 Potsdam-Golm, Germany}
\affiliation[d]{Centro Multidisciplinar de Astrofisica, 
Instituto Superior Tecnico, Universidade de Lisboa,\\ 
Avenida Rovisco Pais 1, 1049-001 Lisboa, Portugal}

\emailAdd{michele.levi@upmc.fr}
\emailAdd{jan.steinhoff@aei.mpg.de}

\abstract{The next-to-next-to-leading order spin-squared interaction potential for generic compact binaries is derived for the first time via the effective field theory for gravitating spinning objects in the post-Newtonian scheme. 
The spin-squared sector is an intricate one, as it requires the consideration of the point particle action beyond minimal coupling, and mainly involves the spin-squared worldline couplings, which are quite complex, compared to the worldline couplings from the minimal coupling part of the action. This sector also involves the linear in spin couplings, as we go up in the nonlinearity of the interaction, and in the loop order. 
Hence, there is an excessive increase in the number of Feynman diagrams, of which more are higher loop ones. We provide all the Feynman diagrams and their values. The beneficial ``nonrelativistic gravitational'' fields are employed in the computation. This spin-squared correction, which enters at the fourth post-Newtonian order for rapidly rotating compact objects, completes the conservative sector up to the fourth post-Newtonian accuracy. The robustness of the effective field theory for gravitating spinning objects is shown here once again, as demonstrated in a recent series of papers by the authors, which obtained all spin dependent sectors, required up to the fourth post-Newtonian accuracy. The effective field theory of spinning objects allows to directly obtain the equations of motion, and the Hamiltonians, and these will be derived for the potential obtained here in a forthcoming paper.}

\begin{document}

\maketitle

\flushbottom

\section{Introduction} \label{intro}

The anticipated direct detection of gravitational waves (GW) may become a 
realistic prospect in view of the imminent worldwide operation of second-generation ground-based interferometers, such as the two of Advanced LIGO \cite{LIGO}, Advanced Virgo \cite{Virgo}, and KAGRA \cite{Kagra}. 
This will transform GW science into a powerful observational tool.
Inspiralling binaries of compact objects are considered as most promising sources for the detection of the GW signal. 
The post-Newtonian (PN) approximation of General Relativity stands out among the various and complementary approaches to model these systems, as it enables an analytical treatment of the inspiral phase of their evolution \cite{Blanchet:2013haa}.
Using the matched filtering technique to search the GW signal results in
a pressing demand to supply accurate theoretical waveforms, 
where the continuous signal is modeled via the Effective-One-Body (EOB) approach \cite{Buonanno:1998gg}. It turns out that even relative high order corrections beyond Newtonian gravity, such as up to the sixth PN (6PN) order, are required to obtain such waveforms, and furthermore to gain information about the inner structure of the constituents of the binary \cite{Yagi:2013baa}. 

Of these compact objects black holes are expected to be near-extremal, i.e.~ to rotate rapidly. Hence, considering the recent completion of the 4PN order point-mass correction \cite{Damour:2014jta}, it is essential to obtain the same PN accuracy for binaries with spinning objects. To this end finite size effects with spins should also be taken into account. The leading order (LO) finite size spin effects, which are of the LO spin-squared interaction at the quadrupole level, and enter already at the 2PN order, were first derived for black holes in 1975 in \cite{D'Eath:1975vw}, see also \cite{Thorne:1984mz}. Generic quadrupoles, required to describe neutron stars, were already included in \cite{Barker:1975ae,Thorne:1984mz}, and the proportionality of the quadrupole to spin-squared was introduced in \cite{Poisson:1997ha}. However, the next-to-leading order (NLO) spin-squared interaction was treated much later in the following series of works 
\cite{Porto:2008jj, Steinhoff:2008ji, Hergt:2008jn, Hergt:2010pa}, and in \cite{Levi:2015msa} within an effective field theory (EFT) for spinning objects, which was formulated there. Furthermore, the LO cubic and quartic in spin interactions, which enter at the 3.5PN and 4PN orders, respectively, were first approached in parts for black hole binaries in \cite{Hergt:2007ha,Hergt:2008jn}. Recently, these have been obtained for generic compact objects for the cubic in spin interaction in \cite{Levi:2014gsa, Marsat:2014xea}, and also for the quartic in spin interaction in \cite{Levi:2014gsa}, where the latter was using the EFT for spinning objects \cite{Levi:2015msa}. These have been also partially treated in \cite{Vaidya:2014kza}.

In this paper we derive the next-to-next-to-leading order (NNLO) spin-squared interaction potential via the EFT for gravitating spinning 
objects in the PN scheme \cite{Levi:2015msa}. The NNLO level in the spin dependent case, which was already tackled using EFT techniques in \cite{Levi:2011eq}, was also obtained for the spin-orbit interaction at the 3.5PN order for rapidly rotating compact objects \cite{Levi:2015uxa}. The NNLO quadratic in spin potentials here and in \cite{Levi:2011eq} enter at the 4PN order. The EFT for gravitating spinning objects \cite{Levi:2015msa} is based on the original and innovative EFT approach for the binary inspiral, which was presented in \cite{Goldberger:2004jt,Goldberger:2007hy}. The seminal works in \cite{Hanson:1974qy, Bailey:1975fe} treated spin in an 
action approach in special and general relativity, respectively, and an extension of EFT techniques to the spinning case was approached in \cite{Porto:2005ac}, which later adopted a Routhian approach from \cite{Yee:1993ya}.
The EFT for gravitating spinning objects \cite{Levi:2015msa} however actually obtains an effective action at the orbital scale by integrating out \textit{all} the field modes below this scale, and enables the relation to physical observables. 
That is, the line of work in \cite{Porto:2005ac,Porto:2008jj} leaves the temporal components of the spin
tensor in the final results, which must be eliminated in order to
obtain, e.g.~the gauge invariant binding energy, while these components
depend on field modes at the orbital scale. 
We show here the robustness of the EFT for gravitating spinning objects \cite{Levi:2015msa}, as was already demonstrated in \cite{Levi:2014gsa,Levi:2015msa,Levi:2015uxa}, and allows to obtain straightforwardly the physical EOM \cite{Levi:2014sba,Levi:2015msa}, as well as the Hamiltonians from the effective action \cite{Levi:2015msa}.
Indeed, from the potential, which we obtain here, we derive the EOM and corresponding NNLO spin-squared Hamiltonian in a forthcoming paper.

The spin-squared sector is an intricate one, all the more so considering its LO and NLO levels \cite{Levi:2015msa}, compared to the NNLO here. This sector requires the consideration of the point particle action beyond minimal coupling, and involves mainly contributions with the spin-squared worldline couplings, which are quite complex, compared to the worldline couplings, that follow from the minimal coupling part of the action. Yet, we also have contributions to this sector, involving the linear in spin couplings, as we go up in the nonlinearity of the interaction, and in the loop order. Thus, there is an increase in the number of Feynman diagrams and in complexity with respect to the NLO, which is even more excessive than that, which occurs in the spin1-spin2 sector \cite{Levi:2008nh,Levi:2011eq,Levi:2015msa}. We have here 64 diagrams, of which more are higher loop ones, where we have 11 two-loops, and tensor integrals up to order 6. 
However, the application of the 
``nonrelativistic gravitational'' (NRG) fields \cite{Kol:2007bc,Kol:2010ze} in the spinning case, as demonstrated in \cite{Levi:2008nh,Levi:2010zu,Levi:2011eq,Levi:2014gsa,Levi:2015msa,Levi:2015uxa}, makes the computations more efficient. This ingredient is actually included together with other beneficial gauge choices in the EFT for gravitating spinning objects \cite{Levi:2015msa}. 
Finally, the EFT computations are also automatized here, using the xAct free packages for Mathematica \cite{xAct, Martin-Garcia:2008aa}. 

This paper is organized as follows. In section \ref{bmcfr} we recall the required effective action beyond minimal coupling, following the EFT for gravitating spinning objects\cite{Levi:2015msa}, and provide 
the Feynman rules relevant for this sector. In section \ref{ssfd} 
we provide the Feynman diagrams of the NNLO spin-squared interaction, and their values. In section \ref{result} we introduce the NNLO spin-squared potential, and explain how to obtain from it the EOM,and the Hamiltonian. Finally, our conclusions are presented in section \ref{endfriend}.

\section{The EFT for gravitating spinning objects in the PN scheme} \label{bmcfr}

In this section we briefly review the effective action, which should be considered in this sector beyond minimal coupling, following the EFT for spinning objects, which was formulated in \cite{Levi:2015msa}. From this action the Feynman rules, which are required for the EFT computation of the NNLO spin-squared 
interaction, are derived. 
We build on the series of works in
\cite{Levi:2008nh,Levi:2010zu,Levi:2011eq,Levi:2014sba,Levi:2014gsa,Levi:2015msa,Levi:2015uxa}, where we follow similar notations and conventions, in particular, to those in \cite{Levi:2015msa,Levi:2014gsa,Levi:2015uxa}. 
We also use the ``NRG'' fields, which were already shown to perform well in spin dependent sectors in 
\cite{Levi:2008nh,Levi:2010zu,Levi:2011eq,Levi:2014gsa,Levi:2015msa,Levi:2015uxa}. Similarly, in what follows $d$ is set to 3, as only the dependence in the generic $d$ dimensional, related with the loop Feynman integrals, see appendix~A in \cite{Levi:2011eq}, should be followed.

Let us first write the approximate metric in terms of the ``NRG'' fields to the orders required in this work:
\begin{align} \label{eq:gkk}
g_{\mu\nu}&=
\left(\begin{array}{cc} 
e^{2\phi}      & \quad -e^{2\phi} A_j \\
-e^{2\phi} A_i & \quad -e^{-2\phi}\gamma_{ij}+e^{2\phi} A_i A_j
\end{array}\right)\nn\\
&\simeq
\left(\begin{array}{cc} 
1+2\phi+2\phi^2+\frac{4}{3}\phi^3 & \quad-A_j-2A_j\phi
 \\
-A_i-2A_i\phi  
 & \quad-\delta_{ij}+2\phi\delta_{ij}-\sigma_{ij}-2\phi^2\delta_{ij}
 +2\phi\sigma_{ij}+A_iA_j+\frac{4}{3}\phi^3\delta_{ij}
\end{array}\right). 
\end{align}

The effective action for the binary system reads
\be \label{totact}
S=S_{\text{g}}+\sum_{I=1}^{2}S_{\text{(I)pp}},
\ee
where $S_{\text{g}}$ is the pure gravitational action, and $S_{\text{(I)pp}}$ is the worldline point particle action for each of the two components of the binary. The gravitational action comprises the Einstein-Hilbert action plus 
the harmonic gauge-fixing term, that is
\be
S_{\text{g}}=S_{\text{EH}} + S_{\text{GF}} 
= -\frac{1}{16\pi G} \int d^4x \sqrt{g} \,R + \frac{1}{32\pi G} \int d^4x\sqrt{g}
\,g_{\mu\nu}\Gamma^\mu\Gamma^\nu, 
\ee
where $\Gamma^\mu\equiv\Gamma^\mu_{\rho\sigma}g^{\rho\sigma}$.

From the gravitational action the propagators, and the self-gravitational 
vertices are derived. The ``NRG'' scalar, vector, and tensor field propagators, and the corresponding time dependent propagator vertices can be found, e.g., in section 2 of \cite{Levi:2015uxa}.

As for the three-graviton vertices, required for the NNLO of the 
spin-squared interaction, they are similar to those, required in \cite{Levi:2015uxa}, except for the following vertex:
\begin{align}
\label{eq:sigmaphi2} \parbox{18mm}{\includegraphics[scale=0.6]{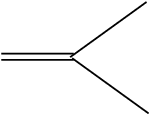}}
 & = ~ \frac{1}{16\pi G}\int  d^4x~\left(2\sigma_{ij}
 \partial_i\phi\partial_j\phi
 -\sigma_{jj}\partial_i\phi\partial_i\phi\right),
\end{align}
in which only the stationary part of the vertex is required here, unlike in the NNLO spin-orbit sector, where its time dependence should be included \cite{Levi:2015uxa}.

Also, the only four-graviton vertex, required at the NNLO level, is the stationary one, which is found in \cite{Levi:2011eq,Levi:2015uxa}.

We proceed to the point particle action of each of the spinning objects. 
The minimal coupling part of the covariant action, which is also invariant under rotational variables gauge transformations reads \cite{Levi:2015msa}
\begin{align} \label{mcact}
S_{\text{pp}}=&\int 
d\lambda\left[-m \sqrt{u^2}-\frac{1}{2} \hat{S}_{\mu\nu} \hat{\Omega}^{\mu\nu}
	-\frac{\hat{S}^{\mu\nu} p_{\nu}}{p^2} \frac{D p_{\mu}}{D \lambda}\right],
\end{align}
with the affine parameter $\lambda$, parametrized here as the coordinate time, 
i.e.~$\lambda=t=x^0$, the 4-velocity $u^{\mu}\equiv dx^\mu/d\lambda$, and 
the generic angular velocity and spin variables of the object 
$\hat{\Omega}^{\mu\nu}$, $\hat{S}_{\mu\nu}$, respectively \cite{Levi:2015msa}. 
Notice that henceforth we drop the hat notation on the rotational variables. 

Yet, in the spin-squared interaction, we consider finite size effects of the objects, i.e.~also the nonminimal coupling part of the action. Therefore, the action in eq.~\eqref{mcact} should be supplemented here with its spin-induced part, which is quadratic in the spin \cite{Levi:2015msa}. From the symmetry considerations outlined in \cite{Levi:2015msa}, in particular parity invariance, which holds for macroscopic objects in General Relativity, and the $SO(3)$ invariance of the body-fixed spatial triad, it follows that the only such operator, which appears in the action \cite{Levi:2015msa}, is given by 
\begin{align} \label{es2}
L_{ES^2} =& -\frac{C_{ES^2}}{2m} \frac{E_{\mu\nu}}{\sqrt{u^2}} 
S^{\mu} S^{\nu},
\end{align}
where we have here the Wilson coefficient $C_{ES^2}$, that is the quadrupolar deformation constant due to spin from \cite{Poisson:1997ha}, and similarly introduced in \cite{Porto:2008jj}. This coefficient equals unity in the black hole case. This LO nonminimal coupling for the spin-induced quadrupole, which also contributes the LO finite size effects, is the only operator, which contributes to the conservative spin-squared sector. We expect then to have in the spin-squared interaction both contributions from spin-squared couplings from eq.~\eqref{es2}, and from nonlinear effects of the linear in spin couplings, which arise from the minimal coupling part of the action in eq.~\eqref{mcact}, that would play a smaller role in the interaction. The mass couplings also play a smaller role in this interaction. 

We begin then with the mass couplings, where we present their expansion in the velocity, to the order required here. The one-graviton couplings to the worldline mass read
\begin{align}
\label{eq:mphi} \parbox{12mm}{\includegraphics[scale=0.6]{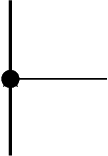}}
 & = - m \int dt~\phi~\left[1+\frac{3}{2}v^2+\frac{7}{8}v^4+\cdots\right], \\ 
\label{eq:mA} \parbox{12mm}{\includegraphics[scale=0.6]{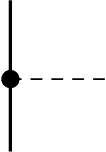}}
 & = ~m \int dt~A_iv^i~\left[1+\frac{1}{2}v^2+\cdots\right],\\ 
\label{eq:msigma} \parbox{12mm}{\includegraphics[scale=0.6]{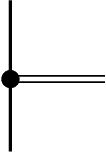}}
 & = \frac{1}{2}m \int dt~\sigma_{ij}v^iv^j~\left[1+\cdots\right].
\end{align}
with similar notations as in, e.g., \cite{Levi:2015uxa}. 

The two-graviton mass couplings, required here, read 
\begin{align}
\label{eq:mphi2}  \parbox{12mm}{\includegraphics[scale=0.6]{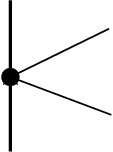}}
 & = -\frac{1}{2}m \int dt~\phi^2~\left[1-\frac{9}{2}v^2+\cdots\right], \\ 
\label{eq:mphiA}   \parbox{12mm}{\includegraphics[scale=0.6]{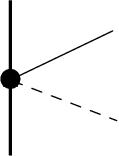}}
 & = ~~m \int dt~\phi \,A_iv^i~\biggl[1+\cdots\biggr].
\end{align}

The three-graviton mass coupling, that is require here, reads 
\begin{align}
\label{eq:mphi3}  \parbox{12mm}{\includegraphics[scale=0.6]{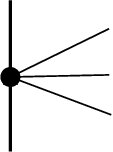}}
 & = -\frac{1}{6}m \int dt~\phi^3~\left[1+\cdots\right].
\end{align}

Next, we consider the spin couplings, required for this sector, given in the canonical gauge \cite{Levi:2015msa}. Yet, we recall the kinematic terms, which involve spin with no field coupling \cite{Levi:2015msa}, to the order required here
\be \label{frskin}
L_{\text{kin}}=-\vec{S}\cdot \vec{\Omega}
   + \frac{1}{2} \vec{S}\cdot\vec{v}\times\vec{a}\left(1+\frac{3}{4}v^2
   \right) ,  
\ee
with similar notations as in \cite{Levi:2015uxa}. 

Then, the one-graviton linear in spin couplings, required in this work, read 
\begin{align}
\label{eq:sA}  \parbox{12mm}{\includegraphics[scale=0.6]{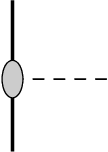}}
 & = \int dt \,\,\epsilon_{ijk}S_{k}\left(\frac{1}{2}\partial_iA_j + \frac{3}{4} v^iv^l 
 \left(\partial_lA_j-\partial_jA_l\right)+v^i\partial_tA_j\right),\\
\label{eq:sphi}   \parbox{12mm}{\includegraphics[scale=0.6]{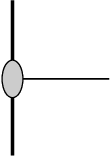}}
 & = \int dt \,\,2\epsilon_{ijk}S_{k}v^i\partial_j\phi ,\\
\label{eq:ssigma}   \parbox{12mm}{\includegraphics[scale=0.6]{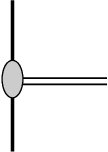}}
 & = \int dt \,\,\frac{1}{2}\epsilon_{ijk}S_{k}v^l\partial_i\sigma_{jl},
\end{align}
with similar notations as in \cite{Levi:2015uxa}. 

The two-graviton spin coupling, which we require here, reads
\begin{align}
\label{eq:sphiA}  \parbox{12mm}{\includegraphics[scale=0.6]{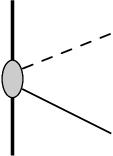}}
 & = \int dt \,\,2\epsilon_{ijk}S_{k}\partial_iA_j\phi. 
\end{align}

Finally, we turn to the spin-squared couplings from eq.~\eqref{es2}, which play the major role in this interaction.

The one-graviton spin-squared couplings read
\begin{align}
\label{eq:sqphi}   \parbox{12mm}{\includegraphics[scale=0.6]{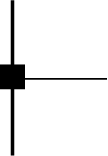}}
 & = \int dt \,\,\frac{C_{ES^2}}{2m}\left[S^{i}S^{j}\left(
 \partial_i\partial_j\phi \left(1+\frac{3}{2}v^2+\frac{7}{8}v^4\right)
 -3\partial_i\partial_k\phi\,v^jv^k\left(1+\frac{7}{12}v^2\right)
 \right.\right.\nn\\
 &\qquad\qquad\qquad\qquad\quad\left.
 -2\partial_t\partial_i\phi\,v^j
 +\frac{1}{4}\partial_k\partial_l\phi\,v^i v^j v^k v^l
 -\partial_t\partial_k\phi\, v^i v^j v^k 
 -\partial_t^2\phi\,v^i v^j\right)\nn\\
 &\qquad\qquad\qquad\quad-S^2\left(
 \partial_i\partial_i\phi \left(1+\frac{3}{2}v^2+\frac{7}{8}v^4\right)
 -\partial_i\partial_j\phi\, v^iv^j \left(1+\frac{1}{2}v^2\right)\right.\nn\\
 &\qquad\qquad\qquad\qquad\qquad\left.\left.
 +2\partial_t\partial_i\phi\,v^i\left(1+\frac{1}{2}v^2\right)
 +2\partial_t^2\phi\right)\right],\\ 
\label{eq:sqA}  \parbox{12mm}{\includegraphics[scale=0.6]{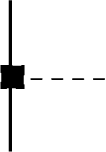}}
 & = \int dt \,\, -\frac{C_{ES^2}}{2m}\left[S^{i}S^{j}
 \left(\left(\partial_i\partial_j A_k v^k-\partial_i\partial_k A_j v^k
 -\partial_t\partial_i A_j\right)\left(1+\frac{1}{2}v^2\right)\right.\right.\nn\\
 &\left.\qquad\qquad\qquad\qquad\qquad+\frac{1}{2}\left(\partial_k
 \left(\partial_l A_i-\partial_i A_l\right) v^j v^k v^l
 +\partial_t\left(\partial_i A_k+\partial_k A_i\right) v^j v^k
 \right)\right)\nn\\
 &\left.\qquad\qquad\qquad\quad-S^2\left(\left(
 \partial_i\partial_i A_k v^k
 -\partial_i\partial_j A_i v^j
 -\partial_t\partial_iA_i\right)\left(1+\frac{1}{2}v^2\right)
 +\partial_t\partial_i A_j v^i v^j\right)\right],\\ 
\label{eq:sqsigma}  \parbox{12mm}{\includegraphics[scale=0.6]{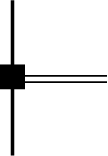}}
 & = \int dt \,\, -\frac{C_{ES^2}}{4m}\left[S^{i}S^{j}\left(
 \partial_i\partial_j \sigma_{kl} v^k v^l
 +\partial_k\partial_l\sigma_{ij} v^k v^l
 -2\partial_i\partial_k \sigma_{jl} v^k v^l\right.\right.\nn\\
 &\left.\qquad\qquad\qquad\qquad\qquad
 -2\partial_t\partial_i \sigma_{jk} v^k
 +2\partial_t\partial_k \sigma_{ij} v^k
 +\partial_t^2\sigma_{ij}\right)\nn\\
 &\qquad\qquad\qquad-S^2\left(
 \partial_k\partial_k\sigma_{ij} v^i v^j
 +\partial_i\partial_j\sigma_{kk} v^i v^j
 -2\partial_i\partial_j \sigma_{ik} v^j v^k\right.\nn\\
 &\left.\left.\qquad\qquad\qquad\qquad
 -2\partial_t\partial_i \sigma_{ij} v^j
 +2\partial_t\partial_i \sigma_{jj} v^i
 +\partial_t^2\sigma_{ii}\right)
 \right],
\end{align}
where the black square boxes represent the spin-squared operator on the 
worldlines, as in \cite{Levi:2014gsa,Levi:2015msa}. 

The two-graviton spin-squared couplings, which are required here, read
\begin{align}
\label{eq:sqphi2}   \parbox{14mm}{\includegraphics[scale=0.6]{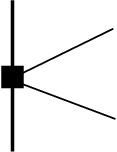}}
 & = \int dt\,\,\frac{C_{ES^2}}{2m}\left[S^{i}S^{j}\left(
 3\partial_i\phi\,\partial_j\phi\left(1+\frac{5}{6}v^2\right)
 -5 \partial_i\phi\,\partial_k\phi \,v^j v^k
 +2\left(\partial_k\phi\right)^2 v^i v^j
 \right.\right.\nn\\
 &\qquad\qquad\qquad\quad\left.
 +2 \partial_t\phi\,\partial_i\phi\, v^j
 +3\phi\,\partial_i\partial_j\phi\left(1-\frac{1}{2}v^2\right)
 +3\phi\,\partial_i\partial_k\phi \, v^j v^k
 +2\phi\,\partial_t\partial_i\phi \, v^j\right)\nn\\
 &\qquad\qquad\qquad\quad-S^2\left(\left(\partial_i\phi\right)^2\left(1+\frac{3}{2}v^2\right)
 -3\partial_i\phi\,\partial_j\phi \,v^i v^j
 -2\partial_t\phi\,\partial_i\phi \,v^i
 -4\left(\partial_t\phi\right)^2\right.\nn\\
 &\qquad\qquad\qquad\quad\left.\left.
 +3\phi\,\partial_i\partial_i\phi\left(1-\frac{1}{2}v^2\right)
 +\phi\,\partial_i\partial_j\phi\,v^i v^j
 -2\phi\,\partial_t\partial_i\phi\,v^i
 -2\phi\,\partial_t^2\phi\right)\right],\\ 
 \label{eq:sqphiA}   \parbox{14mm}{\includegraphics[scale=0.6]{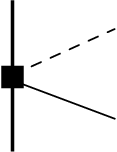}}
 & = \int dt\,\,-\frac{C_{ES^2}}{2m}\left[
 S^{i}S^{j}\left(
 \partial_i\partial_j\phi\,A_k v^k
 -2\partial_t\partial_i\phi\,A_j 
 +6\partial_i\phi\,\partial_j A_k v^k
 -6\partial_i\phi\,\partial_k A_j v^k\right.\right.\nn\\
 &\qquad\qquad\qquad\qquad\qquad
 -\partial_k\phi\,\partial_i A_k\,v^j
 +\partial_k\phi\,\partial_k A_i\,v^j
 -\partial_t\phi\,\partial_i A_j
 -4\partial_i\phi\,\partial_t A_j\nn\\
  &\left.\qquad\qquad\qquad\qquad\qquad
 +3\phi\,\partial_i\partial_j A_k v^k
 -3\phi\,\partial_i\partial_k A_j v^k 
 -3\phi\,\partial_t\partial_i A_j\right)\nn\\
 &\qquad\qquad\qquad+S^2\left(
 2\partial_t\partial_i\phi\,A_i
 -\partial_k\partial_k\phi\,A_l v^l
 -5 \partial_i\phi\,\partial_i A_j v^j
 +5 \partial_i\phi\,\partial_j A_i v^j
 +\partial_t\phi\,\partial_i A_i\right.\nn\\
 &\left.\left.\qquad\qquad\qquad\qquad\qquad
 +2\partial_i\phi\,\partial_t A_i
 -3\phi\,\partial_k\partial_k A_l v^l 
 +3\phi\,\partial_i\partial_j A_i v^j 
 +3\phi\,\partial_t\partial_i A_i\right)\right],\\
\label{eq:sqphisigma}   \parbox{14mm}{\includegraphics[scale=0.6]
 {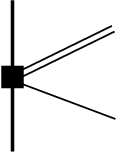}}
 & = \int dt\,\,-\frac{C_{ES^2}}{2m}\left[S^{i}S^{j}\left(\sigma_{ik}
 \partial_j\partial_k\phi+\partial_k\phi\left(\partial_i\sigma_{jk}-\frac{1}{2}\partial_k\sigma_{ij}\right)\right)\right.\nn\\
 &\qquad\qquad\qquad\qquad\left.-S^2\left(\sigma_{ij}
 \partial_i\partial_j\phi+\partial_i\phi\left(\partial_j\sigma_{ij}-\frac{1}{2}\partial_i\sigma_{jj}\right)\right)\right],\\
 \label{eq:sqAA}   \parbox{14mm}{\includegraphics[scale=0.6]
 	{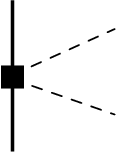}}
 & = \int dt\,\,\frac{C_{ES^2}}{8m}\left[S^{i}S^{j}\left(
 \partial_k A_i\partial_k A_j +\partial_i A_k \partial_j A_k 
 -2\partial_i A_k\partial_k A_j \right)\right.\nn\\
 &\qquad\qquad\qquad\left.-2S^2\left(\left(\partial_i A_j\right)^2-\partial_i A_j \partial_j A_i\right)\right].
\end{align}

Finally, the required three-graviton spin-squared coupling reads
\begin{align}
\label{eq:sqphi3}   \parbox{14mm}{\includegraphics[scale=0.6]{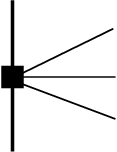}}
 & = \int dt\,\,\frac{3C_{ES^2}}{2m}\left[
 3S^{i} S^{j} \left(\phi\,\partial_i\phi\,\partial_j\phi
 +\frac{1}{2}\phi^2\,\partial_i\,\partial_j\phi\right)
 -S^2\left(\phi\left(\partial_i\phi\right)^2+\frac{3}{2}\phi^2\partial_i\partial_i\phi\right)\right].
\end{align}

Note the complexity of the spin-squared couplings with respect to the other worldline couplings.

\section{Next-to-next-to-leading order spin-squared interaction} \label{ssfd}

In this section we present the perturbative expansion of the NNLO spin-squared two-body effective action in terms of Feynman diagrams, and we provide the value for each diagram. At the NNLO level, which is up to order 
$G^3$, the classification of the Feynman diagrams is similar to that in \cite{Levi:2011eq,Levi:2015uxa}, as can be seen in figures \ref{fig:ssnnlo1g}-\ref{fig:ssnnlog3twoloop} below. There is a total of 64 diagrams, which make up the NNLO spin-squared interaction. We note that whereas the linear interaction involves exclusively the spin-squared couplings, as we go up in the nonlinearity of the interaction, that is in the loop order, but even at the two-graviton exchange diagrams, we start to have more diagrams, which involve the linear in spin couplings. Lastly, we also have take into account higher order time derivatives, which appear in the 2PN order \cite{Gilmore:2008gq}, and in up to NLO spin-orbit, and spin-squared \cite{Levi:2015msa} sectors. We refer the reader to \cite{Levi:2010zu, Levi:2011eq, Levi:2015uxa} for further details on the generation of the Feynman diagrams from the Feynman rules, and the evaluation of such Feynman diagrams. In particular, we also refer to \cite{Levi:2015uxa} for the conventions and notations, used here.

\subsection{One-graviton exchange}

\begin{figure}[t]
\includegraphics[scale=0.93]{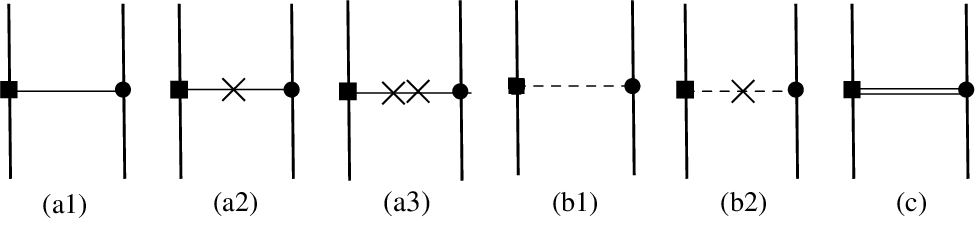}
\caption{NNLO spin-squared one-graviton exchange Feynman diagrams.}
\label{fig:ssnnlo1g}
\end{figure}

There are 6 one-graviton exchange diagrams in this sector, which can be seen in figure \ref{fig:ssnnlo1g}. We note that all diagrams here involve only the spin-squared couplings, and add up to the ones, which already entered in the NLO spin-squared sector \cite{Levi:2015msa}. Here we require tensor Fourier integrals of up to order 6 as in \cite{Levi:2011eq}, see appendix~A. Our treatment of time derivatives, and higher order time derivative terms is similar to that in \cite{Levi:2014sba,Levi:2015msa,Levi:2015uxa}. 

These diagrams are evaluated by: 
\begin{align}
\text{Fig.~1(a1)} =&
\frac{G C_{1ES^2} m_{2}}{2 m_{1} r^3} \Big[ S_{1}^2
	 - 3 ( \vec{S}_{1}\cdot\vec{n} )^{2} \Big]
+ \frac{G C_{1ES^2} m_{2}}{4 m_{1} r^3} \Big[ 5 S_{1}^2 v_{1}^2
	 + 3 S_{1}^2 v_{2}^2
	 - 2 ( \vec{S}_{1}\cdot\vec{v}_{1} )^{2} \nl
	 + 6 \vec{S}_{1}\cdot\vec{n} \vec{v}_{1}\cdot\vec{n} \vec{S}_{1}\cdot\vec{v}_{1}
	 - 9 v_{1}^2 ( \vec{S}_{1}\cdot\vec{n} )^{2}
	 - 9 v_{2}^2 ( \vec{S}_{1}\cdot\vec{n} )^{2}
	 - 6 S_{1}^2 ( \vec{v}_{1}\cdot\vec{n} )^{2} \Big] \nl
+ \frac{G C_{1ES^2} m_{2}}{m_{1} r^2} \vec{S}_{1}\cdot\vec{a}_{1} \vec{S}_{1}\cdot\vec{n}
- \frac{G C_{1ES^2} m_{2}}{m_{1} r^2} \Big[ 2 \vec{S}_{1}\cdot\dot{\vec{S}}_{1} \vec{v}_{1}\cdot\vec{n}
	 - \dot{\vec{S}}_{1}\cdot\vec{n} \vec{S}_{1}\cdot\vec{v}_{1} \nl
	 - \vec{S}_{1}\cdot\vec{n} \dot{\vec{S}}_{1}\cdot\vec{v}_{1} \Big]
+ \frac{G C_{1ES^2} m_{2}}{m_{1} r} \ddot{S^2_{1}}
+ \frac{G C_{1ES^2} m_{2}}{16 m_{1} r^3} \Big[ 30 S_{1}^2 v_{1}^2 v_{2}^2
	 - 12 v_{1}^2 ( \vec{S}_{1}\cdot\vec{v}_{1} )^{2} \nl
	 - 12 v_{2}^2 ( \vec{S}_{1}\cdot\vec{v}_{1} )^{2}
	 + 19 S_{1}^2 v_{1}^{4}
	 + 7 S_{1}^2 v_{2}^{4}
	 + 42 \vec{S}_{1}\cdot\vec{n} \vec{v}_{1}\cdot\vec{n} \vec{S}_{1}\cdot\vec{v}_{1} v_{1}^2 \nl
	 + 36 \vec{S}_{1}\cdot\vec{n} \vec{v}_{1}\cdot\vec{n} \vec{S}_{1}\cdot\vec{v}_{1} v_{2}^2
	 - 54 v_{1}^2 v_{2}^2 ( \vec{S}_{1}\cdot\vec{n} )^{2}
	 - 6 ( \vec{v}_{1}\cdot\vec{n} )^{2} ( \vec{S}_{1}\cdot\vec{v}_{1} )^{2} \nl
	 - 36 S_{1}^2 v_{1}^2 ( \vec{v}_{1}\cdot\vec{n} )^{2}
	 - 36 S_{1}^2 v_{2}^2 ( \vec{v}_{1}\cdot\vec{n} )^{2}
	 - 21 ( \vec{S}_{1}\cdot\vec{n} )^{2} v_{1}^{4}
	 - 21 ( \vec{S}_{1}\cdot\vec{n} )^{2} v_{2}^{4} \Big] \nl
+ \frac{G C_{1ES^2} m_{2}}{2 m_{1} r^2} \Big[ S_{1}^2 v_{1}^2 \vec{a}_{1}\cdot\vec{n}
	 - 2 \vec{v}_{1}\cdot\vec{n} \vec{S}_{1}\cdot\vec{v}_{1} \vec{S}_{1}\cdot\vec{a}_{1}
	 + 2 S_{1}^2 \vec{v}_{1}\cdot\vec{n} \vec{v}_{1}\cdot\vec{a}_{1} \nl
	 + 3 \vec{S}_{1}\cdot\vec{n} \vec{S}_{1}\cdot\vec{a}_{1} v_{2}^2 \Big]
- \frac{G C_{1ES^2} m_{2}}{2 m_{1} r^2} \Big[ 2 \vec{v}_{1}\cdot\vec{n} \vec{S}_{1}\cdot\vec{v}_{1} \dot{\vec{S}}_{1}\cdot\vec{v}_{1}
	 - 2 \vec{S}_{1}\cdot\dot{\vec{S}}_{1} \vec{v}_{1}\cdot\vec{n} v_{1}^2 \nl
	 + 6 \vec{S}_{1}\cdot\dot{\vec{S}}_{1} \vec{v}_{1}\cdot\vec{n} v_{2}^2
	 - 3 \dot{\vec{S}}_{1}\cdot\vec{n} \vec{S}_{1}\cdot\vec{v}_{1} v_{2}^2
	 - 3 \vec{S}_{1}\cdot\vec{n} \dot{\vec{S}}_{1}\cdot\vec{v}_{1} v_{2}^2 \Big] \nl
+ \frac{G C_{1ES^2} m_{2}}{2 m_{1} r} \Big[ 2 \vec{S}_{1}\cdot\vec{v}_{1} \vec{S}_{1}\cdot\dot{\vec{a}}_{1}
	 + \big( 2 \vec{S}_{1}\cdot\vec{v}_{1} \ddot{\vec{S}}_{1}\cdot\vec{v}_{1}
	 - 2 \vec{S}_{1}\cdot\ddot{\vec{S}}_{1} v_{1}^2
	 + v_{1}^2 \ddot{S^2_{1}}
	 + 3 v_{2}^2 \ddot{S^2_{1}} \big) \nl
	 + 2 ( \vec{S}_{1}\cdot\vec{a}_{1} )^{2}
	 + 4 \big( \dot{\vec{S}}_{1}\cdot\vec{v}_{1} \vec{S}_{1}\cdot\vec{a}_{1}
	 + \vec{S}_{1}\cdot\vec{v}_{1} \dot{\vec{S}}_{1}\cdot\vec{a}_{1}
	 - 2 \vec{S}_{1}\cdot\dot{\vec{S}}_{1} \vec{v}_{1}\cdot\vec{a}_{1}
	 + \vec{v}_{1}\cdot\vec{a}_{1} \dot{S^2_{1}} \big) \nl
	 - 2 \big( \dot{S}_{1}^2 v_{1}^2
	 - ( \dot{\vec{S}}_{1}\cdot\vec{v}_{1} )^{2} \big) \Big]\\
\text{Fig.~1(a2)} =&
\frac{G C_{1ES^2} m_{2}}{4 m_{1} r^3} \Big[ 2 \vec{S}_{1}\cdot\vec{v}_{1} \vec{S}_{1}\cdot\vec{v}_{2}
	 - S_{1}^2 \vec{v}_{1}\cdot\vec{v}_{2}
	 + 3 S_{1}^2 \vec{v}_{1}\cdot\vec{n} \vec{v}_{2}\cdot\vec{n}
	 - 6 \vec{S}_{1}\cdot\vec{n} \vec{S}_{1}\cdot\vec{v}_{1} \vec{v}_{2}\cdot\vec{n} \nl
	 - 6 \vec{S}_{1}\cdot\vec{n} \vec{v}_{1}\cdot\vec{n} \vec{S}_{1}\cdot\vec{v}_{2}
	 - 3 \vec{v}_{1}\cdot\vec{v}_{2} ( \vec{S}_{1}\cdot\vec{n} )^{2}
	 + 15 \vec{v}_{1}\cdot\vec{n} \vec{v}_{2}\cdot\vec{n} ( \vec{S}_{1}\cdot\vec{n} )^{2} \Big] \nl
- \frac{G C_{1ES^2} m_{2}}{2 m_{1} r^2} \Big[ \vec{S}_{1}\cdot\dot{\vec{S}}_{1} \vec{v}_{2}\cdot\vec{n}
	 - \dot{\vec{S}}_{1}\cdot\vec{n} \vec{S}_{1}\cdot\vec{v}_{2}
	 - \vec{S}_{1}\cdot\vec{n} \dot{\vec{S}}_{1}\cdot\vec{v}_{2}
	 + 3 \vec{S}_{1}\cdot\vec{n} \dot{\vec{S}}_{1}\cdot\vec{n} \vec{v}_{2}\cdot\vec{n} \Big] \nl
+ \frac{G C_{1ES^2} m_{2}}{8 m_{1} r^3} \Big[ 4 \vec{S}_{1}\cdot\vec{v}_{1} v_{1}^2 \vec{S}_{1}\cdot\vec{v}_{2}
	 + 3 S_{1}^2 v_{1}^2 \vec{v}_{1}\cdot\vec{v}_{2}
	 + 6 \vec{S}_{1}\cdot\vec{v}_{1} \vec{S}_{1}\cdot\vec{v}_{2} v_{2}^2
	 - 3 S_{1}^2 \vec{v}_{1}\cdot\vec{v}_{2} v_{2}^2 \nl
	 - 4 \vec{v}_{1}\cdot\vec{v}_{2} ( \vec{S}_{1}\cdot\vec{v}_{1} )^{2}
	 - 9 S_{1}^2 \vec{v}_{1}\cdot\vec{n} v_{1}^2 \vec{v}_{2}\cdot\vec{n}
	 - 12 \vec{S}_{1}\cdot\vec{n} \vec{S}_{1}\cdot\vec{v}_{1} v_{1}^2 \vec{v}_{2}\cdot\vec{n} \nl
	 - 18 \vec{S}_{1}\cdot\vec{n} \vec{v}_{1}\cdot\vec{n} v_{1}^2 \vec{S}_{1}\cdot\vec{v}_{2}
	 + 12 \vec{S}_{1}\cdot\vec{n} \vec{v}_{1}\cdot\vec{n} \vec{S}_{1}\cdot\vec{v}_{1} \vec{v}_{1}\cdot\vec{v}_{2}
	 + 9 S_{1}^2 \vec{v}_{1}\cdot\vec{n} \vec{v}_{2}\cdot\vec{n} v_{2}^2 \nl
	 - 18 \vec{S}_{1}\cdot\vec{n} \vec{S}_{1}\cdot\vec{v}_{1} \vec{v}_{2}\cdot\vec{n} v_{2}^2
	 - 18 \vec{S}_{1}\cdot\vec{n} \vec{v}_{1}\cdot\vec{n} \vec{S}_{1}\cdot\vec{v}_{2} v_{2}^2
	 - 9 v_{1}^2 \vec{v}_{1}\cdot\vec{v}_{2} ( \vec{S}_{1}\cdot\vec{n} )^{2} \nl
	 - 9 \vec{v}_{1}\cdot\vec{v}_{2} v_{2}^2 ( \vec{S}_{1}\cdot\vec{n} )^{2}
	 + 12 \vec{v}_{1}\cdot\vec{n} \vec{v}_{2}\cdot\vec{n} ( \vec{S}_{1}\cdot\vec{v}_{1} )^{2}
	 + 6 \vec{S}_{1}\cdot\vec{v}_{1} \vec{S}_{1}\cdot\vec{v}_{2} ( \vec{v}_{1}\cdot\vec{n} )^{2} \nl
	 - 18 S_{1}^2 \vec{v}_{1}\cdot\vec{v}_{2} ( \vec{v}_{1}\cdot\vec{n} )^{2}
	 + 45 \vec{v}_{1}\cdot\vec{n} v_{1}^2 \vec{v}_{2}\cdot\vec{n} ( \vec{S}_{1}\cdot\vec{n} )^{2}
	 + 45 \vec{v}_{1}\cdot\vec{n} \vec{v}_{2}\cdot\vec{n} v_{2}^2 ( \vec{S}_{1}\cdot\vec{n} )^{2} \nl
	 + 30 S_{1}^2 \vec{v}_{2}\cdot\vec{n} ( \vec{v}_{1}\cdot\vec{n} )^{3}
	 - 30 \vec{S}_{1}\cdot\vec{n} \vec{S}_{1}\cdot\vec{v}_{1} \vec{v}_{2}\cdot\vec{n} ( \vec{v}_{1}\cdot\vec{n} )^{2} \Big] \nl
- \frac{G C_{1ES^2} m_{2}}{4 m_{1} r^2} \Big[ S_{1}^2 \vec{v}_{1}\cdot\vec{a}_{1} \vec{v}_{2}\cdot\vec{n}
	 + \vec{S}_{1}\cdot\vec{v}_{1} \vec{a}_{1}\cdot\vec{n} \vec{S}_{1}\cdot\vec{v}_{2}
	 - \vec{v}_{1}\cdot\vec{n} \vec{S}_{1}\cdot\vec{a}_{1} \vec{S}_{1}\cdot\vec{v}_{2} \nl
	 - 6 \vec{S}_{1}\cdot\vec{n} \vec{v}_{1}\cdot\vec{a}_{1} \vec{S}_{1}\cdot\vec{v}_{2}
	 - 2 S_{1}^2 \vec{a}_{1}\cdot\vec{n} \vec{v}_{1}\cdot\vec{v}_{2}
	 - \vec{S}_{1}\cdot\vec{n} \vec{S}_{1}\cdot\vec{a}_{1} \vec{v}_{1}\cdot\vec{v}_{2}
	 - 2 S_{1}^2 \vec{v}_{1}\cdot\vec{n} \vec{a}_{1}\cdot\vec{v}_{2} \nl
	 + \vec{S}_{1}\cdot\vec{n} \vec{S}_{1}\cdot\vec{v}_{1} \vec{a}_{1}\cdot\vec{v}_{2}
	 - 3 S_{1}^2 \vec{v}_{1}\cdot\vec{n} \vec{v}_{2}\cdot\vec{a}_{2}
	 + 6 \vec{S}_{1}\cdot\vec{n} \vec{S}_{1}\cdot\vec{v}_{1} \vec{v}_{2}\cdot\vec{a}_{2} \nl
	 + 6 S_{1}^2 \vec{v}_{1}\cdot\vec{n} \vec{a}_{1}\cdot\vec{n} \vec{v}_{2}\cdot\vec{n}
	 - 3 \vec{S}_{1}\cdot\vec{n} \vec{S}_{1}\cdot\vec{v}_{1} \vec{a}_{1}\cdot\vec{n} \vec{v}_{2}\cdot\vec{n}
	 + 3 \vec{S}_{1}\cdot\vec{n} \vec{v}_{1}\cdot\vec{n} \vec{S}_{1}\cdot\vec{a}_{1} \vec{v}_{2}\cdot\vec{n} \nl
	 + 9 \vec{v}_{1}\cdot\vec{a}_{1} \vec{v}_{2}\cdot\vec{n} ( \vec{S}_{1}\cdot\vec{n} )^{2}
	 - 9 \vec{v}_{1}\cdot\vec{n} \vec{v}_{2}\cdot\vec{a}_{2} ( \vec{S}_{1}\cdot\vec{n} )^{2} \Big] \nl
+ \frac{G C_{1ES^2} m_{2}}{4 m_{1} r^2} \Big[ 2 \vec{S}_{1}\cdot\vec{v}_{1} \dot{\vec{S}}_{1}\cdot\vec{v}_{1} \vec{v}_{2}\cdot\vec{n}
	 - 5 \vec{S}_{1}\cdot\dot{\vec{S}}_{1} v_{1}^2 \vec{v}_{2}\cdot\vec{n}
	 + \vec{v}_{1}\cdot\vec{n} \dot{\vec{S}}_{1}\cdot\vec{v}_{1} \vec{S}_{1}\cdot\vec{v}_{2} \nl
	 + 3 \dot{\vec{S}}_{1}\cdot\vec{n} v_{1}^2 \vec{S}_{1}\cdot\vec{v}_{2}
	 + \vec{v}_{1}\cdot\vec{n} \vec{S}_{1}\cdot\vec{v}_{1} \dot{\vec{S}}_{1}\cdot\vec{v}_{2}
	 + 3 \vec{S}_{1}\cdot\vec{n} v_{1}^2 \dot{\vec{S}}_{1}\cdot\vec{v}_{2} \nl
	 - 4 \vec{S}_{1}\cdot\dot{\vec{S}}_{1} \vec{v}_{1}\cdot\vec{n} \vec{v}_{1}\cdot\vec{v}_{2}
	 + \dot{\vec{S}}_{1}\cdot\vec{n} \vec{S}_{1}\cdot\vec{v}_{1} \vec{v}_{1}\cdot\vec{v}_{2}
	 + \vec{S}_{1}\cdot\vec{n} \dot{\vec{S}}_{1}\cdot\vec{v}_{1} \vec{v}_{1}\cdot\vec{v}_{2}
	 - 3 \vec{S}_{1}\cdot\dot{\vec{S}}_{1} \vec{v}_{2}\cdot\vec{n} v_{2}^2 \nl
	 + 3 \dot{\vec{S}}_{1}\cdot\vec{n} \vec{S}_{1}\cdot\vec{v}_{2} v_{2}^2
	 + 3 \vec{S}_{1}\cdot\vec{n} \dot{\vec{S}}_{1}\cdot\vec{v}_{2} v_{2}^2
	 - 3 \dot{\vec{S}}_{1}\cdot\vec{n} \vec{v}_{1}\cdot\vec{n} \vec{S}_{1}\cdot\vec{v}_{1} \vec{v}_{2}\cdot\vec{n} \nl
	 - 3 \vec{S}_{1}\cdot\vec{n} \vec{v}_{1}\cdot\vec{n} \dot{\vec{S}}_{1}\cdot\vec{v}_{1} \vec{v}_{2}\cdot\vec{n}
	 - 9 \vec{S}_{1}\cdot\vec{n} \dot{\vec{S}}_{1}\cdot\vec{n} v_{1}^2 \vec{v}_{2}\cdot\vec{n}
	 - 9 \vec{S}_{1}\cdot\vec{n} \dot{\vec{S}}_{1}\cdot\vec{n} \vec{v}_{2}\cdot\vec{n} v_{2}^2 \nl
	 + 6 \vec{S}_{1}\cdot\dot{\vec{S}}_{1} \vec{v}_{2}\cdot\vec{n} ( \vec{v}_{1}\cdot\vec{n} )^{2} \Big]
- \frac{G C_{1ES^2} m_{2}}{2 m_{1} r} \Big[ \big( \vec{S}_{1}\cdot\dot{\vec{a}}_{1} \vec{S}_{1}\cdot\vec{v}_{2}
	 - \vec{S}_{1}\cdot\vec{n} \vec{S}_{1}\cdot\dot{\vec{a}}_{1} \vec{v}_{2}\cdot\vec{n} \big) \nl
	 + \big( \ddot{\vec{S}}_{1}\cdot\vec{v}_{1} \vec{S}_{1}\cdot\vec{v}_{2}
	 + \vec{S}_{1}\cdot\vec{v}_{1} \ddot{\vec{S}}_{1}\cdot\vec{v}_{2}
	 - 2 \vec{S}_{1}\cdot\ddot{\vec{S}}_{1} \vec{v}_{1}\cdot\vec{v}_{2}
	 - \vec{v}_{1}\cdot\vec{v}_{2} \ddot{S^2_{1}}
	 + 2 \vec{S}_{1}\cdot\ddot{\vec{S}}_{1} \vec{v}_{1}\cdot\vec{n} \vec{v}_{2}\cdot\vec{n} \nl
	 - \ddot{\vec{S}}_{1}\cdot\vec{n} \vec{S}_{1}\cdot\vec{v}_{1} \vec{v}_{2}\cdot\vec{n}
	 - \vec{S}_{1}\cdot\vec{n} \ddot{\vec{S}}_{1}\cdot\vec{v}_{1} \vec{v}_{2}\cdot\vec{n}
	 + \vec{v}_{1}\cdot\vec{n} \vec{v}_{2}\cdot\vec{n} \ddot{S^2_{1}} \big)
	 + \big( 2 \dot{\vec{S}}_{1}\cdot\vec{a}_{1} \vec{S}_{1}\cdot\vec{v}_{2} \nl
	 + 2 \vec{S}_{1}\cdot\vec{a}_{1} \dot{\vec{S}}_{1}\cdot\vec{v}_{2}
	 - 4 \vec{S}_{1}\cdot\dot{\vec{S}}_{1} \vec{a}_{1}\cdot\vec{v}_{2}
	 + 3 \vec{S}_{1}\cdot\dot{\vec{S}}_{1} \vec{v}_{2}\cdot\vec{a}_{2}
	 + \vec{a}_{1}\cdot\vec{v}_{2} \dot{S^2_{1}} \nl
	 + 4 \vec{S}_{1}\cdot\dot{\vec{S}}_{1} \vec{a}_{1}\cdot\vec{n} \vec{v}_{2}\cdot\vec{n}
	 - 2 \dot{\vec{S}}_{1}\cdot\vec{n} \vec{S}_{1}\cdot\vec{a}_{1} \vec{v}_{2}\cdot\vec{n}
	 - 2 \vec{S}_{1}\cdot\vec{n} \dot{\vec{S}}_{1}\cdot\vec{a}_{1} \vec{v}_{2}\cdot\vec{n} \nl
	 + 3 \vec{S}_{1}\cdot\vec{n} \dot{\vec{S}}_{1}\cdot\vec{n} \vec{v}_{2}\cdot\vec{a}_{2}
	 - \vec{a}_{1}\cdot\vec{n} \vec{v}_{2}\cdot\vec{n} \dot{S^2_{1}} \big)
	 + 2 \big( \dot{\vec{S}}_{1}\cdot\vec{v}_{1} \dot{\vec{S}}_{1}\cdot\vec{v}_{2}
	 - \dot{S}_{1}^2 \vec{v}_{1}\cdot\vec{v}_{2} \nl
	 + \dot{S}_{1}^2 \vec{v}_{1}\cdot\vec{n} \vec{v}_{2}\cdot\vec{n}
	 - \dot{\vec{S}}_{1}\cdot\vec{n} \dot{\vec{S}}_{1}\cdot\vec{v}_{1} \vec{v}_{2}\cdot\vec{n} \big) \Big]
+ \frac{G C_{1ES^2} m_{2}}{2 m_{1}} \stackrel{(3)}{S_{1}^2} \vec{v}_{2}\cdot\vec{n} \\
\text{Fig.~1(a3)} =&
\frac{G C_{1ES^2} m_{2}}{16 m_{1} r^3} \Big[ 8 \vec{S}_{1}\cdot\vec{v}_{1} \vec{S}_{1}\cdot\vec{v}_{2} \vec{v}_{1}\cdot\vec{v}_{2}
	 - 3 S_{1}^2 v_{1}^2 v_{2}^2
	 + 2 v_{2}^2 ( \vec{S}_{1}\cdot\vec{v}_{1} )^{2}
	 + 2 v_{1}^2 ( \vec{S}_{1}\cdot\vec{v}_{2} )^{2} \nl
	 - 6 S_{1}^2 ( \vec{v}_{1}\cdot\vec{v}_{2} )^{2}
	 - 24 \vec{v}_{1}\cdot\vec{n} \vec{S}_{1}\cdot\vec{v}_{1} \vec{v}_{2}\cdot\vec{n} \vec{S}_{1}\cdot\vec{v}_{2}
	 - 12 \vec{S}_{1}\cdot\vec{n} v_{1}^2 \vec{v}_{2}\cdot\vec{n} \vec{S}_{1}\cdot\vec{v}_{2} \nl
	 + 36 S_{1}^2 \vec{v}_{1}\cdot\vec{n} \vec{v}_{2}\cdot\vec{n} \vec{v}_{1}\cdot\vec{v}_{2}
	 - 24 \vec{S}_{1}\cdot\vec{n} \vec{S}_{1}\cdot\vec{v}_{1} \vec{v}_{2}\cdot\vec{n} \vec{v}_{1}\cdot\vec{v}_{2}
	 - 24 \vec{S}_{1}\cdot\vec{n} \vec{v}_{1}\cdot\vec{n} \vec{S}_{1}\cdot\vec{v}_{2} \vec{v}_{1}\cdot\vec{v}_{2} \nl
	 - 12 \vec{S}_{1}\cdot\vec{n} \vec{v}_{1}\cdot\vec{n} \vec{S}_{1}\cdot\vec{v}_{1} v_{2}^2
	 - 3 v_{1}^2 v_{2}^2 ( \vec{S}_{1}\cdot\vec{n} )^{2}
	 + 9 S_{1}^2 v_{2}^2 ( \vec{v}_{1}\cdot\vec{n} )^{2}
	 - 6 ( \vec{S}_{1}\cdot\vec{v}_{1} )^{2} ( \vec{v}_{2}\cdot\vec{n} )^{2} \nl
	 + 9 S_{1}^2 v_{1}^2 ( \vec{v}_{2}\cdot\vec{n} )^{2}
	 - 6 ( \vec{v}_{1}\cdot\vec{n} )^{2} ( \vec{S}_{1}\cdot\vec{v}_{2} )^{2}
	 - 6 ( \vec{S}_{1}\cdot\vec{n} )^{2} ( \vec{v}_{1}\cdot\vec{v}_{2} )^{2} \nl
	 + 60 \vec{v}_{1}\cdot\vec{n} \vec{v}_{2}\cdot\vec{n} \vec{v}_{1}\cdot\vec{v}_{2} ( \vec{S}_{1}\cdot\vec{n} )^{2}
	 + 60 \vec{S}_{1}\cdot\vec{n} \vec{v}_{2}\cdot\vec{n} \vec{S}_{1}\cdot\vec{v}_{2} ( \vec{v}_{1}\cdot\vec{n} )^{2} \nl
	 + 15 v_{2}^2 ( \vec{S}_{1}\cdot\vec{n} )^{2} ( \vec{v}_{1}\cdot\vec{n} )^{2}
	 - 45 S_{1}^2 ( \vec{v}_{1}\cdot\vec{n} )^{2} ( \vec{v}_{2}\cdot\vec{n} )^{2}
	 + 60 \vec{S}_{1}\cdot\vec{n} \vec{v}_{1}\cdot\vec{n} \vec{S}_{1}\cdot\vec{v}_{1} ( \vec{v}_{2}\cdot\vec{n} )^{2} \nl
	 + 15 v_{1}^2 ( \vec{S}_{1}\cdot\vec{n} )^{2} ( \vec{v}_{2}\cdot\vec{n} )^{2}
	 - 105 ( \vec{S}_{1}\cdot\vec{n} )^{2} ( \vec{v}_{1}\cdot\vec{n} )^{2} ( \vec{v}_{2}\cdot\vec{n} )^{2} \Big] \nl
+ \frac{G C_{1ES^2} m_{2}}{16 m_{1} r^2} \Big[ 4 \vec{S}_{1}\cdot\vec{a}_{1} \vec{v}_{2}\cdot\vec{n} \vec{S}_{1}\cdot\vec{v}_{2}
	 - 6 S_{1}^2 \vec{v}_{2}\cdot\vec{n} \vec{a}_{1}\cdot\vec{v}_{2}
	 + 4 \vec{S}_{1}\cdot\vec{n} \vec{S}_{1}\cdot\vec{v}_{2} \vec{a}_{1}\cdot\vec{v}_{2} \nl
	 - 3 S_{1}^2 \vec{a}_{1}\cdot\vec{n} v_{2}^2
	 + 2 \vec{S}_{1}\cdot\vec{n} \vec{S}_{1}\cdot\vec{a}_{1} v_{2}^2
	 + 3 S_{1}^2 v_{1}^2 \vec{a}_{2}\cdot\vec{n}
	 - 4 \vec{v}_{1}\cdot\vec{n} \vec{S}_{1}\cdot\vec{v}_{1} \vec{S}_{1}\cdot\vec{a}_{2} \nl
	 - 2 \vec{S}_{1}\cdot\vec{n} v_{1}^2 \vec{S}_{1}\cdot\vec{a}_{2}
	 + 6 S_{1}^2 \vec{v}_{1}\cdot\vec{n} \vec{v}_{1}\cdot\vec{a}_{2}
	 - 4 \vec{S}_{1}\cdot\vec{n} \vec{S}_{1}\cdot\vec{v}_{1} \vec{v}_{1}\cdot\vec{a}_{2}
	 - 2 \vec{a}_{2}\cdot\vec{n} ( \vec{S}_{1}\cdot\vec{v}_{1} )^{2} \nl
	 + 2 \vec{a}_{1}\cdot\vec{n} ( \vec{S}_{1}\cdot\vec{v}_{2} )^{2}
	 - 12 \vec{S}_{1}\cdot\vec{n} \vec{a}_{1}\cdot\vec{n} \vec{v}_{2}\cdot\vec{n} \vec{S}_{1}\cdot\vec{v}_{2}
	 + 12 \vec{S}_{1}\cdot\vec{n} \vec{v}_{1}\cdot\vec{n} \vec{S}_{1}\cdot\vec{v}_{1} \vec{a}_{2}\cdot\vec{n} \nl
	 - 6 \vec{v}_{2}\cdot\vec{n} \vec{a}_{1}\cdot\vec{v}_{2} ( \vec{S}_{1}\cdot\vec{n} )^{2}
	 - 3 \vec{a}_{1}\cdot\vec{n} v_{2}^2 ( \vec{S}_{1}\cdot\vec{n} )^{2}
	 + 3 v_{1}^2 \vec{a}_{2}\cdot\vec{n} ( \vec{S}_{1}\cdot\vec{n} )^{2} \nl
	 + 6 \vec{v}_{1}\cdot\vec{n} \vec{v}_{1}\cdot\vec{a}_{2} ( \vec{S}_{1}\cdot\vec{n} )^{2}
	 - 9 S_{1}^2 \vec{a}_{2}\cdot\vec{n} ( \vec{v}_{1}\cdot\vec{n} )^{2}
	 + 6 \vec{S}_{1}\cdot\vec{n} \vec{S}_{1}\cdot\vec{a}_{2} ( \vec{v}_{1}\cdot\vec{n} )^{2} \nl
	 + 9 S_{1}^2 \vec{a}_{1}\cdot\vec{n} ( \vec{v}_{2}\cdot\vec{n} )^{2}
	 - 6 \vec{S}_{1}\cdot\vec{n} \vec{S}_{1}\cdot\vec{a}_{1} ( \vec{v}_{2}\cdot\vec{n} )^{2}
	 - 15 \vec{a}_{2}\cdot\vec{n} ( \vec{S}_{1}\cdot\vec{n} )^{2} ( \vec{v}_{1}\cdot\vec{n} )^{2} \nl
	 + 15 \vec{a}_{1}\cdot\vec{n} ( \vec{S}_{1}\cdot\vec{n} )^{2} ( \vec{v}_{2}\cdot\vec{n} )^{2} \Big]
+ \frac{G C_{1ES^2} m_{2}}{4 m_{1} r^2} \Big[ 2 \dot{\vec{S}}_{1}\cdot\vec{v}_{1} \vec{v}_{2}\cdot\vec{n} \vec{S}_{1}\cdot\vec{v}_{2} \nl
	 + 2 \vec{S}_{1}\cdot\vec{v}_{1} \vec{v}_{2}\cdot\vec{n} \dot{\vec{S}}_{1}\cdot\vec{v}_{2}
	 + 2 \vec{v}_{1}\cdot\vec{n} \vec{S}_{1}\cdot\vec{v}_{2} \dot{\vec{S}}_{1}\cdot\vec{v}_{2}
	 - 6 \vec{S}_{1}\cdot\dot{\vec{S}}_{1} \vec{v}_{2}\cdot\vec{n} \vec{v}_{1}\cdot\vec{v}_{2} \nl
	 + 2 \dot{\vec{S}}_{1}\cdot\vec{n} \vec{S}_{1}\cdot\vec{v}_{2} \vec{v}_{1}\cdot\vec{v}_{2}
	 + 2 \vec{S}_{1}\cdot\vec{n} \dot{\vec{S}}_{1}\cdot\vec{v}_{2} \vec{v}_{1}\cdot\vec{v}_{2}
	 - 3 \vec{S}_{1}\cdot\dot{\vec{S}}_{1} \vec{v}_{1}\cdot\vec{n} v_{2}^2
	 + \dot{\vec{S}}_{1}\cdot\vec{n} \vec{S}_{1}\cdot\vec{v}_{1} v_{2}^2 \nl
	 + \vec{S}_{1}\cdot\vec{n} \dot{\vec{S}}_{1}\cdot\vec{v}_{1} v_{2}^2
	 - 6 \dot{\vec{S}}_{1}\cdot\vec{n} \vec{v}_{1}\cdot\vec{n} \vec{v}_{2}\cdot\vec{n} \vec{S}_{1}\cdot\vec{v}_{2}
	 - 6 \vec{S}_{1}\cdot\vec{n} \vec{v}_{1}\cdot\vec{n} \vec{v}_{2}\cdot\vec{n} \dot{\vec{S}}_{1}\cdot\vec{v}_{2} \nl
	 - 6 \vec{S}_{1}\cdot\vec{n} \dot{\vec{S}}_{1}\cdot\vec{n} \vec{v}_{2}\cdot\vec{n} \vec{v}_{1}\cdot\vec{v}_{2}
	 - 3 \vec{S}_{1}\cdot\vec{n} \dot{\vec{S}}_{1}\cdot\vec{n} \vec{v}_{1}\cdot\vec{n} v_{2}^2
	 + 9 \vec{S}_{1}\cdot\dot{\vec{S}}_{1} \vec{v}_{1}\cdot\vec{n} ( \vec{v}_{2}\cdot\vec{n} )^{2} \nl
	 - 3 \dot{\vec{S}}_{1}\cdot\vec{n} \vec{S}_{1}\cdot\vec{v}_{1} ( \vec{v}_{2}\cdot\vec{n} )^{2}
	 - 3 \vec{S}_{1}\cdot\vec{n} \dot{\vec{S}}_{1}\cdot\vec{v}_{1} ( \vec{v}_{2}\cdot\vec{n} )^{2}
	 + 15 \vec{S}_{1}\cdot\vec{n} \dot{\vec{S}}_{1}\cdot\vec{n} \vec{v}_{1}\cdot\vec{n} ( \vec{v}_{2}\cdot\vec{n} )^{2} \Big] \nl
- \frac{G C_{1ES^2} m_{2}}{16 m_{1} r} \Big[ 2 \big( 2 \vec{S}_{1}\cdot\vec{v}_{2} \ddot{\vec{S}}_{1}\cdot\vec{v}_{2}
	 - 3 \vec{S}_{1}\cdot\ddot{\vec{S}}_{1} v_{2}^2
	 - 2 \ddot{\vec{S}}_{1}\cdot\vec{n} \vec{v}_{2}\cdot\vec{n} \vec{S}_{1}\cdot\vec{v}_{2} \nl
	 - 2 \vec{S}_{1}\cdot\vec{n} \vec{v}_{2}\cdot\vec{n} \ddot{\vec{S}}_{1}\cdot\vec{v}_{2}
	 - \vec{S}_{1}\cdot\vec{n} \ddot{\vec{S}}_{1}\cdot\vec{n} v_{2}^2
	 + 3 \vec{S}_{1}\cdot\ddot{\vec{S}}_{1} ( \vec{v}_{2}\cdot\vec{n} )^{2}
	 + 3 \vec{S}_{1}\cdot\vec{n} \ddot{\vec{S}}_{1}\cdot\vec{n} ( \vec{v}_{2}\cdot\vec{n} )^{2} \big) \nl
	 - \big( 2 \vec{S}_{1}\cdot\vec{a}_{1} \vec{S}_{1}\cdot\vec{a}_{2}
	 - 3 S_{1}^2 \vec{a}_{1}\cdot\vec{a}_{2}
	 + 3 S_{1}^2 \vec{a}_{1}\cdot\vec{n} \vec{a}_{2}\cdot\vec{n}
	 - 2 \vec{S}_{1}\cdot\vec{n} \vec{S}_{1}\cdot\vec{a}_{1} \vec{a}_{2}\cdot\vec{n} \nl
	 - 2 \vec{S}_{1}\cdot\vec{n} \vec{a}_{1}\cdot\vec{n} \vec{S}_{1}\cdot\vec{a}_{2}
	 - \vec{a}_{1}\cdot\vec{a}_{2} ( \vec{S}_{1}\cdot\vec{n} )^{2}
	 + 3 \vec{a}_{1}\cdot\vec{n} \vec{a}_{2}\cdot\vec{n} ( \vec{S}_{1}\cdot\vec{n} )^{2} \big)
	 - 4 \big( \dot{\vec{S}}_{1}\cdot\vec{v}_{1} \vec{S}_{1}\cdot\vec{a}_{2} \nl
	 + \vec{S}_{1}\cdot\vec{v}_{1} \dot{\vec{S}}_{1}\cdot\vec{a}_{2}
	 - 3 \vec{S}_{1}\cdot\dot{\vec{S}}_{1} \vec{v}_{1}\cdot\vec{a}_{2}
	 + 3 \vec{S}_{1}\cdot\dot{\vec{S}}_{1} \vec{v}_{1}\cdot\vec{n} \vec{a}_{2}\cdot\vec{n}
	 - \dot{\vec{S}}_{1}\cdot\vec{n} \vec{S}_{1}\cdot\vec{v}_{1} \vec{a}_{2}\cdot\vec{n} \nl
	 - \vec{S}_{1}\cdot\vec{n} \dot{\vec{S}}_{1}\cdot\vec{v}_{1} \vec{a}_{2}\cdot\vec{n}
	 - \dot{\vec{S}}_{1}\cdot\vec{n} \vec{v}_{1}\cdot\vec{n} \vec{S}_{1}\cdot\vec{a}_{2}
	 - \vec{S}_{1}\cdot\vec{n} \vec{v}_{1}\cdot\vec{n} \dot{\vec{S}}_{1}\cdot\vec{a}_{2}
	 - \vec{S}_{1}\cdot\vec{n} \dot{\vec{S}}_{1}\cdot\vec{n} \vec{v}_{1}\cdot\vec{a}_{2} \nl
	 + 3 \vec{S}_{1}\cdot\vec{n} \dot{\vec{S}}_{1}\cdot\vec{n} \vec{v}_{1}\cdot\vec{n} \vec{a}_{2}\cdot\vec{n} \big)
	 - 2 \big( 3 \dot{S}_{1}^2 v_{2}^2
	 - 2 ( \dot{\vec{S}}_{1}\cdot\vec{v}_{2} )^{2}
	 + 4 \dot{\vec{S}}_{1}\cdot\vec{n} \vec{v}_{2}\cdot\vec{n} \dot{\vec{S}}_{1}\cdot\vec{v}_{2} \nl
	 + v_{2}^2 ( \dot{\vec{S}}_{1}\cdot\vec{n} )^{2}
	 - 3 \dot{S}_{1}^2 ( \vec{v}_{2}\cdot\vec{n} )^{2}
	 - 3 ( \dot{\vec{S}}_{1}\cdot\vec{n} )^{2} ( \vec{v}_{2}\cdot\vec{n} )^{2} \big) \Big] \nl
- \frac{G C_{1ES^2} m_{2}}{8 m_{1}} \Big[ \big( 3 \vec{S}_{1}\cdot\ddot{\vec{S}}_{1} \vec{a}_{2}\cdot\vec{n}
	 - \ddot{\vec{S}}_{1}\cdot\vec{n} \vec{S}_{1}\cdot\vec{a}_{2}
	 - \vec{S}_{1}\cdot\vec{n} \ddot{\vec{S}}_{1}\cdot\vec{a}_{2}
	 + \vec{S}_{1}\cdot\vec{n} \ddot{\vec{S}}_{1}\cdot\vec{n} \vec{a}_{2}\cdot\vec{n} \big) \nl
	 + \big( 3 \dot{S}_{1}^2 \vec{a}_{2}\cdot\vec{n}
	 - 2 \dot{\vec{S}}_{1}\cdot\vec{n} \dot{\vec{S}}_{1}\cdot\vec{a}_{2}
	 + \vec{a}_{2}\cdot\vec{n} ( \dot{\vec{S}}_{1}\cdot\vec{n} )^{2} \big) \Big]\\
\text{Fig.~1(b1)} =&
- \frac{2 G C_{1ES^2} m_{2}}{m_{1} r^3} \Big[ S_{1}^2 \vec{v}_{1}\cdot\vec{v}_{2}
	 - 3 \vec{v}_{1}\cdot\vec{v}_{2} ( \vec{S}_{1}\cdot\vec{n} )^{2} \Big]
+ \frac{2 G C_{1ES^2} m_{2}}{m_{1} r^2} \Big[ 2 \vec{S}_{1}\cdot\dot{\vec{S}}_{1} \vec{v}_{2}\cdot\vec{n} \nl
	 - \dot{\vec{S}}_{1}\cdot\vec{n} \vec{S}_{1}\cdot\vec{v}_{2}
	 - \vec{S}_{1}\cdot\vec{n} \dot{\vec{S}}_{1}\cdot\vec{v}_{2} \Big]
- \frac{G C_{1ES^2} m_{2}}{m_{1} r^3} \Big[ 3 S_{1}^2 v_{1}^2 \vec{v}_{1}\cdot\vec{v}_{2}
	 + S_{1}^2 \vec{v}_{1}\cdot\vec{v}_{2} v_{2}^2 \nl
	 - 2 \vec{v}_{1}\cdot\vec{v}_{2} ( \vec{S}_{1}\cdot\vec{v}_{1} )^{2}
	 + 6 \vec{S}_{1}\cdot\vec{n} \vec{v}_{1}\cdot\vec{n} \vec{S}_{1}\cdot\vec{v}_{1} \vec{v}_{1}\cdot\vec{v}_{2}
	 - 3 v_{1}^2 \vec{v}_{1}\cdot\vec{v}_{2} ( \vec{S}_{1}\cdot\vec{n} )^{2} \nl
	 - 3 \vec{v}_{1}\cdot\vec{v}_{2} v_{2}^2 ( \vec{S}_{1}\cdot\vec{n} )^{2}
	 - 6 S_{1}^2 \vec{v}_{1}\cdot\vec{v}_{2} ( \vec{v}_{1}\cdot\vec{n} )^{2} \Big]
+ \frac{G C_{1ES^2} m_{2}}{m_{1} r^2} \Big[ 2 S_{1}^2 \vec{v}_{1}\cdot\vec{a}_{1} \vec{v}_{2}\cdot\vec{n} \nl
	 + \vec{S}_{1}\cdot\vec{v}_{1} \vec{a}_{1}\cdot\vec{n} \vec{S}_{1}\cdot\vec{v}_{2}
	 + \vec{v}_{1}\cdot\vec{n} \vec{S}_{1}\cdot\vec{a}_{1} \vec{S}_{1}\cdot\vec{v}_{2}
	 - 2 \vec{S}_{1}\cdot\vec{n} \vec{v}_{1}\cdot\vec{a}_{1} \vec{S}_{1}\cdot\vec{v}_{2} \nl
	 - 2 S_{1}^2 \vec{a}_{1}\cdot\vec{n} \vec{v}_{1}\cdot\vec{v}_{2}
	 + \vec{S}_{1}\cdot\vec{n} \vec{S}_{1}\cdot\vec{a}_{1} \vec{v}_{1}\cdot\vec{v}_{2}
	 - 2 S_{1}^2 \vec{v}_{1}\cdot\vec{n} \vec{a}_{1}\cdot\vec{v}_{2}
	 + \vec{S}_{1}\cdot\vec{n} \vec{S}_{1}\cdot\vec{v}_{1} \vec{a}_{1}\cdot\vec{v}_{2} \Big] \nl
+ \frac{G C_{1ES^2} m_{2}}{m_{1} r^2} \Big[ 2 \vec{S}_{1}\cdot\dot{\vec{S}}_{1} v_{1}^2 \vec{v}_{2}\cdot\vec{n}
	 + \vec{v}_{1}\cdot\vec{n} \dot{\vec{S}}_{1}\cdot\vec{v}_{1} \vec{S}_{1}\cdot\vec{v}_{2}
	 - \dot{\vec{S}}_{1}\cdot\vec{n} v_{1}^2 \vec{S}_{1}\cdot\vec{v}_{2} \nl
	 + \vec{v}_{1}\cdot\vec{n} \vec{S}_{1}\cdot\vec{v}_{1} \dot{\vec{S}}_{1}\cdot\vec{v}_{2}
	 - \vec{S}_{1}\cdot\vec{n} v_{1}^2 \dot{\vec{S}}_{1}\cdot\vec{v}_{2}
	 - 4 \vec{S}_{1}\cdot\dot{\vec{S}}_{1} \vec{v}_{1}\cdot\vec{n} \vec{v}_{1}\cdot\vec{v}_{2}
	 + \dot{\vec{S}}_{1}\cdot\vec{n} \vec{S}_{1}\cdot\vec{v}_{1} \vec{v}_{1}\cdot\vec{v}_{2} \nl
	 + \vec{S}_{1}\cdot\vec{n} \dot{\vec{S}}_{1}\cdot\vec{v}_{1} \vec{v}_{1}\cdot\vec{v}_{2}
	 + 2 \vec{S}_{1}\cdot\dot{\vec{S}}_{1} \vec{v}_{2}\cdot\vec{n} v_{2}^2
	 - \dot{\vec{S}}_{1}\cdot\vec{n} \vec{S}_{1}\cdot\vec{v}_{2} v_{2}^2
	 - \vec{S}_{1}\cdot\vec{n} \dot{\vec{S}}_{1}\cdot\vec{v}_{2} v_{2}^2 \Big]\\
\text{Fig.~1(b2)} =&
- \frac{G C_{1ES^2} m_{2}}{m_{1} r^3} \Big[ 2 \vec{S}_{1}\cdot\vec{v}_{1} \vec{S}_{1}\cdot\vec{v}_{2} \vec{v}_{1}\cdot\vec{v}_{2}
	 - S_{1}^2 ( \vec{v}_{1}\cdot\vec{v}_{2} )^{2}
	 + 3 S_{1}^2 \vec{v}_{1}\cdot\vec{n} \vec{v}_{2}\cdot\vec{n} \vec{v}_{1}\cdot\vec{v}_{2} \nl
	 - 6 \vec{S}_{1}\cdot\vec{n} \vec{S}_{1}\cdot\vec{v}_{1} \vec{v}_{2}\cdot\vec{n} \vec{v}_{1}\cdot\vec{v}_{2}
	 - 6 \vec{S}_{1}\cdot\vec{n} \vec{v}_{1}\cdot\vec{n} \vec{S}_{1}\cdot\vec{v}_{2} \vec{v}_{1}\cdot\vec{v}_{2}
	 - 3 ( \vec{S}_{1}\cdot\vec{n} )^{2} ( \vec{v}_{1}\cdot\vec{v}_{2} )^{2} \nl
	 + 15 \vec{v}_{1}\cdot\vec{n} \vec{v}_{2}\cdot\vec{n} \vec{v}_{1}\cdot\vec{v}_{2} ( \vec{S}_{1}\cdot\vec{n} )^{2} \Big]
+ \frac{G C_{1ES^2} m_{2}}{m_{1} r^2} \Big[ S_{1}^2 \vec{v}_{2}\cdot\vec{n} \vec{a}_{1}\cdot\vec{v}_{2} \nl
	 - 2 \vec{S}_{1}\cdot\vec{n} \vec{S}_{1}\cdot\vec{v}_{2} \vec{a}_{1}\cdot\vec{v}_{2}
	 - S_{1}^2 \vec{v}_{1}\cdot\vec{n} \vec{v}_{1}\cdot\vec{a}_{2}
	 + 2 \vec{S}_{1}\cdot\vec{n} \vec{S}_{1}\cdot\vec{v}_{1} \vec{v}_{1}\cdot\vec{a}_{2} \nl
	 + 3 \vec{v}_{2}\cdot\vec{n} \vec{a}_{1}\cdot\vec{v}_{2} ( \vec{S}_{1}\cdot\vec{n} )^{2}
	 - 3 \vec{v}_{1}\cdot\vec{n} \vec{v}_{1}\cdot\vec{a}_{2} ( \vec{S}_{1}\cdot\vec{n} )^{2} \Big] \nl
- \frac{G C_{1ES^2} m_{2}}{m_{1} r^2} \Big[ \dot{\vec{S}}_{1}\cdot\vec{v}_{1} \vec{v}_{2}\cdot\vec{n} \vec{S}_{1}\cdot\vec{v}_{2}
	 + \vec{S}_{1}\cdot\vec{v}_{1} \vec{v}_{2}\cdot\vec{n} \dot{\vec{S}}_{1}\cdot\vec{v}_{2}
	 + 2 \vec{v}_{1}\cdot\vec{n} \vec{S}_{1}\cdot\vec{v}_{2} \dot{\vec{S}}_{1}\cdot\vec{v}_{2} \nl
	 - 6 \vec{S}_{1}\cdot\dot{\vec{S}}_{1} \vec{v}_{2}\cdot\vec{n} \vec{v}_{1}\cdot\vec{v}_{2}
	 + 3 \dot{\vec{S}}_{1}\cdot\vec{n} \vec{S}_{1}\cdot\vec{v}_{2} \vec{v}_{1}\cdot\vec{v}_{2}
	 + 3 \vec{S}_{1}\cdot\vec{n} \dot{\vec{S}}_{1}\cdot\vec{v}_{2} \vec{v}_{1}\cdot\vec{v}_{2} \nl
	 - 2 \vec{S}_{1}\cdot\dot{\vec{S}}_{1} \vec{v}_{1}\cdot\vec{n} v_{2}^2
	 - 3 \dot{\vec{S}}_{1}\cdot\vec{n} \vec{v}_{1}\cdot\vec{n} \vec{v}_{2}\cdot\vec{n} \vec{S}_{1}\cdot\vec{v}_{2}
	 - 3 \vec{S}_{1}\cdot\vec{n} \vec{v}_{1}\cdot\vec{n} \vec{v}_{2}\cdot\vec{n} \dot{\vec{S}}_{1}\cdot\vec{v}_{2} \nl
	 - 6 \vec{S}_{1}\cdot\vec{n} \dot{\vec{S}}_{1}\cdot\vec{n} \vec{v}_{2}\cdot\vec{n} \vec{v}_{1}\cdot\vec{v}_{2}
	 + 6 \vec{S}_{1}\cdot\dot{\vec{S}}_{1} \vec{v}_{1}\cdot\vec{n} ( \vec{v}_{2}\cdot\vec{n} )^{2} \Big] \nl
+ \frac{G C_{1ES^2} m_{2}}{m_{1} r} \Big[ \big( 2 \vec{S}_{1}\cdot\vec{v}_{2} \ddot{\vec{S}}_{1}\cdot\vec{v}_{2}
	 - 2 \vec{S}_{1}\cdot\ddot{\vec{S}}_{1} v_{2}^2
	 - \ddot{\vec{S}}_{1}\cdot\vec{n} \vec{v}_{2}\cdot\vec{n} \vec{S}_{1}\cdot\vec{v}_{2} \nl
	 - \vec{S}_{1}\cdot\vec{n} \vec{v}_{2}\cdot\vec{n} \ddot{\vec{S}}_{1}\cdot\vec{v}_{2}
	 + 2 \vec{S}_{1}\cdot\ddot{\vec{S}}_{1} ( \vec{v}_{2}\cdot\vec{n} )^{2} \big)
	 + \big( S_{1}^2 \vec{a}_{1}\cdot\vec{a}_{2}
	 + \vec{a}_{1}\cdot\vec{a}_{2} ( \vec{S}_{1}\cdot\vec{n} )^{2} \big) \nl
	 - \big( \dot{\vec{S}}_{1}\cdot\vec{v}_{1} \vec{S}_{1}\cdot\vec{a}_{2}
	 + \vec{S}_{1}\cdot\vec{v}_{1} \dot{\vec{S}}_{1}\cdot\vec{a}_{2}
	 - 4 \vec{S}_{1}\cdot\dot{\vec{S}}_{1} \vec{v}_{1}\cdot\vec{a}_{2}
	 + 2 \vec{S}_{1}\cdot\dot{\vec{S}}_{1} \vec{v}_{1}\cdot\vec{n} \vec{a}_{2}\cdot\vec{n} \nl
	 - \dot{\vec{S}}_{1}\cdot\vec{n} \vec{v}_{1}\cdot\vec{n} \vec{S}_{1}\cdot\vec{a}_{2}
	 - \vec{S}_{1}\cdot\vec{n} \vec{v}_{1}\cdot\vec{n} \dot{\vec{S}}_{1}\cdot\vec{a}_{2}
	 - 2 \vec{S}_{1}\cdot\vec{n} \dot{\vec{S}}_{1}\cdot\vec{n} \vec{v}_{1}\cdot\vec{a}_{2} \big)
	 - 2 \big( \dot{S}_{1}^2 v_{2}^2 \nl
	 - ( \dot{\vec{S}}_{1}\cdot\vec{v}_{2} )^{2}
	 + \dot{\vec{S}}_{1}\cdot\vec{n} \vec{v}_{2}\cdot\vec{n} \dot{\vec{S}}_{1}\cdot\vec{v}_{2}
	 - \dot{S}_{1}^2 ( \vec{v}_{2}\cdot\vec{n} )^{2} \big) \Big]
+ \frac{G C_{1ES^2} m_{2}}{m_{1}} \Big[ \big( 2 \vec{S}_{1}\cdot\ddot{\vec{S}}_{1} \vec{a}_{2}\cdot\vec{n} \nl
	 - \ddot{\vec{S}}_{1}\cdot\vec{n} \vec{S}_{1}\cdot\vec{a}_{2}
	 - \vec{S}_{1}\cdot\vec{n} \ddot{\vec{S}}_{1}\cdot\vec{a}_{2} \big)
	 + 2 \big( \dot{S}_{1}^2 \vec{a}_{2}\cdot\vec{n}
	 - \dot{\vec{S}}_{1}\cdot\vec{n} \dot{\vec{S}}_{1}\cdot\vec{a}_{2} \big) \Big]\\
\text{Fig.~1(c)} =&
- \frac{G C_{1ES^2} m_{2}}{m_{1} r^3} \Big[ S_{1}^2 v_{1}^2 v_{2}^2
	 - S_{1}^2 ( \vec{v}_{1}\cdot\vec{v}_{2} )^{2}
	 - 3 v_{1}^2 v_{2}^2 ( \vec{S}_{1}\cdot\vec{n} )^{2}
	 + 3 ( \vec{S}_{1}\cdot\vec{n} )^{2} ( \vec{v}_{1}\cdot\vec{v}_{2} )^{2} \Big] \nl
- \frac{G C_{1ES^2} m_{2}}{m_{1} r^2} \Big[ 2 S_{1}^2 \vec{v}_{2}\cdot\vec{n} \vec{a}_{1}\cdot\vec{v}_{2}
	 - 2 \vec{S}_{1}\cdot\vec{n} \vec{S}_{1}\cdot\vec{v}_{2} \vec{a}_{1}\cdot\vec{v}_{2}
	 - S_{1}^2 \vec{a}_{1}\cdot\vec{n} v_{2}^2 \nl
	 + 2 \vec{S}_{1}\cdot\vec{n} \vec{S}_{1}\cdot\vec{a}_{1} v_{2}^2
	 + \vec{a}_{1}\cdot\vec{n} ( \vec{S}_{1}\cdot\vec{v}_{2} )^{2} \Big]
- \frac{2 G C_{1ES^2} m_{2}}{m_{1} r^2} \Big[ 2 \vec{S}_{1}\cdot\dot{\vec{S}}_{1} \vec{v}_{2}\cdot\vec{n} \vec{v}_{1}\cdot\vec{v}_{2} \nl
	 - \dot{\vec{S}}_{1}\cdot\vec{n} \vec{S}_{1}\cdot\vec{v}_{2} \vec{v}_{1}\cdot\vec{v}_{2}
	 - \vec{S}_{1}\cdot\vec{n} \dot{\vec{S}}_{1}\cdot\vec{v}_{2} \vec{v}_{1}\cdot\vec{v}_{2}
	 - 2 \vec{S}_{1}\cdot\dot{\vec{S}}_{1} \vec{v}_{1}\cdot\vec{n} v_{2}^2
	 + \dot{\vec{S}}_{1}\cdot\vec{n} \vec{S}_{1}\cdot\vec{v}_{1} v_{2}^2 \nl
	 + \vec{S}_{1}\cdot\vec{n} \dot{\vec{S}}_{1}\cdot\vec{v}_{1} v_{2}^2 \Big]
- \frac{2 G C_{1ES^2} m_{2}}{m_{1} r} \Big[ \big( \vec{S}_{1}\cdot\vec{v}_{2} \ddot{\vec{S}}_{1}\cdot\vec{v}_{2}
	 + \vec{S}_{1}\cdot\ddot{\vec{S}}_{1} v_{2}^2 \big) \nl
	 + \big( \dot{S}_{1}^2 v_{2}^2
	 + ( \dot{\vec{S}}_{1}\cdot\vec{v}_{2} )^{2} \big) \Big]
\end{align}

\subsection{Two-graviton exchange and cubic self-interaction}

\subsubsection{Two-graviton exchange}

\begin{figure}[t]
\includegraphics[scale=0.93]{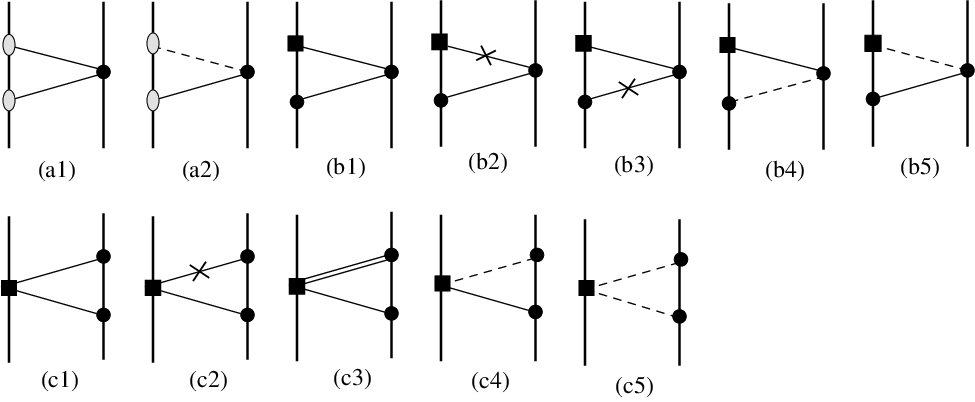}
\caption{NNLO spin-squared two-graviton exchange Feynman diagrams.}
\label{fig:ssnnlo2g}
\end{figure}

There are 12 two-graviton exchange diagrams, which can be seen in figure \ref{fig:ssnnlo2g}. They involve two-graviton mass couplings, which at the NNLO level are also contracted with linear in spin couplings, or two-graviton spin-squared couplings. We note that time derivatives originate here also from the spin-squared couplings. 

These diagrams are evaluated by:
\begin{align}
\text{Fig.~2(a1)} =&
- \frac{2 G^2 m_{2}}{r^4} \Big[ S_{1}^2 v_{1}^2
	 - ( \vec{S}_{1}\cdot\vec{v}_{1} )^{2}
	 + 2 \vec{S}_{1}\cdot\vec{n} \vec{v}_{1}\cdot\vec{n} \vec{S}_{1}\cdot\vec{v}_{1}
	 - v_{1}^2 ( \vec{S}_{1}\cdot\vec{n} )^{2}
	 - S_{1}^2 ( \vec{v}_{1}\cdot\vec{n} )^{2} \Big]\\
\text{Fig.~2(a2)} =&
- \frac{4 G^2 m_{2}}{r^4} \Big[ \vec{S}_{1}\cdot\vec{v}_{1} \vec{S}_{1}\cdot\vec{v}_{2}
	 - S_{1}^2 \vec{v}_{1}\cdot\vec{v}_{2}
	 + S_{1}^2 \vec{v}_{1}\cdot\vec{n} \vec{v}_{2}\cdot\vec{n}
	 - \vec{S}_{1}\cdot\vec{n} \vec{S}_{1}\cdot\vec{v}_{1} \vec{v}_{2}\cdot\vec{n} \nl
	 - \vec{S}_{1}\cdot\vec{n} \vec{v}_{1}\cdot\vec{n} \vec{S}_{1}\cdot\vec{v}_{2}
	 + \vec{v}_{1}\cdot\vec{v}_{2} ( \vec{S}_{1}\cdot\vec{n} )^{2} \Big]\\
\text{Fig.~2(b1)} =&
- \frac{G^2 C_{1ES^2} m_{2}}{2 r^4} \Big[ S_{1}^2
	 - 3 ( \vec{S}_{1}\cdot\vec{n} )^{2} \Big]
- \frac{G^2 C_{1ES^2} m_{2}}{4 r^4} \Big[ 8 S_{1}^2 v_{1}^2
	 - 9 S_{1}^2 v_{2}^2
	 - 2 ( \vec{S}_{1}\cdot\vec{v}_{1} )^{2} \nl
	 + 6 \vec{S}_{1}\cdot\vec{n} \vec{v}_{1}\cdot\vec{n} \vec{S}_{1}\cdot\vec{v}_{1}
	 - 18 v_{1}^2 ( \vec{S}_{1}\cdot\vec{n} )^{2}
	 + 27 v_{2}^2 ( \vec{S}_{1}\cdot\vec{n} )^{2}
	 - 6 S_{1}^2 ( \vec{v}_{1}\cdot\vec{n} )^{2} \Big] \nl
- \frac{G^2 C_{1ES^2} m_{2}}{r^3} \vec{S}_{1}\cdot\vec{a}_{1} \vec{S}_{1}\cdot\vec{n}
+ \frac{G^2 C_{1ES^2} m_{2}}{r^3} \Big[ 2 \vec{S}_{1}\cdot\dot{\vec{S}}_{1} \vec{v}_{1}\cdot\vec{n}
	 - \dot{\vec{S}}_{1}\cdot\vec{n} \vec{S}_{1}\cdot\vec{v}_{1} \nl
	 - \vec{S}_{1}\cdot\vec{n} \dot{\vec{S}}_{1}\cdot\vec{v}_{1} \Big]
- \frac{G^2 C_{1ES^2} m_{2}}{r^2} \ddot{S^2_{1}}\\
\text{Fig.~2(b2)} =&
- \frac{G^2 C_{1ES^2} m_{2}}{4 r^4} \Big[ 2 \vec{S}_{1}\cdot\vec{v}_{1} \vec{S}_{1}\cdot\vec{v}_{2}
	 - S_{1}^2 \vec{v}_{1}\cdot\vec{v}_{2}
	 + 2 \vec{S}_{1}\cdot\vec{n} \vec{v}_{1}\cdot\vec{n} \vec{S}_{1}\cdot\vec{v}_{1}
	 + 4 S_{1}^2 \vec{v}_{1}\cdot\vec{n} \vec{v}_{2}\cdot\vec{n} \nl
	 - 8 \vec{S}_{1}\cdot\vec{n} \vec{S}_{1}\cdot\vec{v}_{1} \vec{v}_{2}\cdot\vec{n}
	 - 6 \vec{S}_{1}\cdot\vec{n} \vec{v}_{1}\cdot\vec{n} \vec{S}_{1}\cdot\vec{v}_{2}
	 - 3 \vec{v}_{1}\cdot\vec{v}_{2} ( \vec{S}_{1}\cdot\vec{n} )^{2}
	 - S_{1}^2 ( \vec{v}_{1}\cdot\vec{n} )^{2} \nl
	 + 18 \vec{v}_{1}\cdot\vec{n} \vec{v}_{2}\cdot\vec{n} ( \vec{S}_{1}\cdot\vec{n} )^{2}
	 - 3 ( \vec{S}_{1}\cdot\vec{n} )^{2} ( \vec{v}_{1}\cdot\vec{n} )^{2} \Big]
- \frac{G^2 C_{1ES^2} m_{2}}{2 r^3} \Big[ \vec{S}_{1}\cdot\dot{\vec{S}}_{1} \vec{v}_{1}\cdot\vec{n} \nl
	 - 2 \vec{S}_{1}\cdot\dot{\vec{S}}_{1} \vec{v}_{2}\cdot\vec{n}
	 + \dot{\vec{S}}_{1}\cdot\vec{n} \vec{S}_{1}\cdot\vec{v}_{2}
	 + \vec{S}_{1}\cdot\vec{n} \dot{\vec{S}}_{1}\cdot\vec{v}_{2}
	 + \vec{S}_{1}\cdot\vec{n} \dot{\vec{S}}_{1}\cdot\vec{n} \vec{v}_{1}\cdot\vec{n} \nl
	 - 4 \vec{S}_{1}\cdot\vec{n} \dot{\vec{S}}_{1}\cdot\vec{n} \vec{v}_{2}\cdot\vec{n} \Big]\\
\text{Fig.~2(b3)} =&
- \frac{G^2 C_{1ES^2} m_{2}}{4 r^4} \Big[ S_{1}^2 \vec{v}_{1}\cdot\vec{v}_{2}
	 + 6 \vec{S}_{1}\cdot\vec{n} \vec{v}_{1}\cdot\vec{n} \vec{S}_{1}\cdot\vec{v}_{1}
	 - 4 S_{1}^2 \vec{v}_{1}\cdot\vec{n} \vec{v}_{2}\cdot\vec{n} \nl
	 - 6 \vec{S}_{1}\cdot\vec{n} \vec{v}_{1}\cdot\vec{n} \vec{S}_{1}\cdot\vec{v}_{2}
	 - 3 \vec{v}_{1}\cdot\vec{v}_{2} ( \vec{S}_{1}\cdot\vec{n} )^{2}
	 + 3 S_{1}^2 ( \vec{v}_{1}\cdot\vec{n} )^{2}
	 + 18 \vec{v}_{1}\cdot\vec{n} \vec{v}_{2}\cdot\vec{n} ( \vec{S}_{1}\cdot\vec{n} )^{2} \nl
	 - 15 ( \vec{S}_{1}\cdot\vec{n} )^{2} ( \vec{v}_{1}\cdot\vec{n} )^{2} \Big]
+ \frac{G^2 C_{1ES^2} m_{2}}{2 r^3} \Big[ \vec{S}_{1}\cdot\dot{\vec{S}}_{1} \vec{v}_{1}\cdot\vec{n}
	 - 3 \vec{S}_{1}\cdot\vec{n} \dot{\vec{S}}_{1}\cdot\vec{n} \vec{v}_{1}\cdot\vec{n} \Big]\\
\text{Fig.~2(b4)} =&
\frac{2 G^2 C_{1ES^2} m_{2}}{r^4} \Big[ S_{1}^2 \vec{v}_{1}\cdot\vec{v}_{2}
	 - 3 \vec{v}_{1}\cdot\vec{v}_{2} ( \vec{S}_{1}\cdot\vec{n} )^{2} \Big]\\
\text{Fig.~2(b5)} =&
\frac{2 G^2 C_{1ES^2} m_{2}}{r^4} \Big[ S_{1}^2 \vec{v}_{1}\cdot\vec{v}_{2}
	 - 3 \vec{v}_{1}\cdot\vec{v}_{2} ( \vec{S}_{1}\cdot\vec{n} )^{2} \Big]
- \frac{2 G^2 C_{1ES^2} m_{2}}{r^3} \Big[ 2 \vec{S}_{1}\cdot\dot{\vec{S}}_{1} \vec{v}_{2}\cdot\vec{n} \nl
	 - \dot{\vec{S}}_{1}\cdot\vec{n} \vec{S}_{1}\cdot\vec{v}_{2}
	 - \vec{S}_{1}\cdot\vec{n} \dot{\vec{S}}_{1}\cdot\vec{v}_{2} \Big]\\
\text{Fig.~2(c1)} =&
- \frac{2 G^2 C_{1ES^2} m_{2}^2}{m_{1} r^4} \Big[ S_{1}^2
	 - 3 ( \vec{S}_{1}\cdot\vec{n} )^{2} \Big]
+ \frac{G^2 C_{1ES^2} m_{2}^2}{2 m_{1} r^4} \Big[ S_{1}^2 v_{1}^2
	 + 2 \vec{S}_{1}\cdot\vec{v}_{1} \vec{S}_{1}\cdot\vec{v}_{2} \nl
	 + 2 S_{1}^2 \vec{v}_{1}\cdot\vec{v}_{2}
	 - 14 S_{1}^2 v_{2}^2
	 - ( \vec{S}_{1}\cdot\vec{v}_{1} )^{2}
	 + 4 \vec{S}_{1}\cdot\vec{n} \vec{v}_{1}\cdot\vec{n} \vec{S}_{1}\cdot\vec{v}_{1}
	 - 8 S_{1}^2 \vec{v}_{1}\cdot\vec{n} \vec{v}_{2}\cdot\vec{n} \nl
	 - 8 \vec{S}_{1}\cdot\vec{n} \vec{S}_{1}\cdot\vec{v}_{1} \vec{v}_{2}\cdot\vec{n}
	 - 2 v_{1}^2 ( \vec{S}_{1}\cdot\vec{n} )^{2}
	 + 36 v_{2}^2 ( \vec{S}_{1}\cdot\vec{n} )^{2}
	 + 10 S_{1}^2 ( \vec{v}_{2}\cdot\vec{n} )^{2} \Big] \nl
+ \frac{G^2 C_{1ES^2} m_{2}^2}{m_{1} r^3} \vec{a}_{2}\cdot\vec{n} S_{1}^2 \\
\text{Fig.~2(c2)} =&
- \frac{G^2 C_{1ES^2} m_{2}^2}{2 m_{1} r^4} \Big[ 6 \vec{S}_{1}\cdot\vec{v}_{1} \vec{S}_{1}\cdot\vec{v}_{2}
	 + S_{1}^2 v_{2}^2
	 - 3 ( \vec{S}_{1}\cdot\vec{v}_{2} )^{2}
	 - 24 \vec{S}_{1}\cdot\vec{n} \vec{S}_{1}\cdot\vec{v}_{1} \vec{v}_{2}\cdot\vec{n} \nl
	 - 24 \vec{S}_{1}\cdot\vec{n} \vec{v}_{1}\cdot\vec{n} \vec{S}_{1}\cdot\vec{v}_{2}
	 + 24 \vec{S}_{1}\cdot\vec{n} \vec{v}_{2}\cdot\vec{n} \vec{S}_{1}\cdot\vec{v}_{2}
	 - 12 \vec{v}_{1}\cdot\vec{v}_{2} ( \vec{S}_{1}\cdot\vec{n} )^{2}
	 + 2 S_{1}^2 ( \vec{v}_{2}\cdot\vec{n} )^{2} \nl
	 + 72 \vec{v}_{1}\cdot\vec{n} \vec{v}_{2}\cdot\vec{n} ( \vec{S}_{1}\cdot\vec{n} )^{2}
	 - 36 ( \vec{S}_{1}\cdot\vec{n} )^{2} ( \vec{v}_{2}\cdot\vec{n} )^{2} \Big]
- \frac{3 G^2 C_{1ES^2} m_{2}^2}{m_{1} r^3} \Big[ \dot{\vec{S}}_{1}\cdot\vec{n} \vec{S}_{1}\cdot\vec{v}_{2} \nl
	 + \vec{S}_{1}\cdot\vec{n} \dot{\vec{S}}_{1}\cdot\vec{v}_{2}
	 - 4 \vec{S}_{1}\cdot\vec{n} \dot{\vec{S}}_{1}\cdot\vec{n} \vec{v}_{2}\cdot\vec{n} \Big]\\
\text{Fig.~2(c3)} =&
\frac{G^2 C_{1ES^2} m_{2}^2}{m_{1} r^4} \Big[ 5 S_{1}^2 v_{2}^2
	 - 3 ( \vec{S}_{1}\cdot\vec{v}_{2} )^{2}
	 + 8 \vec{S}_{1}\cdot\vec{n} \vec{v}_{2}\cdot\vec{n} \vec{S}_{1}\cdot\vec{v}_{2}
	 - 8 v_{2}^2 ( \vec{S}_{1}\cdot\vec{n} )^{2} \nl
	 - 8 S_{1}^2 ( \vec{v}_{2}\cdot\vec{n} )^{2} \Big]\\
\text{Fig.~2(c4)} =&
- \frac{2 G^2 C_{1ES^2} m_{2}^2}{m_{1} r^4} \Big[ 4 \vec{S}_{1}\cdot\vec{v}_{1} \vec{S}_{1}\cdot\vec{v}_{2}
	 - 12 S_{1}^2 \vec{v}_{1}\cdot\vec{v}_{2}
	 + 5 S_{1}^2 v_{2}^2
	 - 5 ( \vec{S}_{1}\cdot\vec{v}_{2} )^{2} \nl
	 + 14 S_{1}^2 \vec{v}_{1}\cdot\vec{n} \vec{v}_{2}\cdot\vec{n}
	 - \vec{S}_{1}\cdot\vec{n} \vec{S}_{1}\cdot\vec{v}_{1} \vec{v}_{2}\cdot\vec{n}
	 - 15 \vec{S}_{1}\cdot\vec{n} \vec{v}_{1}\cdot\vec{n} \vec{S}_{1}\cdot\vec{v}_{2} \nl
	 + 20 \vec{S}_{1}\cdot\vec{n} \vec{v}_{2}\cdot\vec{n} \vec{S}_{1}\cdot\vec{v}_{2}
	 + 18 \vec{v}_{1}\cdot\vec{v}_{2} ( \vec{S}_{1}\cdot\vec{n} )^{2}
	 - 18 S_{1}^2 ( \vec{v}_{2}\cdot\vec{n} )^{2} \Big] \nl
+ \frac{2 G^2 C_{1ES^2} m_{2}^2}{m_{1} r^3} \Big[ 5 S_{1}^2 \vec{a}_{2}\cdot\vec{n}
	 - 7 \vec{S}_{1}\cdot\vec{n} \vec{S}_{1}\cdot\vec{a}_{2} \Big]\\
\text{Fig.~2(c5)} =&
- \frac{2 G^2 C_{1ES^2} m_{2}^2}{m_{1} r^4} \Big[ 2 S_{1}^2 v_{2}^2
	 - ( \vec{S}_{1}\cdot\vec{v}_{2} )^{2}
	 + 2 \vec{S}_{1}\cdot\vec{n} \vec{v}_{2}\cdot\vec{n} \vec{S}_{1}\cdot\vec{v}_{2}
	 - v_{2}^2 ( \vec{S}_{1}\cdot\vec{n} )^{2} \nl
	 - 2 S_{1}^2 ( \vec{v}_{2}\cdot\vec{n} )^{2} \Big]
\end{align}

\subsubsection{Cubic self-interaction} 

There are 27 cubic self-interaction diagrams in this sector, 
which can be seen in figure \ref{fig:ssnnlo1loop}.
Tensor one-loop integrals up to order 4, which appear in appendix~A of \cite{Levi:2011eq}, are required to be initially applied here in addition.

\begin{figure}[t]
\includegraphics[scale=0.93]{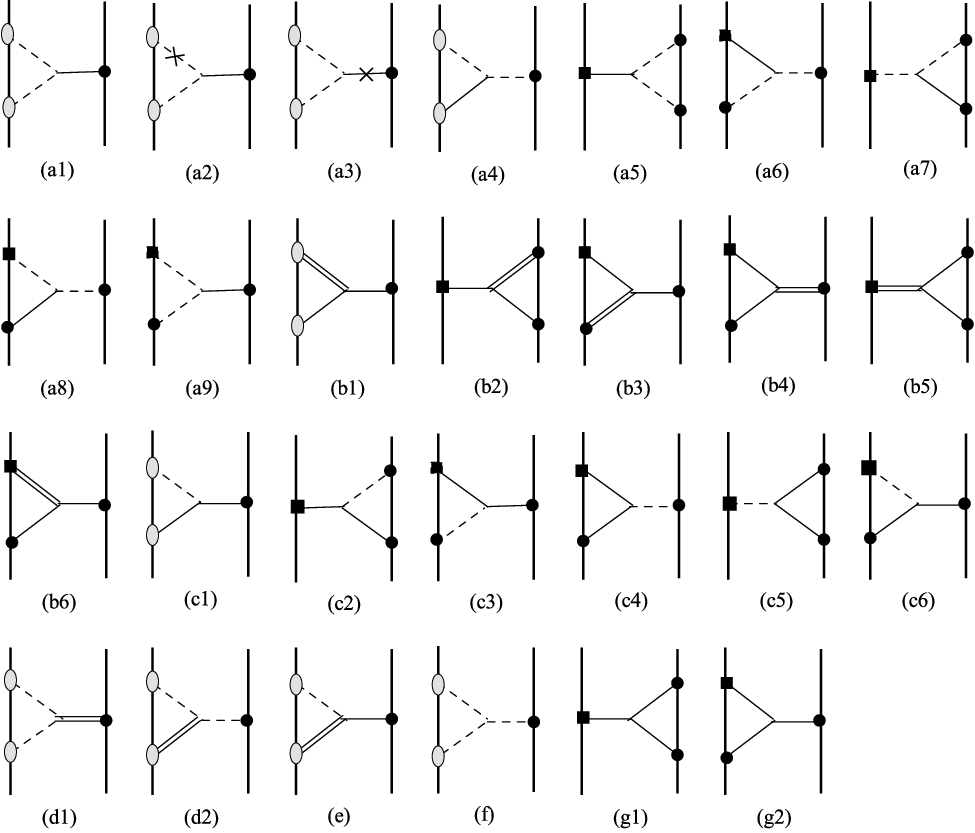}
\caption{NNLO spin-squared one-loop Feynman diagrams.}
\label{fig:ssnnlo1loop}
\end{figure}

These diagrams are evaluated by:
\begin{align}
\text{Fig.~3(a1)} =&
\frac{G^2 m_{2}}{r^4} ( \vec{S}_{1}\cdot\vec{n} )^{2}
+ \frac{G^2 m_{2}}{2 r^4} \Big[ 4 S_{1}^2 v_{1}^2
	 - 4 ( \vec{S}_{1}\cdot\vec{v}_{1} )^{2}
	 + 10 \vec{S}_{1}\cdot\vec{n} \vec{v}_{1}\cdot\vec{n} \vec{S}_{1}\cdot\vec{v}_{1}
	 - 2 v_{1}^2 ( \vec{S}_{1}\cdot\vec{n} )^{2} \nl
	 + 3 v_{2}^2 ( \vec{S}_{1}\cdot\vec{n} )^{2}
	 - 8 S_{1}^2 ( \vec{v}_{1}\cdot\vec{n} )^{2} \Big]
+ \frac{2 G^2 m_{2}}{r^3} \Big[ S_{1}^2 \vec{a}_{1}\cdot\vec{n}
	 - \vec{S}_{1}\cdot\vec{n} \vec{S}_{1}\cdot\vec{a}_{1} \Big] \nl
+ \frac{2 G^2 m_{2}}{r^3} \Big[ \vec{S}_{1}\cdot\dot{\vec{S}}_{1} \vec{v}_{1}\cdot\vec{n}
	 - \dot{\vec{S}}_{1}\cdot\vec{n} \vec{S}_{1}\cdot\vec{v}_{1} \Big]\\
\text{Fig.~3(a2)} =&
- \frac{G^2 m_{2}}{4 r^4} \Big[ 3 S_{1}^2 v_{1}^2
	 - 6 \vec{S}_{1}\cdot\vec{v}_{1} \vec{S}_{1}\cdot\vec{v}_{2}
	 - 2 S_{1}^2 \vec{v}_{1}\cdot\vec{v}_{2}
	 - 12 \vec{S}_{1}\cdot\vec{n} \vec{v}_{1}\cdot\vec{n} \vec{S}_{1}\cdot\vec{v}_{1} \nl
	 + 8 S_{1}^2 \vec{v}_{1}\cdot\vec{n} \vec{v}_{2}\cdot\vec{n}
	 + 24 \vec{S}_{1}\cdot\vec{n} \vec{S}_{1}\cdot\vec{v}_{1} \vec{v}_{2}\cdot\vec{n}
	 + 16 \vec{S}_{1}\cdot\vec{n} \vec{v}_{1}\cdot\vec{n} \vec{S}_{1}\cdot\vec{v}_{2} \nl
	 - 13 v_{1}^2 ( \vec{S}_{1}\cdot\vec{n} )^{2}
	 + 8 \vec{v}_{1}\cdot\vec{v}_{2} ( \vec{S}_{1}\cdot\vec{n} )^{2}
	 - 5 S_{1}^2 ( \vec{v}_{1}\cdot\vec{n} )^{2}
	 - 48 \vec{v}_{1}\cdot\vec{n} \vec{v}_{2}\cdot\vec{n} ( \vec{S}_{1}\cdot\vec{n} )^{2} \nl
	 + 27 ( \vec{S}_{1}\cdot\vec{n} )^{2} ( \vec{v}_{1}\cdot\vec{n} )^{2} \Big]
- \frac{G^2 m_{2}}{2 r^3} \Big[ 6 \vec{S}_{1}\cdot\dot{\vec{S}}_{1} \vec{v}_{1}\cdot\vec{n}
	 + 2 \dot{\vec{S}}_{1}\cdot\vec{n} \vec{S}_{1}\cdot\vec{v}_{1} \nl
	 + 8 \vec{S}_{1}\cdot\vec{n} \dot{\vec{S}}_{1}\cdot\vec{v}_{1}
	 - 6 \vec{S}_{1}\cdot\dot{\vec{S}}_{1} \vec{v}_{2}\cdot\vec{n}
	 - 5 \dot{\vec{S}}_{1}\cdot\vec{n} \vec{S}_{1}\cdot\vec{v}_{2}
	 - 5 \vec{S}_{1}\cdot\vec{n} \dot{\vec{S}}_{1}\cdot\vec{v}_{2} \nl
	 - 20 \vec{S}_{1}\cdot\vec{n} \dot{\vec{S}}_{1}\cdot\vec{n} \vec{v}_{1}\cdot\vec{n}
	 + 20 \vec{S}_{1}\cdot\vec{n} \dot{\vec{S}}_{1}\cdot\vec{n} \vec{v}_{2}\cdot\vec{n} \Big]
+ \frac{G^2 m_{2}}{2 r^2} \Big[ 3 \dot{S}_{1}^2
	 - 5 ( \dot{\vec{S}}_{1}\cdot\vec{n} )^{2} \Big]\\
\text{Fig.~3(a3)} =&
- \frac{G^2 m_{2}}{2 r^4} \Big[ \vec{S}_{1}\cdot\vec{v}_{1} \vec{S}_{1}\cdot\vec{v}_{2}
	 + S_{1}^2 \vec{v}_{1}\cdot\vec{v}_{2}
	 - 4 S_{1}^2 \vec{v}_{1}\cdot\vec{n} \vec{v}_{2}\cdot\vec{n}
	 - 4 \vec{S}_{1}\cdot\vec{n} \vec{S}_{1}\cdot\vec{v}_{1} \vec{v}_{2}\cdot\vec{n} \nl
	 - 4 \vec{S}_{1}\cdot\vec{n} \vec{v}_{1}\cdot\vec{n} \vec{S}_{1}\cdot\vec{v}_{2}
	 - 2 \vec{v}_{1}\cdot\vec{v}_{2} ( \vec{S}_{1}\cdot\vec{n} )^{2}
	 + 12 \vec{v}_{1}\cdot\vec{n} \vec{v}_{2}\cdot\vec{n} ( \vec{S}_{1}\cdot\vec{n} )^{2} \Big] \nl
- \frac{G^2 m_{2}}{2 r^3} \Big[ 2 \vec{S}_{1}\cdot\dot{\vec{S}}_{1} \vec{v}_{2}\cdot\vec{n}
	 + \dot{\vec{S}}_{1}\cdot\vec{n} \vec{S}_{1}\cdot\vec{v}_{2}
	 + \vec{S}_{1}\cdot\vec{n} \dot{\vec{S}}_{1}\cdot\vec{v}_{2}
	 - 4 \vec{S}_{1}\cdot\vec{n} \dot{\vec{S}}_{1}\cdot\vec{n} \vec{v}_{2}\cdot\vec{n} \Big]\\
\text{Fig.~3(a4)} =&
\frac{4 G^2 m_{2}}{r^4} \Big[ \vec{S}_{1}\cdot\vec{v}_{1} \vec{S}_{1}\cdot\vec{v}_{2}
	 - S_{1}^2 \vec{v}_{1}\cdot\vec{v}_{2}
	 + 2 S_{1}^2 \vec{v}_{1}\cdot\vec{n} \vec{v}_{2}\cdot\vec{n}
	 - 2 \vec{S}_{1}\cdot\vec{n} \vec{S}_{1}\cdot\vec{v}_{1} \vec{v}_{2}\cdot\vec{n} \Big]\\
\text{Fig.~3(a5)} =&
\frac{8 G^2 C_{1ES^2} m_{2}^2}{m_{1} r^4} \Big[ S_{1}^2 v_{2}^2
	 - 2 v_{2}^2 ( \vec{S}_{1}\cdot\vec{n} )^{2} \Big]\\
\text{Fig.~3(a6)} =&
- \frac{4 G^2 C_{1ES^2} m_{2}}{r^4} \Big[ S_{1}^2 \vec{v}_{1}\cdot\vec{v}_{2}
	 - 2 \vec{S}_{1}\cdot\vec{n} \vec{S}_{1}\cdot\vec{v}_{1} \vec{v}_{2}\cdot\vec{n}
	 + 2 \vec{S}_{1}\cdot\vec{n} \vec{v}_{1}\cdot\vec{n} \vec{S}_{1}\cdot\vec{v}_{2} \nl
	 - 3 \vec{v}_{1}\cdot\vec{v}_{2} ( \vec{S}_{1}\cdot\vec{n} )^{2} \Big]\\
\text{Fig.~3(a7)} =&
- \frac{16 G^2 C_{1ES^2} m_{2}^2}{m_{1} r^4} \Big[ S_{1}^2 \vec{v}_{1}\cdot\vec{v}_{2}
	 - 2 \vec{v}_{1}\cdot\vec{v}_{2} ( \vec{S}_{1}\cdot\vec{n} )^{2} \Big]
+ \frac{8 G^2 C_{1ES^2} m_{2}^2}{m_{1} r^3} \Big[ 2 \vec{S}_{1}\cdot\dot{\vec{S}}_{1} \vec{v}_{2}\cdot\vec{n} \nl
	 - \dot{\vec{S}}_{1}\cdot\vec{n} \vec{S}_{1}\cdot\vec{v}_{2}
	 - \vec{S}_{1}\cdot\vec{n} \dot{\vec{S}}_{1}\cdot\vec{v}_{2} \Big]\\
\text{Fig.~3(a8)} =&
- \frac{4 G^2 C_{1ES^2} m_{2}}{r^4} \Big[ S_{1}^2 \vec{v}_{1}\cdot\vec{v}_{2}
	 + 2 \vec{S}_{1}\cdot\vec{n} \vec{S}_{1}\cdot\vec{v}_{1} \vec{v}_{2}\cdot\vec{n}
	 - 2 \vec{S}_{1}\cdot\vec{n} \vec{v}_{1}\cdot\vec{n} \vec{S}_{1}\cdot\vec{v}_{2} \nl
	 - 3 \vec{v}_{1}\cdot\vec{v}_{2} ( \vec{S}_{1}\cdot\vec{n} )^{2} \Big]
+ \frac{2 G^2 C_{1ES^2} m_{2}}{r^3} \Big[ 2 \vec{S}_{1}\cdot\dot{\vec{S}}_{1} \vec{v}_{2}\cdot\vec{n}
	 - 3 \dot{\vec{S}}_{1}\cdot\vec{n} \vec{S}_{1}\cdot\vec{v}_{2} \nl
	 - 3 \vec{S}_{1}\cdot\vec{n} \dot{\vec{S}}_{1}\cdot\vec{v}_{2} \Big]\\
\text{Fig.~3(a9)} =&
\frac{4 G^2 C_{1ES^2} m_{2}}{r^4} \Big[ S_{1}^2 v_{1}^2
	 - 3 v_{1}^2 ( \vec{S}_{1}\cdot\vec{n} )^{2} \Big]
- \frac{2 G^2 C_{1ES^2} m_{2}}{r^3} \Big[ 2 \vec{S}_{1}\cdot\dot{\vec{S}}_{1} \vec{v}_{1}\cdot\vec{n} \nl
	 - 3 \dot{\vec{S}}_{1}\cdot\vec{n} \vec{S}_{1}\cdot\vec{v}_{1}
	 - 3 \vec{S}_{1}\cdot\vec{n} \dot{\vec{S}}_{1}\cdot\vec{v}_{1} \Big]\\
\text{Fig.~3(b1)} =&
- \frac{G^2 m_{2}}{r^4} \Big[ 3 S_{1}^2 v_{1}^2
	 - 3 ( \vec{S}_{1}\cdot\vec{v}_{1} )^{2}
	 + 10 \vec{S}_{1}\cdot\vec{n} \vec{v}_{1}\cdot\vec{n} \vec{S}_{1}\cdot\vec{v}_{1}
	 - 2 v_{1}^2 ( \vec{S}_{1}\cdot\vec{n} )^{2} \nl
	 - 8 S_{1}^2 ( \vec{v}_{1}\cdot\vec{n} )^{2} \Big]\\
\text{Fig.~3(b2)} =&
\frac{4 G^2 C_{1ES^2} m_{2}^2}{m_{1} r^4} \Big[ ( \vec{S}_{1}\cdot\vec{v}_{2} )^{2}
	 - 8 \vec{S}_{1}\cdot\vec{n} \vec{v}_{2}\cdot\vec{n} \vec{S}_{1}\cdot\vec{v}_{2}
	 - 2 v_{2}^2 ( \vec{S}_{1}\cdot\vec{n} )^{2} \nl
	 + 12 ( \vec{S}_{1}\cdot\vec{n} )^{2} ( \vec{v}_{2}\cdot\vec{n} )^{2} \Big]\\
\text{Fig.~3(b3)} =&
\frac{G^2 C_{1ES^2} m_{2}}{r^4} \Big[ S_{1}^2 v_{1}^2
	 + 2 ( \vec{S}_{1}\cdot\vec{v}_{1} )^{2}
	 - 18 \vec{S}_{1}\cdot\vec{n} \vec{v}_{1}\cdot\vec{n} \vec{S}_{1}\cdot\vec{v}_{1}
	 - 5 v_{1}^2 ( \vec{S}_{1}\cdot\vec{n} )^{2} \nl
	 - 4 S_{1}^2 ( \vec{v}_{1}\cdot\vec{n} )^{2}
	 + 30 ( \vec{S}_{1}\cdot\vec{n} )^{2} ( \vec{v}_{1}\cdot\vec{n} )^{2} \Big]\\
\text{Fig.~3(b4)} =&
- \frac{G^2 C_{1ES^2} m_{2}}{4 r^4} \Big[ S_{1}^2 v_{2}^2
	 + 2 ( \vec{S}_{1}\cdot\vec{v}_{2} )^{2}
	 - 12 \vec{S}_{1}\cdot\vec{n} \vec{v}_{2}\cdot\vec{n} \vec{S}_{1}\cdot\vec{v}_{2}
	 - 5 v_{2}^2 ( \vec{S}_{1}\cdot\vec{n} )^{2} \nl
	 - S_{1}^2 ( \vec{v}_{2}\cdot\vec{n} )^{2}
	 + 15 ( \vec{S}_{1}\cdot\vec{n} )^{2} ( \vec{v}_{2}\cdot\vec{n} )^{2} \Big]\\
\text{Fig.~3(b5)} =&
- \frac{G^2 C_{1ES^2} m_{2}^2}{2 m_{1} r^4} \Big[ S_{1}^2 v_{1}^2
	 + ( \vec{S}_{1}\cdot\vec{v}_{1} )^{2}
	 - 8 \vec{S}_{1}\cdot\vec{n} \vec{v}_{1}\cdot\vec{n} \vec{S}_{1}\cdot\vec{v}_{1}
	 - 4 v_{1}^2 ( \vec{S}_{1}\cdot\vec{n} )^{2} \nl
	 + 12 ( \vec{S}_{1}\cdot\vec{n} )^{2} ( \vec{v}_{1}\cdot\vec{n} )^{2} \Big]
- \frac{G^2 C_{1ES^2} m_{2}^2}{m_{1} r^3} \Big[ \vec{S}_{1}\cdot\vec{n} \vec{S}_{1}\cdot\vec{a}_{1}
	 - \vec{a}_{1}\cdot\vec{n} ( \vec{S}_{1}\cdot\vec{n} )^{2} \Big] \nl
+ \frac{G^2 C_{1ES^2} m_{2}^2}{2 m_{1} r^3} \Big[ 2 \vec{S}_{1}\cdot\dot{\vec{S}}_{1} \vec{v}_{1}\cdot\vec{n}
	 - 3 \dot{\vec{S}}_{1}\cdot\vec{n} \vec{S}_{1}\cdot\vec{v}_{1}
	 - 3 \vec{S}_{1}\cdot\vec{n} \dot{\vec{S}}_{1}\cdot\vec{v}_{1} \nl
	 + 8 \vec{S}_{1}\cdot\vec{n} \dot{\vec{S}}_{1}\cdot\vec{n} \vec{v}_{1}\cdot\vec{n} \Big]
- \frac{G^2 C_{1ES^2} m_{2}^2}{2 m_{1} r^2} \Big[ \big( \vec{S}_{1}\cdot\ddot{\vec{S}}_{1}
	 + \vec{S}_{1}\cdot\vec{n} \ddot{\vec{S}}_{1}\cdot\vec{n} \big) \nl
	 + \big( \dot{S}_{1}^2
	 + ( \dot{\vec{S}}_{1}\cdot\vec{n} )^{2} \big) \Big]\\
\text{Fig.~3(b6)} =&
\frac{G^2 C_{1ES^2} m_{2}}{r^4} \Big[ ( \vec{S}_{1}\cdot\vec{v}_{1} )^{2}
	 - 6 \vec{S}_{1}\cdot\vec{n} \vec{v}_{1}\cdot\vec{n} \vec{S}_{1}\cdot\vec{v}_{1}
	 - v_{1}^2 ( \vec{S}_{1}\cdot\vec{n} )^{2}
	 + 6 ( \vec{S}_{1}\cdot\vec{n} )^{2} ( \vec{v}_{1}\cdot\vec{n} )^{2} \Big] \nl
- \frac{G^2 C_{1ES^2} m_{2}}{r^3} \Big[ S_{1}^2 \vec{a}_{1}\cdot\vec{n}
	 - \vec{S}_{1}\cdot\vec{n} \vec{S}_{1}\cdot\vec{a}_{1}
	 + 2 \vec{a}_{1}\cdot\vec{n} ( \vec{S}_{1}\cdot\vec{n} )^{2} \Big] \nl
- \frac{G^2 C_{1ES^2} m_{2}}{2 r^3} \Big[ 6 \vec{S}_{1}\cdot\dot{\vec{S}}_{1} \vec{v}_{1}\cdot\vec{n}
	 - 5 \dot{\vec{S}}_{1}\cdot\vec{n} \vec{S}_{1}\cdot\vec{v}_{1}
	 - 5 \vec{S}_{1}\cdot\vec{n} \dot{\vec{S}}_{1}\cdot\vec{v}_{1} \nl
	 + 16 \vec{S}_{1}\cdot\vec{n} \dot{\vec{S}}_{1}\cdot\vec{n} \vec{v}_{1}\cdot\vec{n} \Big]
+ \frac{4 G^2 C_{1ES^2} m_{2}}{r^2} \Big[ \vec{S}_{1}\cdot\vec{n} \ddot{\vec{S}}_{1}\cdot\vec{n}
	 + ( \dot{\vec{S}}_{1}\cdot\vec{n} )^{2} \Big]\\
\text{Fig.~3(c1)} =&
\frac{G^2 m_{2}}{r^4} \Big[ 2 S_{1}^2 v_{1}^2
	 - \vec{S}_{1}\cdot\vec{v}_{1} \vec{S}_{1}\cdot\vec{v}_{2}
	 + S_{1}^2 \vec{v}_{1}\cdot\vec{v}_{2}
	 - 2 ( \vec{S}_{1}\cdot\vec{v}_{1} )^{2}
	 + 6 \vec{S}_{1}\cdot\vec{n} \vec{v}_{1}\cdot\vec{n} \vec{S}_{1}\cdot\vec{v}_{1} \nl
	 - 4 S_{1}^2 \vec{v}_{1}\cdot\vec{n} \vec{v}_{2}\cdot\vec{n}
	 + 4 \vec{S}_{1}\cdot\vec{n} \vec{S}_{1}\cdot\vec{v}_{1} \vec{v}_{2}\cdot\vec{n}
	 - 2 v_{1}^2 ( \vec{S}_{1}\cdot\vec{n} )^{2}
	 - 4 S_{1}^2 ( \vec{v}_{1}\cdot\vec{n} )^{2} \Big] \nl
+ \frac{G^2 m_{2}}{r^3} \Big[ S_{1}^2 \vec{a}_{1}\cdot\vec{n}
	 - \vec{S}_{1}\cdot\vec{n} \vec{S}_{1}\cdot\vec{a}_{1} \Big]
+ \frac{G^2 m_{2}}{r^3} \Big[ \vec{S}_{1}\cdot\dot{\vec{S}}_{1} \vec{v}_{1}\cdot\vec{n}
	 - \dot{\vec{S}}_{1}\cdot\vec{n} \vec{S}_{1}\cdot\vec{v}_{1} \Big]\\
\text{Fig.~3(c2)} =&
- \frac{4 G^2 C_{1ES^2} m_{2}^2}{m_{1} r^4} \Big[ \vec{S}_{1}\cdot\vec{v}_{1} \vec{S}_{1}\cdot\vec{v}_{2}
	 + ( \vec{S}_{1}\cdot\vec{v}_{2} )^{2}
	 - 4 \vec{S}_{1}\cdot\vec{n} \vec{S}_{1}\cdot\vec{v}_{1} \vec{v}_{2}\cdot\vec{n} \nl
	 - 4 \vec{S}_{1}\cdot\vec{n} \vec{v}_{1}\cdot\vec{n} \vec{S}_{1}\cdot\vec{v}_{2}
	 - 8 \vec{S}_{1}\cdot\vec{n} \vec{v}_{2}\cdot\vec{n} \vec{S}_{1}\cdot\vec{v}_{2}
	 - 2 \vec{v}_{1}\cdot\vec{v}_{2} ( \vec{S}_{1}\cdot\vec{n} )^{2} \nl
	 - 2 v_{2}^2 ( \vec{S}_{1}\cdot\vec{n} )^{2}
	 + 12 \vec{v}_{1}\cdot\vec{n} \vec{v}_{2}\cdot\vec{n} ( \vec{S}_{1}\cdot\vec{n} )^{2}
	 + 12 ( \vec{S}_{1}\cdot\vec{n} )^{2} ( \vec{v}_{2}\cdot\vec{n} )^{2} \Big] \nl
- \frac{4 G^2 C_{1ES^2} m_{2}^2}{m_{1} r^3} \Big[ \dot{\vec{S}}_{1}\cdot\vec{n} \vec{S}_{1}\cdot\vec{v}_{2}
	 + \vec{S}_{1}\cdot\vec{n} \dot{\vec{S}}_{1}\cdot\vec{v}_{2}
	 - 4 \vec{S}_{1}\cdot\vec{n} \dot{\vec{S}}_{1}\cdot\vec{n} \vec{v}_{2}\cdot\vec{n} \Big]\\
\text{Fig.~3(c3)} =&
- \frac{G^2 C_{1ES^2} m_{2}}{r^4} \Big[ S_{1}^2 v_{1}^2
	 + 2 \vec{S}_{1}\cdot\vec{v}_{1} \vec{S}_{1}\cdot\vec{v}_{2}
	 + S_{1}^2 \vec{v}_{1}\cdot\vec{v}_{2}
	 + 2 ( \vec{S}_{1}\cdot\vec{v}_{1} )^{2} \nl
	 - 18 \vec{S}_{1}\cdot\vec{n} \vec{v}_{1}\cdot\vec{n} \vec{S}_{1}\cdot\vec{v}_{1}
	 - 4 S_{1}^2 \vec{v}_{1}\cdot\vec{n} \vec{v}_{2}\cdot\vec{n}
	 - 8 \vec{S}_{1}\cdot\vec{n} \vec{S}_{1}\cdot\vec{v}_{1} \vec{v}_{2}\cdot\vec{n} \nl
	 - 10 \vec{S}_{1}\cdot\vec{n} \vec{v}_{1}\cdot\vec{n} \vec{S}_{1}\cdot\vec{v}_{2}
	 - 5 v_{1}^2 ( \vec{S}_{1}\cdot\vec{n} )^{2}
	 - 5 \vec{v}_{1}\cdot\vec{v}_{2} ( \vec{S}_{1}\cdot\vec{n} )^{2}
	 - 4 S_{1}^2 ( \vec{v}_{1}\cdot\vec{n} )^{2} \nl
	 + 30 \vec{v}_{1}\cdot\vec{n} \vec{v}_{2}\cdot\vec{n} ( \vec{S}_{1}\cdot\vec{n} )^{2}
	 + 30 ( \vec{S}_{1}\cdot\vec{n} )^{2} ( \vec{v}_{1}\cdot\vec{n} )^{2} \Big]
- \frac{G^2 C_{1ES^2} m_{2}}{r^3} \Big[ 2 \vec{S}_{1}\cdot\dot{\vec{S}}_{1} \vec{v}_{1}\cdot\vec{n} \nl
	 + 3 \dot{\vec{S}}_{1}\cdot\vec{n} \vec{S}_{1}\cdot\vec{v}_{1}
	 + 3 \vec{S}_{1}\cdot\vec{n} \dot{\vec{S}}_{1}\cdot\vec{v}_{1}
	 - 12 \vec{S}_{1}\cdot\vec{n} \dot{\vec{S}}_{1}\cdot\vec{n} \vec{v}_{1}\cdot\vec{n} \Big]\\
\text{Fig.~3(c4)} =&
\frac{G^2 C_{1ES^2} m_{2}}{2 r^4} \Big[ 2 \vec{S}_{1}\cdot\vec{v}_{1} \vec{S}_{1}\cdot\vec{v}_{2}
	 + S_{1}^2 \vec{v}_{1}\cdot\vec{v}_{2}
	 - S_{1}^2 \vec{v}_{1}\cdot\vec{n} \vec{v}_{2}\cdot\vec{n}
	 - 6 \vec{S}_{1}\cdot\vec{n} \vec{S}_{1}\cdot\vec{v}_{1} \vec{v}_{2}\cdot\vec{n} \nl
	 - 6 \vec{S}_{1}\cdot\vec{n} \vec{v}_{1}\cdot\vec{n} \vec{S}_{1}\cdot\vec{v}_{2}
	 - 5 \vec{v}_{1}\cdot\vec{v}_{2} ( \vec{S}_{1}\cdot\vec{n} )^{2}
	 + 15 \vec{v}_{1}\cdot\vec{n} \vec{v}_{2}\cdot\vec{n} ( \vec{S}_{1}\cdot\vec{n} )^{2} \Big] \nl
+ \frac{G^2 C_{1ES^2} m_{2}}{r^3} \Big[ \dot{\vec{S}}_{1}\cdot\vec{n} \vec{S}_{1}\cdot\vec{v}_{2}
	 + \vec{S}_{1}\cdot\vec{n} \dot{\vec{S}}_{1}\cdot\vec{v}_{2}
	 - 2 \vec{S}_{1}\cdot\vec{n} \dot{\vec{S}}_{1}\cdot\vec{n} \vec{v}_{2}\cdot\vec{n} \Big]\\
\text{Fig.~3(c5)} =&
\frac{G^2 C_{1ES^2} m_{2}^2}{m_{1} r^4} \Big[ \vec{S}_{1}\cdot\vec{v}_{1} \vec{S}_{1}\cdot\vec{v}_{2}
	 + S_{1}^2 \vec{v}_{1}\cdot\vec{v}_{2}
	 - 4 \vec{S}_{1}\cdot\vec{n} \vec{S}_{1}\cdot\vec{v}_{1} \vec{v}_{2}\cdot\vec{n} \nl
	 - 4 \vec{S}_{1}\cdot\vec{n} \vec{v}_{1}\cdot\vec{n} \vec{S}_{1}\cdot\vec{v}_{2}
	 - 4 \vec{v}_{1}\cdot\vec{v}_{2} ( \vec{S}_{1}\cdot\vec{n} )^{2}
	 + 12 \vec{v}_{1}\cdot\vec{n} \vec{v}_{2}\cdot\vec{n} ( \vec{S}_{1}\cdot\vec{n} )^{2} \Big] \nl
- \frac{G^2 C_{1ES^2} m_{2}^2}{2 m_{1} r^3} \Big[ 2 \vec{S}_{1}\cdot\dot{\vec{S}}_{1} \vec{v}_{2}\cdot\vec{n}
	 - 3 \dot{\vec{S}}_{1}\cdot\vec{n} \vec{S}_{1}\cdot\vec{v}_{2}
	 - 3 \vec{S}_{1}\cdot\vec{n} \dot{\vec{S}}_{1}\cdot\vec{v}_{2} \nl
	 + 8 \vec{S}_{1}\cdot\vec{n} \dot{\vec{S}}_{1}\cdot\vec{n} \vec{v}_{2}\cdot\vec{n} \Big]\\
\text{Fig.~3(c6)} =&
- \frac{G^2 C_{1ES^2} m_{2}}{r^4} \Big[ \vec{S}_{1}\cdot\vec{v}_{1} \vec{S}_{1}\cdot\vec{v}_{2}
	 + ( \vec{S}_{1}\cdot\vec{v}_{1} )^{2}
	 - 6 \vec{S}_{1}\cdot\vec{n} \vec{v}_{1}\cdot\vec{n} \vec{S}_{1}\cdot\vec{v}_{1} \nl
	 - 4 \vec{S}_{1}\cdot\vec{n} \vec{S}_{1}\cdot\vec{v}_{1} \vec{v}_{2}\cdot\vec{n}
	 - 2 \vec{S}_{1}\cdot\vec{n} \vec{v}_{1}\cdot\vec{n} \vec{S}_{1}\cdot\vec{v}_{2}
	 - v_{1}^2 ( \vec{S}_{1}\cdot\vec{n} )^{2}
	 - \vec{v}_{1}\cdot\vec{v}_{2} ( \vec{S}_{1}\cdot\vec{n} )^{2} \nl
	 + 6 \vec{v}_{1}\cdot\vec{n} \vec{v}_{2}\cdot\vec{n} ( \vec{S}_{1}\cdot\vec{n} )^{2}
	 + 6 ( \vec{S}_{1}\cdot\vec{n} )^{2} ( \vec{v}_{1}\cdot\vec{n} )^{2} \Big]
+ \frac{G^2 C_{1ES^2} m_{2}}{2 r^3} \Big[ 2 \vec{S}_{1}\cdot\dot{\vec{S}}_{1} \vec{v}_{1}\cdot\vec{n} \nl
	 - 3 \dot{\vec{S}}_{1}\cdot\vec{n} \vec{S}_{1}\cdot\vec{v}_{1}
	 - 3 \vec{S}_{1}\cdot\vec{n} \dot{\vec{S}}_{1}\cdot\vec{v}_{1}
	 + 4 \vec{S}_{1}\cdot\dot{\vec{S}}_{1} \vec{v}_{2}\cdot\vec{n}
	 - 2 \dot{\vec{S}}_{1}\cdot\vec{n} \vec{S}_{1}\cdot\vec{v}_{2} \nl
	 - 2 \vec{S}_{1}\cdot\vec{n} \dot{\vec{S}}_{1}\cdot\vec{v}_{2}
	 + 8 \vec{S}_{1}\cdot\vec{n} \dot{\vec{S}}_{1}\cdot\vec{n} \vec{v}_{1}\cdot\vec{n}
	 + 8 \vec{S}_{1}\cdot\vec{n} \dot{\vec{S}}_{1}\cdot\vec{n} \vec{v}_{2}\cdot\vec{n} \Big]\\
\text{Fig.~3(d1)} =&
\frac{G^2 m_{2}}{4 r^4} \Big[ S_{1}^2 v_{2}^2
	 - 4 \vec{S}_{1}\cdot\vec{n} \vec{v}_{2}\cdot\vec{n} \vec{S}_{1}\cdot\vec{v}_{2}
	 - 3 v_{2}^2 ( \vec{S}_{1}\cdot\vec{n} )^{2}
	 - 3 S_{1}^2 ( \vec{v}_{2}\cdot\vec{n} )^{2} \nl
	 + 9 ( \vec{S}_{1}\cdot\vec{n} )^{2} ( \vec{v}_{2}\cdot\vec{n} )^{2} \Big]\\
\text{Fig.~3(d2)} =&
- \frac{G^2 m_{2}}{r^4} \Big[ 3 \vec{S}_{1}\cdot\vec{v}_{1} \vec{S}_{1}\cdot\vec{v}_{2}
	 - 3 S_{1}^2 \vec{v}_{1}\cdot\vec{v}_{2}
	 + 10 S_{1}^2 \vec{v}_{1}\cdot\vec{n} \vec{v}_{2}\cdot\vec{n}
	 - 6 \vec{S}_{1}\cdot\vec{n} \vec{S}_{1}\cdot\vec{v}_{1} \vec{v}_{2}\cdot\vec{n} \nl
	 - 4 \vec{S}_{1}\cdot\vec{n} \vec{v}_{1}\cdot\vec{n} \vec{S}_{1}\cdot\vec{v}_{2}
	 + 2 \vec{v}_{1}\cdot\vec{v}_{2} ( \vec{S}_{1}\cdot\vec{n} )^{2} \Big]\\
\text{Fig.~3(e)} =&
\frac{G^2 m_{2}}{r^4} \Big[ 3 S_{1}^2 v_{1}^2
	 + 2 \vec{S}_{1}\cdot\vec{v}_{1} \vec{S}_{1}\cdot\vec{v}_{2}
	 - 2 S_{1}^2 \vec{v}_{1}\cdot\vec{v}_{2}
	 - 3 ( \vec{S}_{1}\cdot\vec{v}_{1} )^{2}
	 + 14 \vec{S}_{1}\cdot\vec{n} \vec{v}_{1}\cdot\vec{n} \vec{S}_{1}\cdot\vec{v}_{1} \nl
	 + 8 S_{1}^2 \vec{v}_{1}\cdot\vec{n} \vec{v}_{2}\cdot\vec{n}
	 - 8 \vec{S}_{1}\cdot\vec{n} \vec{S}_{1}\cdot\vec{v}_{1} \vec{v}_{2}\cdot\vec{n}
	 - 4 \vec{S}_{1}\cdot\vec{n} \vec{v}_{1}\cdot\vec{n} \vec{S}_{1}\cdot\vec{v}_{2} \nl
	 - 2 \vec{v}_{1}\cdot\vec{v}_{2} ( \vec{S}_{1}\cdot\vec{n} )^{2}
	 - 8 S_{1}^2 ( \vec{v}_{1}\cdot\vec{n} )^{2}
	 + 12 \vec{v}_{1}\cdot\vec{n} \vec{v}_{2}\cdot\vec{n} ( \vec{S}_{1}\cdot\vec{n} )^{2}
	 - 12 ( \vec{S}_{1}\cdot\vec{n} )^{2} ( \vec{v}_{1}\cdot\vec{n} )^{2} \Big] \nl
+ \frac{G^2 m_{2}}{r^3} \Big[ 3 \vec{S}_{1}\cdot\dot{\vec{S}}_{1} \vec{v}_{1}\cdot\vec{n}
	 - 2 \dot{\vec{S}}_{1}\cdot\vec{n} \vec{S}_{1}\cdot\vec{v}_{1}
	 - \vec{S}_{1}\cdot\vec{n} \dot{\vec{S}}_{1}\cdot\vec{v}_{1}
	 + 4 \vec{S}_{1}\cdot\vec{n} \dot{\vec{S}}_{1}\cdot\vec{n} \vec{v}_{1}\cdot\vec{n} \Big]\\
\text{Fig.~3(f)} =&
\frac{G^2 m_{2}}{2 r^4} \Big[ 4 \vec{S}_{1}\cdot\vec{v}_{1} \vec{S}_{1}\cdot\vec{v}_{2}
	 - 5 S_{1}^2 \vec{v}_{1}\cdot\vec{v}_{2}
	 - 2 S_{1}^2 v_{2}^2
	 + 2 ( \vec{S}_{1}\cdot\vec{v}_{2} )^{2}
	 + 7 S_{1}^2 \vec{v}_{1}\cdot\vec{n} \vec{v}_{2}\cdot\vec{n} \nl
	 - 2 \vec{S}_{1}\cdot\vec{n} \vec{S}_{1}\cdot\vec{v}_{1} \vec{v}_{2}\cdot\vec{n}
	 - 6 \vec{S}_{1}\cdot\vec{n} \vec{v}_{1}\cdot\vec{n} \vec{S}_{1}\cdot\vec{v}_{2}
	 - 8 \vec{S}_{1}\cdot\vec{n} \vec{v}_{2}\cdot\vec{n} \vec{S}_{1}\cdot\vec{v}_{2} \nl
	 + 7 \vec{v}_{1}\cdot\vec{v}_{2} ( \vec{S}_{1}\cdot\vec{n} )^{2}
	 + 8 S_{1}^2 ( \vec{v}_{2}\cdot\vec{n} )^{2}
	 - 9 \vec{v}_{1}\cdot\vec{n} \vec{v}_{2}\cdot\vec{n} ( \vec{S}_{1}\cdot\vec{n} )^{2} \Big]
+ \frac{G^2 m_{2}}{r^3} \Big[ S_{1}^2 \vec{a}_{2}\cdot\vec{n} \nl
	 - \vec{S}_{1}\cdot\vec{n} \vec{S}_{1}\cdot\vec{a}_{2} \Big]
+ \frac{2 G^2 m_{2}}{r^3} \dot{\vec{S}}_{1}\cdot\vec{n} \vec{v}_{2}\cdot\vec{n} \vec{S}_{1}\cdot\vec{n} \\
\text{Fig.~3(g1)} =&
\frac{G^2 C_{1ES^2} m_{2}^2}{2 m_{1} r^4} \Big[ 8 \vec{S}_{1}\cdot\vec{v}_{1} \vec{S}_{1}\cdot\vec{v}_{2}
	 - S_{1}^2 v_{2}^2
	 - ( \vec{S}_{1}\cdot\vec{v}_{2} )^{2}
	 - 32 \vec{S}_{1}\cdot\vec{n} \vec{S}_{1}\cdot\vec{v}_{1} \vec{v}_{2}\cdot\vec{n} \nl
	 - 32 \vec{S}_{1}\cdot\vec{n} \vec{v}_{1}\cdot\vec{n} \vec{S}_{1}\cdot\vec{v}_{2}
	 + 8 \vec{S}_{1}\cdot\vec{n} \vec{v}_{2}\cdot\vec{n} \vec{S}_{1}\cdot\vec{v}_{2}
	 - 16 \vec{v}_{1}\cdot\vec{v}_{2} ( \vec{S}_{1}\cdot\vec{n} )^{2} \nl
	 + 4 v_{2}^2 ( \vec{S}_{1}\cdot\vec{n} )^{2}
	 + 96 \vec{v}_{1}\cdot\vec{n} \vec{v}_{2}\cdot\vec{n} ( \vec{S}_{1}\cdot\vec{n} )^{2}
	 - 12 ( \vec{S}_{1}\cdot\vec{n} )^{2} ( \vec{v}_{2}\cdot\vec{n} )^{2} \Big] \nl
+ \frac{4 G^2 C_{1ES^2} m_{2}^2}{m_{1} r^3} \Big[ \dot{\vec{S}}_{1}\cdot\vec{n} \vec{S}_{1}\cdot\vec{v}_{2}
	 + \vec{S}_{1}\cdot\vec{n} \dot{\vec{S}}_{1}\cdot\vec{v}_{2}
	 - 4 \vec{S}_{1}\cdot\vec{n} \dot{\vec{S}}_{1}\cdot\vec{n} \vec{v}_{2}\cdot\vec{n} \Big]\\
\text{Fig.~3(g2)} =&
- \frac{G^2 C_{1ES^2} m_{2}}{4 r^4} \Big[ S_{1}^2 v_{1}^2
	 - 12 \vec{S}_{1}\cdot\vec{v}_{1} \vec{S}_{1}\cdot\vec{v}_{2}
	 - 4 S_{1}^2 \vec{v}_{1}\cdot\vec{v}_{2}
	 + 2 ( \vec{S}_{1}\cdot\vec{v}_{1} )^{2} \nl
	 - 12 \vec{S}_{1}\cdot\vec{n} \vec{v}_{1}\cdot\vec{n} \vec{S}_{1}\cdot\vec{v}_{1}
	 + 16 S_{1}^2 \vec{v}_{1}\cdot\vec{n} \vec{v}_{2}\cdot\vec{n}
	 + 48 \vec{S}_{1}\cdot\vec{n} \vec{S}_{1}\cdot\vec{v}_{1} \vec{v}_{2}\cdot\vec{n} \nl
	 + 48 \vec{S}_{1}\cdot\vec{n} \vec{v}_{1}\cdot\vec{n} \vec{S}_{1}\cdot\vec{v}_{2}
	 - 5 v_{1}^2 ( \vec{S}_{1}\cdot\vec{n} )^{2}
	 + 24 \vec{v}_{1}\cdot\vec{v}_{2} ( \vec{S}_{1}\cdot\vec{n} )^{2}
	 - S_{1}^2 ( \vec{v}_{1}\cdot\vec{n} )^{2} \nl
	 - 144 \vec{v}_{1}\cdot\vec{n} \vec{v}_{2}\cdot\vec{n} ( \vec{S}_{1}\cdot\vec{n} )^{2}
	 + 15 ( \vec{S}_{1}\cdot\vec{n} )^{2} ( \vec{v}_{1}\cdot\vec{n} )^{2} \Big]
- \frac{G^2 C_{1ES^2} m_{2}}{r^3} \Big[ \dot{\vec{S}}_{1}\cdot\vec{n} \vec{S}_{1}\cdot\vec{v}_{1} \nl
	 + \vec{S}_{1}\cdot\vec{n} \dot{\vec{S}}_{1}\cdot\vec{v}_{1}
	 - 2 \vec{S}_{1}\cdot\dot{\vec{S}}_{1} \vec{v}_{2}\cdot\vec{n}
	 - 3 \dot{\vec{S}}_{1}\cdot\vec{n} \vec{S}_{1}\cdot\vec{v}_{2}
	 - 3 \vec{S}_{1}\cdot\vec{n} \dot{\vec{S}}_{1}\cdot\vec{v}_{2} \nl
	 - 2 \vec{S}_{1}\cdot\vec{n} \dot{\vec{S}}_{1}\cdot\vec{n} \vec{v}_{1}\cdot\vec{n}
	 + 12 \vec{S}_{1}\cdot\vec{n} \dot{\vec{S}}_{1}\cdot\vec{n} \vec{v}_{2}\cdot\vec{n} \Big]
\end{align}

\subsection{Cubic in G interaction}

\subsubsection{Three-graviton exchange}

\begin{figure}[t]
\begin{center}
\includegraphics[scale=0.6]{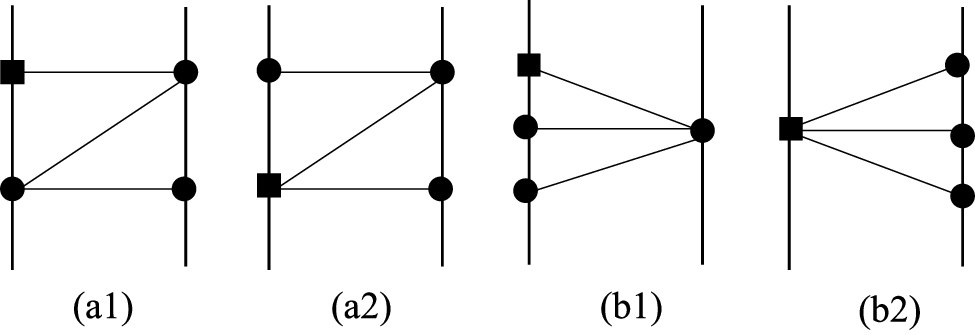}
\end{center}
\caption{NNLO spin-squared cubic in $G$ Feynman diagrams with no loops.}
\label{fig:ssnnlog3nonloop}
\end{figure}

There are 4 diagrams at order $G^3$ with three-graviton exchange in this sector,
which can be seen in figure \ref{fig:ssnnlog3nonloop}. They do not contain linear in spin couplings, but rather up to three-graviton spin-squared ones.

These diagrams are evaluated by:
\begin{align}
\text{Fig.~4(a1)}=& \frac{C_{1ES^2}}{2}\frac{G^3 m_2^2}{r^5}\left[S_1^2
-3\left(\vec{S}_1\cdot\vec{n}\right)^2\right],\\
\text{Fig.~4(a2)}=& 4C_{1ES^2}\frac{G^3 m_2^2}{r^5}\left[S_1^2
-3\left(\vec{S}_1\cdot\vec{n}\right)^2\right],\\
\text{Fig.~4(b1)}=& \frac{C_{1ES^2}}{4}\frac{G^3 m_1 m_2}{r^5}\left[S_1^2
-3\left(\vec{S}_1\cdot\vec{n}\right)^2\right],\\
\text{Fig.~4(b2)}=& \frac{15 C_{1ES^2}}{4}\frac{G^3 m_2^3}{m_1 r^5}\left[S_1^2
-3\left(\vec{S}_1\cdot\vec{n}\right)^2\right].
\end{align}

\subsubsection{Cubic self-interaction with two-graviton exchange}

\begin{figure}[t]
\begin{center}
\includegraphics[scale=0.6]{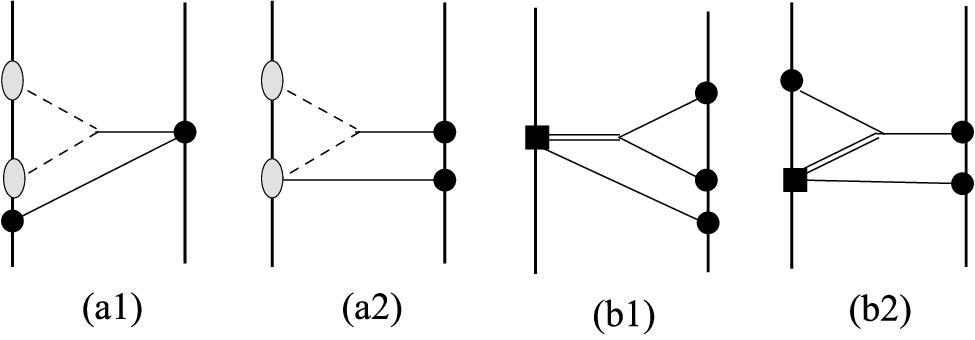}
\end{center}
\caption{NNLO spin-squared cubic in $G$ one-loop Feynman diagrams.}
\label{fig:ssnnlog3oneloop}
\end{figure}

There are 4 one-loop diagrams at order $G^3$ in this sector, which can be seen in figure \ref{fig:ssnnlog3oneloop}. Two-graviton spin-squared couplings also appear here.

These diagrams are evaluated by:
\begin{align}
\text{Fig.~5(a1)}=& -\frac{G^3 m_1 m_2} {r^5} \left(\vec{S}_1\cdot\vec{n}\right)^2,\\
\text{Fig.~5(a2)}=& -8\frac{G^3 m_2^2}{r^5}\left(\vec{S}_1\cdot\vec{n}\right)^2,\\
\text{Fig.~5(b1)}=& -\frac{C_{1ES^2}}{2}\frac{G^3 m_2^3}{m_1 r^5}\left[S_1^2+\left(\vec{S}_1\cdot\vec{n}\right)^2\right],\\
\text{Fig.~5(b2)}=& C_{1ES^2}\frac{G^3 m_2^2}{r^5}\left[5S_1^2
-3\left(\vec{S}_1\cdot\vec{n}\right)^2\right].
\end{align}

\subsubsection{Two-loop interaction}

\begin{figure}[t]
\includegraphics[scale=0.93]{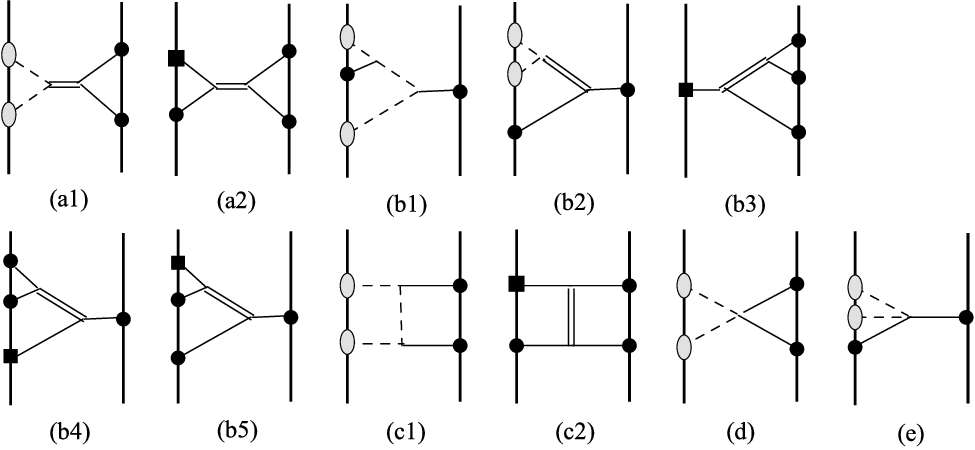}
\caption{NNLO spin-squared two-loop Feynman diagrams.}
\label{fig:ssnnlog3twoloop}
\end{figure}

There are 11 two-loop diagrams at order $G^3$ in this sector, which can be seen in figure \ref{fig:ssnnlog3twoloop}. Here, the cubic vertices have no time dependence, as in the spin1-spin2 sector \cite{Levi:2011eq}, but unlike the spin-orbit sector \cite{Levi:2015uxa}. This is similar to what happens in one-loop diagrams at the NLO level \cite{Levi:2008nh,Levi:2010zu,Levi:2015msa}. 

The irreducible two-loop tensor integrals, which are required here up to order 4, are reduced as in Appendix~A of \cite{Levi:2011eq}, see eqs.~(A12), (A13), and in Appendix~A of \cite{Levi:2015uxa}, see eq.~(A2).

These diagrams are evaluated by:
\begin{align}
\text{Fig.~6(a1)}=& 0,\\
\text{Fig.~6(a2)}=& 0,\\
\text{Fig.~6(b1)}=& \frac{4}{35}\frac{G^3 m_1 m_2}{r^5}\left[3S_1^2+5
\left(\vec{S}_1\cdot\vec{n}\right)^2\right],\\
\text{Fig.~6(b2)}=& -\frac{1}{70}\frac{G^3 m_1 m_2}{r^5}\left[13S_1^2
-25\left(\vec{S}_1\cdot\vec{n}\right)^2\right],\\
\text{Fig.~6(b3)}=& \frac{C_{1ES^2}}{2}\frac{G^3 m_2^3}{m_1 r^5}\left[3S_1^2
-5\left(\vec{S}_1\cdot\vec{n}\right)^2\right],\\
\text{Fig.~6(b4)}=& \frac{3C_{1ES^2}}{7}\frac{G^3 m_1 m_2}{r^5}\left[S_1^2
-3\left(\vec{S}_1\cdot\vec{n}\right)^2\right],\\
\text{Fig.~6(b5)}=& \frac{C_{1ES^2}}{7}\frac{G^3 m_1 m_2}{r^5}\left[S_1^2
-3\left(\vec{S}_1\cdot\vec{n}\right)^2\right],\\
\text{Fig.~6(c1)}=& 0,\\
\text{Fig.~6(c2)}=& C_{1ES^2}\frac{G^3 m_2^2}{r^5}\left[S_1^2
-15\left(\vec{S}_1\cdot\vec{n}\right)^2\right],\\
\text{Fig.~6(d)}=& 0,\\
\text{Fig.~6(e)}=& -\frac{4}{35}\frac{G^3 m_1 m_2}{r^5}\left[2S_1^2
+15\left(\vec{S}_1\cdot\vec{n}\right)^2\right].
\end{align}
Notice that the diagram in figure 6(c1) is equivalent to 0, despite its being irreducible, rather than of the factorizable kind.

\section{Next-to-next-to-leading order spin-squared potential} \label{result}

The NNLO spin-squared potential for a binary with compact spinning 
components is split according to the number and order of higher-order time derivatives:
\begin{align}
V_{\text{NNLO}}^{\text{SS}} = \stackrel{(0)}{V}
         + \stackrel{(1)}{V} + \stackrel{(2)}{V} + \stackrel{(3)}{V}. 
\end{align}
The first part is independent of higher-order time derivatives and is thus
an ordinary potential,
\begin{align}
& \stackrel{(0)}{V} =
- \frac{G C_{1ES^2} m_{2}}{16 m_{1} r^3} \Big[ 8 \vec{S}_{1}\cdot\vec{v}_{1} v_{1}^2 \vec{S}_{1}\cdot\vec{v}_{2}
	 - 42 S_{1}^2 v_{1}^2 \vec{v}_{1}\cdot\vec{v}_{2}
	 - 24 \vec{S}_{1}\cdot\vec{v}_{1} \vec{S}_{1}\cdot\vec{v}_{2} \vec{v}_{1}\cdot\vec{v}_{2}
	 + 11 S_{1}^2 v_{1}^2 v_{2}^2 \nl
	 + 12 \vec{S}_{1}\cdot\vec{v}_{1} \vec{S}_{1}\cdot\vec{v}_{2} v_{2}^2
	 - 22 S_{1}^2 \vec{v}_{1}\cdot\vec{v}_{2} v_{2}^2
	 - 12 v_{1}^2 ( \vec{S}_{1}\cdot\vec{v}_{1} )^{2}
	 + 24 \vec{v}_{1}\cdot\vec{v}_{2} ( \vec{S}_{1}\cdot\vec{v}_{1} )^{2}
	 - 10 v_{2}^2 ( \vec{S}_{1}\cdot\vec{v}_{1} )^{2} \nl
	 + 2 v_{1}^2 ( \vec{S}_{1}\cdot\vec{v}_{2} )^{2}
	 + 26 S_{1}^2 ( \vec{v}_{1}\cdot\vec{v}_{2} )^{2}
	 + 19 S_{1}^2 v_{1}^{4}
	 + 7 S_{1}^2 v_{2}^{4}
	 + 42 \vec{S}_{1}\cdot\vec{n} \vec{v}_{1}\cdot\vec{n} \vec{S}_{1}\cdot\vec{v}_{1} v_{1}^2 \nl
	 - 18 S_{1}^2 \vec{v}_{1}\cdot\vec{n} v_{1}^2 \vec{v}_{2}\cdot\vec{n}
	 - 24 \vec{S}_{1}\cdot\vec{n} \vec{S}_{1}\cdot\vec{v}_{1} v_{1}^2 \vec{v}_{2}\cdot\vec{n}
	 - 36 \vec{S}_{1}\cdot\vec{n} \vec{v}_{1}\cdot\vec{n} v_{1}^2 \vec{S}_{1}\cdot\vec{v}_{2} \nl
	 - 24 \vec{v}_{1}\cdot\vec{n} \vec{S}_{1}\cdot\vec{v}_{1} \vec{v}_{2}\cdot\vec{n} \vec{S}_{1}\cdot\vec{v}_{2}
	 - 12 \vec{S}_{1}\cdot\vec{n} v_{1}^2 \vec{v}_{2}\cdot\vec{n} \vec{S}_{1}\cdot\vec{v}_{2}
	 - 72 \vec{S}_{1}\cdot\vec{n} \vec{v}_{1}\cdot\vec{n} \vec{S}_{1}\cdot\vec{v}_{1} \vec{v}_{1}\cdot\vec{v}_{2} \nl
	 - 12 S_{1}^2 \vec{v}_{1}\cdot\vec{n} \vec{v}_{2}\cdot\vec{n} \vec{v}_{1}\cdot\vec{v}_{2}
	 + 72 \vec{S}_{1}\cdot\vec{n} \vec{S}_{1}\cdot\vec{v}_{1} \vec{v}_{2}\cdot\vec{n} \vec{v}_{1}\cdot\vec{v}_{2}
	 + 72 \vec{S}_{1}\cdot\vec{n} \vec{v}_{1}\cdot\vec{n} \vec{S}_{1}\cdot\vec{v}_{2} \vec{v}_{1}\cdot\vec{v}_{2} \nl
	 + 24 \vec{S}_{1}\cdot\vec{n} \vec{v}_{1}\cdot\vec{n} \vec{S}_{1}\cdot\vec{v}_{1} v_{2}^2
	 + 18 S_{1}^2 \vec{v}_{1}\cdot\vec{n} \vec{v}_{2}\cdot\vec{n} v_{2}^2
	 - 36 \vec{S}_{1}\cdot\vec{n} \vec{S}_{1}\cdot\vec{v}_{1} \vec{v}_{2}\cdot\vec{n} v_{2}^2
	 - 36 \vec{S}_{1}\cdot\vec{n} \vec{v}_{1}\cdot\vec{n} \vec{S}_{1}\cdot\vec{v}_{2} v_{2}^2 \nl
	 + 30 v_{1}^2 \vec{v}_{1}\cdot\vec{v}_{2} ( \vec{S}_{1}\cdot\vec{n} )^{2}
	 - 9 v_{1}^2 v_{2}^2 ( \vec{S}_{1}\cdot\vec{n} )^{2}
	 + 30 \vec{v}_{1}\cdot\vec{v}_{2} v_{2}^2 ( \vec{S}_{1}\cdot\vec{n} )^{2}
	 - 6 ( \vec{v}_{1}\cdot\vec{n} )^{2} ( \vec{S}_{1}\cdot\vec{v}_{1} )^{2} \nl
	 - 36 S_{1}^2 v_{1}^2 ( \vec{v}_{1}\cdot\vec{n} )^{2}
	 + 24 \vec{v}_{1}\cdot\vec{n} \vec{v}_{2}\cdot\vec{n} ( \vec{S}_{1}\cdot\vec{v}_{1} )^{2}
	 + 12 \vec{S}_{1}\cdot\vec{v}_{1} \vec{S}_{1}\cdot\vec{v}_{2} ( \vec{v}_{1}\cdot\vec{n} )^{2}
	 + 60 S_{1}^2 \vec{v}_{1}\cdot\vec{v}_{2} ( \vec{v}_{1}\cdot\vec{n} )^{2} \nl
	 - 27 S_{1}^2 v_{2}^2 ( \vec{v}_{1}\cdot\vec{n} )^{2}
	 - 6 ( \vec{S}_{1}\cdot\vec{v}_{1} )^{2} ( \vec{v}_{2}\cdot\vec{n} )^{2}
	 + 9 S_{1}^2 v_{1}^2 ( \vec{v}_{2}\cdot\vec{n} )^{2}
	 - 6 ( \vec{v}_{1}\cdot\vec{n} )^{2} ( \vec{S}_{1}\cdot\vec{v}_{2} )^{2} \nl
	 - 6 ( \vec{S}_{1}\cdot\vec{n} )^{2} ( \vec{v}_{1}\cdot\vec{v}_{2} )^{2}
	 - 21 ( \vec{S}_{1}\cdot\vec{n} )^{2} v_{1}^{4}
	 - 21 ( \vec{S}_{1}\cdot\vec{n} )^{2} v_{2}^{4}
	 + 90 \vec{v}_{1}\cdot\vec{n} v_{1}^2 \vec{v}_{2}\cdot\vec{n} ( \vec{S}_{1}\cdot\vec{n} )^{2} \nl
	 - 180 \vec{v}_{1}\cdot\vec{n} \vec{v}_{2}\cdot\vec{n} \vec{v}_{1}\cdot\vec{v}_{2} ( \vec{S}_{1}\cdot\vec{n} )^{2}
	 + 90 \vec{v}_{1}\cdot\vec{n} \vec{v}_{2}\cdot\vec{n} v_{2}^2 ( \vec{S}_{1}\cdot\vec{n} )^{2}
	 + 60 S_{1}^2 \vec{v}_{2}\cdot\vec{n} ( \vec{v}_{1}\cdot\vec{n} )^{3} \nl
	 - 60 \vec{S}_{1}\cdot\vec{n} \vec{S}_{1}\cdot\vec{v}_{1} \vec{v}_{2}\cdot\vec{n} ( \vec{v}_{1}\cdot\vec{n} )^{2}
	 + 60 \vec{S}_{1}\cdot\vec{n} \vec{v}_{2}\cdot\vec{n} \vec{S}_{1}\cdot\vec{v}_{2} ( \vec{v}_{1}\cdot\vec{n} )^{2}
	 + 15 v_{2}^2 ( \vec{S}_{1}\cdot\vec{n} )^{2} ( \vec{v}_{1}\cdot\vec{n} )^{2} \nl
	 - 45 S_{1}^2 ( \vec{v}_{1}\cdot\vec{n} )^{2} ( \vec{v}_{2}\cdot\vec{n} )^{2}
	 + 60 \vec{S}_{1}\cdot\vec{n} \vec{v}_{1}\cdot\vec{n} \vec{S}_{1}\cdot\vec{v}_{1} ( \vec{v}_{2}\cdot\vec{n} )^{2}
	 + 15 v_{1}^2 ( \vec{S}_{1}\cdot\vec{n} )^{2} ( \vec{v}_{2}\cdot\vec{n} )^{2} \nl
	 - 105 ( \vec{S}_{1}\cdot\vec{n} )^{2} ( \vec{v}_{1}\cdot\vec{n} )^{2} ( \vec{v}_{2}\cdot\vec{n} )^{2} \Big] \nl
+ \frac{G^2 C_{1ES^2} m_{2}^2}{m_{1} r^4} \Big[ 9 \vec{S}_{1}\cdot\vec{v}_{1} \vec{S}_{1}\cdot\vec{v}_{2}
	 - 10 S_{1}^2 \vec{v}_{1}\cdot\vec{v}_{2}
	 + 9 S_{1}^2 v_{2}^2
	 + ( \vec{S}_{1}\cdot\vec{v}_{1} )^{2}
	 - 10 ( \vec{S}_{1}\cdot\vec{v}_{2} )^{2} \nl
	 - 6 \vec{S}_{1}\cdot\vec{n} \vec{v}_{1}\cdot\vec{n} \vec{S}_{1}\cdot\vec{v}_{1}
	 + 32 S_{1}^2 \vec{v}_{1}\cdot\vec{n} \vec{v}_{2}\cdot\vec{n}
	 - 6 \vec{S}_{1}\cdot\vec{n} \vec{S}_{1}\cdot\vec{v}_{1} \vec{v}_{2}\cdot\vec{n}
	 - 38 \vec{S}_{1}\cdot\vec{n} \vec{v}_{1}\cdot\vec{n} \vec{S}_{1}\cdot\vec{v}_{2} \nl
	 + 44 \vec{S}_{1}\cdot\vec{n} \vec{v}_{2}\cdot\vec{n} \vec{S}_{1}\cdot\vec{v}_{2}
	 - v_{1}^2 ( \vec{S}_{1}\cdot\vec{n} )^{2}
	 + 2 \vec{v}_{1}\cdot\vec{v}_{2} ( \vec{S}_{1}\cdot\vec{n} )^{2}
	 + 2 v_{2}^2 ( \vec{S}_{1}\cdot\vec{n} )^{2}
	 - 36 S_{1}^2 ( \vec{v}_{2}\cdot\vec{n} )^{2} \nl
	 + 24 \vec{v}_{1}\cdot\vec{n} \vec{v}_{2}\cdot\vec{n} ( \vec{S}_{1}\cdot\vec{n} )^{2}
	 + 6 ( \vec{S}_{1}\cdot\vec{n} )^{2} ( \vec{v}_{1}\cdot\vec{n} )^{2}
	 - 12 ( \vec{S}_{1}\cdot\vec{n} )^{2} ( \vec{v}_{2}\cdot\vec{n} )^{2} \Big] \nl
	 - \frac{G^2 m_{2}}{4 r^4} \Big[ 5 S_{1}^2 v_{1}^2
	 + 4 \vec{S}_{1}\cdot\vec{v}_{1} \vec{S}_{1}\cdot\vec{v}_{2}
	 - 2 S_{1}^2 \vec{v}_{1}\cdot\vec{v}_{2}
	 - 3 S_{1}^2 v_{2}^2
	 - 8 ( \vec{S}_{1}\cdot\vec{v}_{1} )^{2}
	 + 4 ( \vec{S}_{1}\cdot\vec{v}_{2} )^{2} \nl
	 + 56 \vec{S}_{1}\cdot\vec{n} \vec{v}_{1}\cdot\vec{n} \vec{S}_{1}\cdot\vec{v}_{1}
	 + 6 S_{1}^2 \vec{v}_{1}\cdot\vec{n} \vec{v}_{2}\cdot\vec{n}
	 - 28 \vec{S}_{1}\cdot\vec{n} \vec{S}_{1}\cdot\vec{v}_{1} \vec{v}_{2}\cdot\vec{n}
	 - 4 \vec{S}_{1}\cdot\vec{n} \vec{v}_{1}\cdot\vec{n} \vec{S}_{1}\cdot\vec{v}_{2} \nl
	 - 20 \vec{S}_{1}\cdot\vec{n} \vec{v}_{2}\cdot\vec{n} \vec{S}_{1}\cdot\vec{v}_{2}
	 + 17 v_{1}^2 ( \vec{S}_{1}\cdot\vec{n} )^{2}
	 - 22 \vec{v}_{1}\cdot\vec{v}_{2} ( \vec{S}_{1}\cdot\vec{n} )^{2}
	 + 3 v_{2}^2 ( \vec{S}_{1}\cdot\vec{n} )^{2}
	 - 19 S_{1}^2 ( \vec{v}_{1}\cdot\vec{n} )^{2} \nl
	 + 13 S_{1}^2 ( \vec{v}_{2}\cdot\vec{n} )^{2}
	 + 54 \vec{v}_{1}\cdot\vec{n} \vec{v}_{2}\cdot\vec{n} ( \vec{S}_{1}\cdot\vec{n} )^{2}
	 - 75 ( \vec{S}_{1}\cdot\vec{n} )^{2} ( \vec{v}_{1}\cdot\vec{n} )^{2}
	 + 9 ( \vec{S}_{1}\cdot\vec{n} )^{2} ( \vec{v}_{2}\cdot\vec{n} )^{2} \Big] \nl
	 - \frac{G^2 C_{1ES^2} m_{2}}{4 r^4} \Big[ 7 S_{1}^2 v_{1}^2
	 + 2 \vec{S}_{1}\cdot\vec{v}_{1} \vec{S}_{1}\cdot\vec{v}_{2}
	 - 14 S_{1}^2 \vec{v}_{1}\cdot\vec{v}_{2}
	 + 8 S_{1}^2 v_{2}^2
	 - 2 ( \vec{S}_{1}\cdot\vec{v}_{2} )^{2} \nl
	 - 2 \vec{S}_{1}\cdot\vec{n} \vec{v}_{1}\cdot\vec{n} \vec{S}_{1}\cdot\vec{v}_{1}
	 - 2 S_{1}^2 \vec{v}_{1}\cdot\vec{n} \vec{v}_{2}\cdot\vec{n}
	 - 4 \vec{S}_{1}\cdot\vec{n} \vec{S}_{1}\cdot\vec{v}_{1} \vec{v}_{2}\cdot\vec{n}
	 + 12 \vec{S}_{1}\cdot\vec{n} \vec{v}_{2}\cdot\vec{n} \vec{S}_{1}\cdot\vec{v}_{2} \nl
	 - 25 v_{1}^2 ( \vec{S}_{1}\cdot\vec{n} )^{2}
	 + 44 \vec{v}_{1}\cdot\vec{v}_{2} ( \vec{S}_{1}\cdot\vec{n} )^{2}
	 - 22 v_{2}^2 ( \vec{S}_{1}\cdot\vec{n} )^{2}
	 + 5 S_{1}^2 ( \vec{v}_{1}\cdot\vec{n} )^{2}
	 + S_{1}^2 ( \vec{v}_{2}\cdot\vec{n} )^{2} \nl
	 - 6 \vec{v}_{1}\cdot\vec{n} \vec{v}_{2}\cdot\vec{n} ( \vec{S}_{1}\cdot\vec{n} )^{2}
	 + 3 ( \vec{S}_{1}\cdot\vec{n} )^{2} ( \vec{v}_{1}\cdot\vec{n} )^{2}
	 - 15 ( \vec{S}_{1}\cdot\vec{n} )^{2} ( \vec{v}_{2}\cdot\vec{n} )^{2} \Big] \nl
+ \frac{G^3 m_{1} m_{2}}{14 r^5} \Big[ S_{1}^2
 	 + 25 ( \vec{S}_{1}\cdot\vec{n} )^{2} \Big]	 
 + \frac{8 G^3 m_{2}^2}{r^5} ( \vec{S}_{1}\cdot\vec{n} )^{2} 
 - \frac{23 G^3 C_{1ES^2} m_{1} m_{2}}{28 r^5} \Big[ S_{1}^2
 	 - 3 ( \vec{S}_{1}\cdot\vec{n} )^{2} \Big] \nl
 - \frac{21 G^3 C_{1ES^2} m_{2}^2}{2 r^5} \Big[ S_{1}^2
 	 - 3 ( \vec{S}_{1}\cdot\vec{n} )^{2} \Big]
 - \frac{19 G^3 C_{1ES^2} m_{2}^3}{4 m_{1} r^5} \Big[ S_{1}^2
 	 - 3 ( \vec{S}_{1}\cdot\vec{n} )^{2} \Big]
+ (1 \leftrightarrow 2).
\end{align}
The part linear in higher-order time derivatives is further split due
to its length as
\begin{align}
\stackrel{(1)}{V} = V_{a} + V_{\dot{S}} ,
\end{align}
where
\begin{align}
& V_{a} =
- \frac{G C_{1ES^2} m_{2}}{16 m_{1} r^2} \Big[ 8 S_{1}^2 v_{1}^2 \vec{a}_{1}\cdot\vec{n}
	 - 16 \vec{v}_{1}\cdot\vec{n} \vec{S}_{1}\cdot\vec{v}_{1} \vec{S}_{1}\cdot\vec{a}_{1}
	 + 16 S_{1}^2 \vec{v}_{1}\cdot\vec{n} \vec{v}_{1}\cdot\vec{a}_{1}
	 + 28 S_{1}^2 \vec{v}_{1}\cdot\vec{a}_{1} \vec{v}_{2}\cdot\vec{n} \nl
	 + 12 \vec{S}_{1}\cdot\vec{v}_{1} \vec{a}_{1}\cdot\vec{n} \vec{S}_{1}\cdot\vec{v}_{2}
	 + 20 \vec{v}_{1}\cdot\vec{n} \vec{S}_{1}\cdot\vec{a}_{1} \vec{S}_{1}\cdot\vec{v}_{2}
	 - 8 \vec{S}_{1}\cdot\vec{n} \vec{v}_{1}\cdot\vec{a}_{1} \vec{S}_{1}\cdot\vec{v}_{2}
	 + 4 \vec{S}_{1}\cdot\vec{a}_{1} \vec{v}_{2}\cdot\vec{n} \vec{S}_{1}\cdot\vec{v}_{2} \nl
	 - 24 S_{1}^2 \vec{a}_{1}\cdot\vec{n} \vec{v}_{1}\cdot\vec{v}_{2}
	 + 20 \vec{S}_{1}\cdot\vec{n} \vec{S}_{1}\cdot\vec{a}_{1} \vec{v}_{1}\cdot\vec{v}_{2}
	 - 24 S_{1}^2 \vec{v}_{1}\cdot\vec{n} \vec{a}_{1}\cdot\vec{v}_{2}
	 + 12 \vec{S}_{1}\cdot\vec{n} \vec{S}_{1}\cdot\vec{v}_{1} \vec{a}_{1}\cdot\vec{v}_{2} \nl
	 - 22 S_{1}^2 \vec{v}_{2}\cdot\vec{n} \vec{a}_{1}\cdot\vec{v}_{2}
	 + 4 \vec{S}_{1}\cdot\vec{n} \vec{S}_{1}\cdot\vec{v}_{2} \vec{a}_{1}\cdot\vec{v}_{2}
	 + 13 S_{1}^2 \vec{a}_{1}\cdot\vec{n} v_{2}^2
	 - 6 \vec{S}_{1}\cdot\vec{n} \vec{S}_{1}\cdot\vec{a}_{1} v_{2}^2
	 + 3 S_{1}^2 v_{1}^2 \vec{a}_{2}\cdot\vec{n} \nl
	 - 4 \vec{v}_{1}\cdot\vec{n} \vec{S}_{1}\cdot\vec{v}_{1} \vec{S}_{1}\cdot\vec{a}_{2}
	 - 2 \vec{S}_{1}\cdot\vec{n} v_{1}^2 \vec{S}_{1}\cdot\vec{a}_{2}
	 - 10 S_{1}^2 \vec{v}_{1}\cdot\vec{n} \vec{v}_{1}\cdot\vec{a}_{2}
	 + 28 \vec{S}_{1}\cdot\vec{n} \vec{S}_{1}\cdot\vec{v}_{1} \vec{v}_{1}\cdot\vec{a}_{2} \nl
	 + 12 S_{1}^2 \vec{v}_{1}\cdot\vec{n} \vec{v}_{2}\cdot\vec{a}_{2}
	 - 24 \vec{S}_{1}\cdot\vec{n} \vec{S}_{1}\cdot\vec{v}_{1} \vec{v}_{2}\cdot\vec{a}_{2}
	 - 2 \vec{a}_{2}\cdot\vec{n} ( \vec{S}_{1}\cdot\vec{v}_{1} )^{2}
	 - 14 \vec{a}_{1}\cdot\vec{n} ( \vec{S}_{1}\cdot\vec{v}_{2} )^{2} \nl
	 - 24 S_{1}^2 \vec{v}_{1}\cdot\vec{n} \vec{a}_{1}\cdot\vec{n} \vec{v}_{2}\cdot\vec{n}
	 + 12 \vec{S}_{1}\cdot\vec{n} \vec{S}_{1}\cdot\vec{v}_{1} \vec{a}_{1}\cdot\vec{n} \vec{v}_{2}\cdot\vec{n}
	 - 12 \vec{S}_{1}\cdot\vec{n} \vec{v}_{1}\cdot\vec{n} \vec{S}_{1}\cdot\vec{a}_{1} \vec{v}_{2}\cdot\vec{n} \nl
	 - 12 \vec{S}_{1}\cdot\vec{n} \vec{a}_{1}\cdot\vec{n} \vec{v}_{2}\cdot\vec{n} \vec{S}_{1}\cdot\vec{v}_{2}
	 + 12 \vec{S}_{1}\cdot\vec{n} \vec{v}_{1}\cdot\vec{n} \vec{S}_{1}\cdot\vec{v}_{1} \vec{a}_{2}\cdot\vec{n}
	 - 36 \vec{v}_{1}\cdot\vec{a}_{1} \vec{v}_{2}\cdot\vec{n} ( \vec{S}_{1}\cdot\vec{n} )^{2} \nl
	 + 42 \vec{v}_{2}\cdot\vec{n} \vec{a}_{1}\cdot\vec{v}_{2} ( \vec{S}_{1}\cdot\vec{n} )^{2}
	 - 3 \vec{a}_{1}\cdot\vec{n} v_{2}^2 ( \vec{S}_{1}\cdot\vec{n} )^{2}
	 + 3 v_{1}^2 \vec{a}_{2}\cdot\vec{n} ( \vec{S}_{1}\cdot\vec{n} )^{2}
	 - 42 \vec{v}_{1}\cdot\vec{n} \vec{v}_{1}\cdot\vec{a}_{2} ( \vec{S}_{1}\cdot\vec{n} )^{2} \nl
	 + 36 \vec{v}_{1}\cdot\vec{n} \vec{v}_{2}\cdot\vec{a}_{2} ( \vec{S}_{1}\cdot\vec{n} )^{2}
	 - 9 S_{1}^2 \vec{a}_{2}\cdot\vec{n} ( \vec{v}_{1}\cdot\vec{n} )^{2}
	 + 6 \vec{S}_{1}\cdot\vec{n} \vec{S}_{1}\cdot\vec{a}_{2} ( \vec{v}_{1}\cdot\vec{n} )^{2}
	 + 9 S_{1}^2 \vec{a}_{1}\cdot\vec{n} ( \vec{v}_{2}\cdot\vec{n} )^{2} \nl
	 - 6 \vec{S}_{1}\cdot\vec{n} \vec{S}_{1}\cdot\vec{a}_{1} ( \vec{v}_{2}\cdot\vec{n} )^{2}
	 - 15 \vec{a}_{2}\cdot\vec{n} ( \vec{S}_{1}\cdot\vec{n} )^{2} ( \vec{v}_{1}\cdot\vec{n} )^{2}
	 + 15 \vec{a}_{1}\cdot\vec{n} ( \vec{S}_{1}\cdot\vec{n} )^{2} ( \vec{v}_{2}\cdot\vec{n} )^{2} \Big] \nl
+ \frac{G^2 C_{1ES^2} m_{2}^2}{m_{1} r^3} \Big[ \vec{S}_{1}\cdot\vec{n} \vec{S}_{1}\cdot\vec{a}_{1}
	 - 11 S_{1}^2 \vec{a}_{2}\cdot\vec{n}
	 + 14 \vec{S}_{1}\cdot\vec{n} \vec{S}_{1}\cdot\vec{a}_{2}
	 - \vec{a}_{1}\cdot\vec{n} ( \vec{S}_{1}\cdot\vec{n} )^{2} \Big] \nl
+ \frac{G^2 C_{1ES^2} m_{2}}{r^3} \Big[ S_{1}^2 \vec{a}_{1}\cdot\vec{n}
	 + 2 \vec{a}_{1}\cdot\vec{n} ( \vec{S}_{1}\cdot\vec{n} )^{2} \Big] \nl
- \frac{G^2 m_{2}}{r^3} \Big[ 3 S_{1}^2 \vec{a}_{1}\cdot\vec{n}
	 - 3 \vec{S}_{1}\cdot\vec{n} \vec{S}_{1}\cdot\vec{a}_{1}
	 + S_{1}^2 \vec{a}_{2}\cdot\vec{n}
	 - \vec{S}_{1}\cdot\vec{n} \vec{S}_{1}\cdot\vec{a}_{2} \Big]
+ (1 \leftrightarrow 2), 
\end{align}
and
\begin{align}
& V_{\dot{S}} =
\frac{G C_{1ES^2} m_{2}}{4 m_{1} r^2} \Big[ 4 \vec{v}_{1}\cdot\vec{n} \vec{S}_{1}\cdot\vec{v}_{1} \dot{\vec{S}}_{1}\cdot\vec{v}_{1}
	 - 4 \vec{S}_{1}\cdot\dot{\vec{S}}_{1} \vec{v}_{1}\cdot\vec{n} v_{1}^2
	 - 2 \vec{S}_{1}\cdot\vec{v}_{1} \dot{\vec{S}}_{1}\cdot\vec{v}_{1} \vec{v}_{2}\cdot\vec{n} \nl
	 - 3 \vec{S}_{1}\cdot\dot{\vec{S}}_{1} v_{1}^2 \vec{v}_{2}\cdot\vec{n}
	 - 5 \vec{v}_{1}\cdot\vec{n} \dot{\vec{S}}_{1}\cdot\vec{v}_{1} \vec{S}_{1}\cdot\vec{v}_{2}
	 + \dot{\vec{S}}_{1}\cdot\vec{n} v_{1}^2 \vec{S}_{1}\cdot\vec{v}_{2}
	 + 2 \dot{\vec{S}}_{1}\cdot\vec{v}_{1} \vec{v}_{2}\cdot\vec{n} \vec{S}_{1}\cdot\vec{v}_{2} \nl
	 - 5 \vec{v}_{1}\cdot\vec{n} \vec{S}_{1}\cdot\vec{v}_{1} \dot{\vec{S}}_{1}\cdot\vec{v}_{2}
	 + \vec{S}_{1}\cdot\vec{n} v_{1}^2 \dot{\vec{S}}_{1}\cdot\vec{v}_{2}
	 + 2 \vec{S}_{1}\cdot\vec{v}_{1} \vec{v}_{2}\cdot\vec{n} \dot{\vec{S}}_{1}\cdot\vec{v}_{2}
	 + 6 \vec{v}_{1}\cdot\vec{n} \vec{S}_{1}\cdot\vec{v}_{2} \dot{\vec{S}}_{1}\cdot\vec{v}_{2} \nl
	 + 20 \vec{S}_{1}\cdot\dot{\vec{S}}_{1} \vec{v}_{1}\cdot\vec{n} \vec{v}_{1}\cdot\vec{v}_{2}
	 - 5 \dot{\vec{S}}_{1}\cdot\vec{n} \vec{S}_{1}\cdot\vec{v}_{1} \vec{v}_{1}\cdot\vec{v}_{2}
	 - 5 \vec{S}_{1}\cdot\vec{n} \dot{\vec{S}}_{1}\cdot\vec{v}_{1} \vec{v}_{1}\cdot\vec{v}_{2}
	 - 2 \vec{S}_{1}\cdot\dot{\vec{S}}_{1} \vec{v}_{2}\cdot\vec{n} \vec{v}_{1}\cdot\vec{v}_{2} \nl
	 + 2 \dot{\vec{S}}_{1}\cdot\vec{n} \vec{S}_{1}\cdot\vec{v}_{2} \vec{v}_{1}\cdot\vec{v}_{2}
	 + 2 \vec{S}_{1}\cdot\vec{n} \dot{\vec{S}}_{1}\cdot\vec{v}_{2} \vec{v}_{1}\cdot\vec{v}_{2}
	 - 9 \vec{S}_{1}\cdot\dot{\vec{S}}_{1} \vec{v}_{1}\cdot\vec{n} v_{2}^2
	 + \dot{\vec{S}}_{1}\cdot\vec{n} \vec{S}_{1}\cdot\vec{v}_{1} v_{2}^2 \nl
	 + \vec{S}_{1}\cdot\vec{n} \dot{\vec{S}}_{1}\cdot\vec{v}_{1} v_{2}^2
	 - 5 \vec{S}_{1}\cdot\dot{\vec{S}}_{1} \vec{v}_{2}\cdot\vec{n} v_{2}^2
	 + \dot{\vec{S}}_{1}\cdot\vec{n} \vec{S}_{1}\cdot\vec{v}_{2} v_{2}^2
	 + \vec{S}_{1}\cdot\vec{n} \dot{\vec{S}}_{1}\cdot\vec{v}_{2} v_{2}^2 \nl
	 + 3 \dot{\vec{S}}_{1}\cdot\vec{n} \vec{v}_{1}\cdot\vec{n} \vec{S}_{1}\cdot\vec{v}_{1} \vec{v}_{2}\cdot\vec{n}
	 + 3 \vec{S}_{1}\cdot\vec{n} \vec{v}_{1}\cdot\vec{n} \dot{\vec{S}}_{1}\cdot\vec{v}_{1} \vec{v}_{2}\cdot\vec{n}
	 + 9 \vec{S}_{1}\cdot\vec{n} \dot{\vec{S}}_{1}\cdot\vec{n} v_{1}^2 \vec{v}_{2}\cdot\vec{n} \nl
	 - 6 \dot{\vec{S}}_{1}\cdot\vec{n} \vec{v}_{1}\cdot\vec{n} \vec{v}_{2}\cdot\vec{n} \vec{S}_{1}\cdot\vec{v}_{2}
	 - 6 \vec{S}_{1}\cdot\vec{n} \vec{v}_{1}\cdot\vec{n} \vec{v}_{2}\cdot\vec{n} \dot{\vec{S}}_{1}\cdot\vec{v}_{2}
	 - 18 \vec{S}_{1}\cdot\vec{n} \dot{\vec{S}}_{1}\cdot\vec{n} \vec{v}_{2}\cdot\vec{n} \vec{v}_{1}\cdot\vec{v}_{2} \nl
	 + 3 \vec{S}_{1}\cdot\vec{n} \dot{\vec{S}}_{1}\cdot\vec{n} \vec{v}_{1}\cdot\vec{n} v_{2}^2
	 + 9 \vec{S}_{1}\cdot\vec{n} \dot{\vec{S}}_{1}\cdot\vec{n} \vec{v}_{2}\cdot\vec{n} v_{2}^2
	 - 6 \vec{S}_{1}\cdot\dot{\vec{S}}_{1} \vec{v}_{2}\cdot\vec{n} ( \vec{v}_{1}\cdot\vec{n} )^{2} \nl
	 + 15 \vec{S}_{1}\cdot\dot{\vec{S}}_{1} \vec{v}_{1}\cdot\vec{n} ( \vec{v}_{2}\cdot\vec{n} )^{2}
	 + 3 \dot{\vec{S}}_{1}\cdot\vec{n} \vec{S}_{1}\cdot\vec{v}_{1} ( \vec{v}_{2}\cdot\vec{n} )^{2}
	 + 3 \vec{S}_{1}\cdot\vec{n} \dot{\vec{S}}_{1}\cdot\vec{v}_{1} ( \vec{v}_{2}\cdot\vec{n} )^{2} \nl
	 - 15 \vec{S}_{1}\cdot\vec{n} \dot{\vec{S}}_{1}\cdot\vec{n} \vec{v}_{1}\cdot\vec{n} ( \vec{v}_{2}\cdot\vec{n} )^{2} \Big] \nl
- \frac{G^2 m_{2}}{r^3} \Big[ 3 \vec{S}_{1}\cdot\dot{\vec{S}}_{1} \vec{v}_{1}\cdot\vec{n}
	 - 6 \dot{\vec{S}}_{1}\cdot\vec{n} \vec{S}_{1}\cdot\vec{v}_{1}
	 - 5 \vec{S}_{1}\cdot\vec{n} \dot{\vec{S}}_{1}\cdot\vec{v}_{1}
	 + 2 \vec{S}_{1}\cdot\dot{\vec{S}}_{1} \vec{v}_{2}\cdot\vec{n}
	 + 2 \dot{\vec{S}}_{1}\cdot\vec{n} \vec{S}_{1}\cdot\vec{v}_{2} \nl
	 + 2 \vec{S}_{1}\cdot\vec{n} \dot{\vec{S}}_{1}\cdot\vec{v}_{2}
	 + 14 \vec{S}_{1}\cdot\vec{n} \dot{\vec{S}}_{1}\cdot\vec{n} \vec{v}_{1}\cdot\vec{n}
	 - 6 \vec{S}_{1}\cdot\vec{n} \dot{\vec{S}}_{1}\cdot\vec{n} \vec{v}_{2}\cdot\vec{n} \Big] \nl
+ \frac{G^2 C_{1ES^2} m_{2}}{2 r^3} \Big[ 12 \vec{S}_{1}\cdot\dot{\vec{S}}_{1} \vec{v}_{1}\cdot\vec{n}
	 - 4 \dot{\vec{S}}_{1}\cdot\vec{n} \vec{S}_{1}\cdot\vec{v}_{1}
	 - 4 \vec{S}_{1}\cdot\vec{n} \dot{\vec{S}}_{1}\cdot\vec{v}_{1}
	 - 10 \vec{S}_{1}\cdot\dot{\vec{S}}_{1} \vec{v}_{2}\cdot\vec{n} \nl
	 + 3 \dot{\vec{S}}_{1}\cdot\vec{n} \vec{S}_{1}\cdot\vec{v}_{2}
	 + 3 \vec{S}_{1}\cdot\vec{n} \dot{\vec{S}}_{1}\cdot\vec{v}_{2}
	 - 16 \vec{S}_{1}\cdot\vec{n} \dot{\vec{S}}_{1}\cdot\vec{n} \vec{v}_{1}\cdot\vec{n}
	 + 16 \vec{S}_{1}\cdot\vec{n} \dot{\vec{S}}_{1}\cdot\vec{n} \vec{v}_{2}\cdot\vec{n} \Big] \nl
	 - \frac{G^2 C_{1ES^2} m_{2}^2}{2 m_{1} r^3} \Big[ 2 \vec{S}_{1}\cdot\dot{\vec{S}}_{1} \vec{v}_{1}\cdot\vec{n}
	 - 3 \dot{\vec{S}}_{1}\cdot\vec{n} \vec{S}_{1}\cdot\vec{v}_{1}
	 - 3 \vec{S}_{1}\cdot\vec{n} \dot{\vec{S}}_{1}\cdot\vec{v}_{1}
	 + 30 \vec{S}_{1}\cdot\dot{\vec{S}}_{1} \vec{v}_{2}\cdot\vec{n} \nl
	 - 19 \dot{\vec{S}}_{1}\cdot\vec{n} \vec{S}_{1}\cdot\vec{v}_{2}
	 - 19 \vec{S}_{1}\cdot\vec{n} \dot{\vec{S}}_{1}\cdot\vec{v}_{2}
	 + 8 \vec{S}_{1}\cdot\vec{n} \dot{\vec{S}}_{1}\cdot\vec{n} \vec{v}_{1}\cdot\vec{n}
	 + 16 \vec{S}_{1}\cdot\vec{n} \dot{\vec{S}}_{1}\cdot\vec{n} \vec{v}_{2}\cdot\vec{n} \Big] \nl
+ (1 \leftrightarrow 2).
\end{align}
The contribution with two higher-order time derivatives reads
\begin{align}
& \stackrel{(2)}{V} =
- \frac{G C_{1ES^2} m_{2}}{16 m_{1} r} \Big[ 8 \big( 2 \vec{S}_{1}\cdot\vec{v}_{1} \vec{S}_{1}\cdot\dot{\vec{a}}_{1}
	 - \vec{S}_{1}\cdot\dot{\vec{a}}_{1} \vec{S}_{1}\cdot\vec{v}_{2}
	 + \vec{S}_{1}\cdot\vec{n} \vec{S}_{1}\cdot\dot{\vec{a}}_{1} \vec{v}_{2}\cdot\vec{n} \big)
	 + 2 \big( 8 \vec{S}_{1}\cdot\vec{v}_{1} \ddot{\vec{S}}_{1}\cdot\vec{v}_{1} \nl
	 - 8 \vec{S}_{1}\cdot\ddot{\vec{S}}_{1} v_{1}^2
	 - 4 \ddot{\vec{S}}_{1}\cdot\vec{v}_{1} \vec{S}_{1}\cdot\vec{v}_{2}
	 - 4 \vec{S}_{1}\cdot\vec{v}_{1} \ddot{\vec{S}}_{1}\cdot\vec{v}_{2}
	 - 2 \vec{S}_{1}\cdot\vec{v}_{2} \ddot{\vec{S}}_{1}\cdot\vec{v}_{2}
	 + 8 \vec{S}_{1}\cdot\ddot{\vec{S}}_{1} \vec{v}_{1}\cdot\vec{v}_{2}
	 - 29 \vec{S}_{1}\cdot\ddot{\vec{S}}_{1} v_{2}^2 \nl
	 + 4 v_{1}^2 \ddot{S^2_{1}}
	 + 4 \vec{v}_{1}\cdot\vec{v}_{2} \ddot{S^2_{1}}
	 + 12 v_{2}^2 \ddot{S^2_{1}}
	 - 8 \vec{S}_{1}\cdot\ddot{\vec{S}}_{1} \vec{v}_{1}\cdot\vec{n} \vec{v}_{2}\cdot\vec{n}
	 + 4 \ddot{\vec{S}}_{1}\cdot\vec{n} \vec{S}_{1}\cdot\vec{v}_{1} \vec{v}_{2}\cdot\vec{n} \nl
	 + 4 \vec{S}_{1}\cdot\vec{n} \ddot{\vec{S}}_{1}\cdot\vec{v}_{1} \vec{v}_{2}\cdot\vec{n}
	 - 6 \ddot{\vec{S}}_{1}\cdot\vec{n} \vec{v}_{2}\cdot\vec{n} \vec{S}_{1}\cdot\vec{v}_{2}
	 - 6 \vec{S}_{1}\cdot\vec{n} \vec{v}_{2}\cdot\vec{n} \ddot{\vec{S}}_{1}\cdot\vec{v}_{2}
	 + \vec{S}_{1}\cdot\vec{n} \ddot{\vec{S}}_{1}\cdot\vec{n} v_{2}^2 \nl
	 + 13 \vec{S}_{1}\cdot\ddot{\vec{S}}_{1} ( \vec{v}_{2}\cdot\vec{n} )^{2}
	 - 4 \vec{v}_{1}\cdot\vec{n} \vec{v}_{2}\cdot\vec{n} \ddot{S^2_{1}}
	 - 3 \vec{S}_{1}\cdot\vec{n} \ddot{\vec{S}}_{1}\cdot\vec{n} ( \vec{v}_{2}\cdot\vec{n} )^{2} \big)
	 + \big( 2 \vec{S}_{1}\cdot\vec{a}_{1} \vec{S}_{1}\cdot\vec{a}_{2}
	 + 13 S_{1}^2 \vec{a}_{1}\cdot\vec{a}_{2} \nl
	 + 16 ( \vec{S}_{1}\cdot\vec{a}_{1} )^{2}
	 + 3 S_{1}^2 \vec{a}_{1}\cdot\vec{n} \vec{a}_{2}\cdot\vec{n}
	 - 2 \vec{S}_{1}\cdot\vec{n} \vec{S}_{1}\cdot\vec{a}_{1} \vec{a}_{2}\cdot\vec{n}
	 - 2 \vec{S}_{1}\cdot\vec{n} \vec{a}_{1}\cdot\vec{n} \vec{S}_{1}\cdot\vec{a}_{2} \nl
	 + 15 \vec{a}_{1}\cdot\vec{a}_{2} ( \vec{S}_{1}\cdot\vec{n} )^{2}
	 + 3 \vec{a}_{1}\cdot\vec{n} \vec{a}_{2}\cdot\vec{n} ( \vec{S}_{1}\cdot\vec{n} )^{2} \big)
	 + 4 \big( 8 \dot{\vec{S}}_{1}\cdot\vec{v}_{1} \vec{S}_{1}\cdot\vec{a}_{1}
	 + 8 \vec{S}_{1}\cdot\vec{v}_{1} \dot{\vec{S}}_{1}\cdot\vec{a}_{1} \nl
	 - 16 \vec{S}_{1}\cdot\dot{\vec{S}}_{1} \vec{v}_{1}\cdot\vec{a}_{1}
	 - 4 \dot{\vec{S}}_{1}\cdot\vec{a}_{1} \vec{S}_{1}\cdot\vec{v}_{2}
	 - 4 \vec{S}_{1}\cdot\vec{a}_{1} \dot{\vec{S}}_{1}\cdot\vec{v}_{2}
	 + 8 \vec{S}_{1}\cdot\dot{\vec{S}}_{1} \vec{a}_{1}\cdot\vec{v}_{2}
	 - 3 \dot{\vec{S}}_{1}\cdot\vec{v}_{1} \vec{S}_{1}\cdot\vec{a}_{2} \nl
	 - 3 \vec{S}_{1}\cdot\vec{v}_{1} \dot{\vec{S}}_{1}\cdot\vec{a}_{2}
	 + 13 \vec{S}_{1}\cdot\dot{\vec{S}}_{1} \vec{v}_{1}\cdot\vec{a}_{2}
	 - 6 \vec{S}_{1}\cdot\dot{\vec{S}}_{1} \vec{v}_{2}\cdot\vec{a}_{2}
	 + 8 \vec{v}_{1}\cdot\vec{a}_{1} \dot{S^2_{1}}
	 - 2 \vec{a}_{1}\cdot\vec{v}_{2} \dot{S^2_{1}} \nl
	 - 8 \vec{S}_{1}\cdot\dot{\vec{S}}_{1} \vec{a}_{1}\cdot\vec{n} \vec{v}_{2}\cdot\vec{n}
	 + 4 \dot{\vec{S}}_{1}\cdot\vec{n} \vec{S}_{1}\cdot\vec{a}_{1} \vec{v}_{2}\cdot\vec{n}
	 + 4 \vec{S}_{1}\cdot\vec{n} \dot{\vec{S}}_{1}\cdot\vec{a}_{1} \vec{v}_{2}\cdot\vec{n}
	 - 5 \vec{S}_{1}\cdot\dot{\vec{S}}_{1} \vec{v}_{1}\cdot\vec{n} \vec{a}_{2}\cdot\vec{n} \nl
	 - \dot{\vec{S}}_{1}\cdot\vec{n} \vec{S}_{1}\cdot\vec{v}_{1} \vec{a}_{2}\cdot\vec{n}
	 - \vec{S}_{1}\cdot\vec{n} \dot{\vec{S}}_{1}\cdot\vec{v}_{1} \vec{a}_{2}\cdot\vec{n}
	 + 3 \dot{\vec{S}}_{1}\cdot\vec{n} \vec{v}_{1}\cdot\vec{n} \vec{S}_{1}\cdot\vec{a}_{2}
	 + 3 \vec{S}_{1}\cdot\vec{n} \vec{v}_{1}\cdot\vec{n} \dot{\vec{S}}_{1}\cdot\vec{a}_{2} \nl
	 + 7 \vec{S}_{1}\cdot\vec{n} \dot{\vec{S}}_{1}\cdot\vec{n} \vec{v}_{1}\cdot\vec{a}_{2}
	 - 6 \vec{S}_{1}\cdot\vec{n} \dot{\vec{S}}_{1}\cdot\vec{n} \vec{v}_{2}\cdot\vec{a}_{2}
	 + 2 \vec{a}_{1}\cdot\vec{n} \vec{v}_{2}\cdot\vec{n} \dot{S^2_{1}}
	 + 3 \vec{S}_{1}\cdot\vec{n} \dot{\vec{S}}_{1}\cdot\vec{n} \vec{v}_{1}\cdot\vec{n} \vec{a}_{2}\cdot\vec{n} \big) \nl
	 - 2 \big( 8 \dot{S}_{1}^2 v_{1}^2
	 + 8 \dot{\vec{S}}_{1}\cdot\vec{v}_{1} \dot{\vec{S}}_{1}\cdot\vec{v}_{2}
	 - 8 \dot{S}_{1}^2 \vec{v}_{1}\cdot\vec{v}_{2}
	 + 29 \dot{S}_{1}^2 v_{2}^2
	 - 8 ( \dot{\vec{S}}_{1}\cdot\vec{v}_{1} )^{2}
	 + 2 ( \dot{\vec{S}}_{1}\cdot\vec{v}_{2} )^{2}
	 + 8 \dot{S}_{1}^2 \vec{v}_{1}\cdot\vec{n} \vec{v}_{2}\cdot\vec{n} \nl
	 - 8 \dot{\vec{S}}_{1}\cdot\vec{n} \dot{\vec{S}}_{1}\cdot\vec{v}_{1} \vec{v}_{2}\cdot\vec{n}
	 + 12 \dot{\vec{S}}_{1}\cdot\vec{n} \vec{v}_{2}\cdot\vec{n} \dot{\vec{S}}_{1}\cdot\vec{v}_{2}
	 - v_{2}^2 ( \dot{\vec{S}}_{1}\cdot\vec{n} )^{2}
	 - 13 \dot{S}_{1}^2 ( \vec{v}_{2}\cdot\vec{n} )^{2} \nl
	 + 3 ( \dot{\vec{S}}_{1}\cdot\vec{n} )^{2} ( \vec{v}_{2}\cdot\vec{n} )^{2} \big) \Big] \nl
+ \frac{G^2 C_{1ES^2} m_{2}^2}{2 m_{1} r^2} \Big[ \big( \vec{S}_{1}\cdot\ddot{\vec{S}}_{1}
	 + \vec{S}_{1}\cdot\vec{n} \ddot{\vec{S}}_{1}\cdot\vec{n} \big)
	 + \big( \dot{S}_{1}^2
	 + ( \dot{\vec{S}}_{1}\cdot\vec{n} )^{2} \big) \Big] \nl
+ \frac{G^2 C_{1ES^2} m_{2}}{r^2} \Big[ \big( \ddot{S^2_{1}}
	 - 4 \vec{S}_{1}\cdot\vec{n} \ddot{\vec{S}}_{1}\cdot\vec{n} \big)
	 - 4 ( \dot{\vec{S}}_{1}\cdot\vec{n} )^{2} \Big]
- \frac{G^2 m_{2}}{2 r^2} \Big[ 3 \dot{S}_{1}^2
	 - 5 ( \dot{\vec{S}}_{1}\cdot\vec{n} )^{2} \Big] \nl
+ (1 \leftrightarrow 2) .
\end{align}
Finally, several terms with three higher-order time derivatives appear in the potential as well,
\begin{align}
& \stackrel{(3)}{V} =
- \frac{G C_{1ES^2} m_{2}}{8 m_{1}} \Big[ 4 \vec{v}_{2}\cdot\vec{n} \dddot{S_{1}^2}
	 + \big( 13 \vec{S}_{1}\cdot\ddot{\vec{S}}_{1} \vec{a}_{2}\cdot\vec{n}
	 - 7 \ddot{\vec{S}}_{1}\cdot\vec{n} \vec{S}_{1}\cdot\vec{a}_{2}
	 - 7 \vec{S}_{1}\cdot\vec{n} \ddot{\vec{S}}_{1}\cdot\vec{a}_{2} \nl
	 - \vec{S}_{1}\cdot\vec{n} \ddot{\vec{S}}_{1}\cdot\vec{n} \vec{a}_{2}\cdot\vec{n} \big)
	 + \big( 13 \dot{S}_{1}^2 \vec{a}_{2}\cdot\vec{n}
	 - 14 \dot{\vec{S}}_{1}\cdot\vec{n} \dot{\vec{S}}_{1}\cdot\vec{a}_{2}
	 - \vec{a}_{2}\cdot\vec{n} ( \dot{\vec{S}}_{1}\cdot\vec{n} )^{2} \big) \Big]
+ (1 \leftrightarrow 2) .
\end{align}
As explained in \cite{Levi:2015msa}, since this potential contains higher order time derivatives of the variables, the EOM, which are derived from it via a variation of the action,
also contain higher order time derivatives, which must be eliminated, using lower order EOM. In a forthcoming work we are going to reduce these higher order time derivatives at the level of the potential, following the procedure of redefinition of variables, outlined in \cite{Damour:1990jh, Levi:2014sba}, where it was extended for spin variables in \cite{Levi:2014sba}. We are going to provide the EOM, and to transform this potential to a Hamiltonian, from which gauge-invariant relations for the binding energy can be obtained in a straightforward way. These relations are of great value for refining, for instance, the EOB Hamiltonian, and the gravitational
waveforms derived from it.

\section{Conclusions} \label{endfriend}

In this work we derived for the first time the NNLO spin-squared interaction potential for generic compact binaries via the EFT for gravitating spinning objects in the PN scheme, which was formulated in \cite{Levi:2015msa}. This correction, which enters at the 4PN order for rapidly rotating compact objects, completes the conservative sector up to the 4PN accuracy. This high level of PN accuracy is necessary in order to detect gravitational waves. The robustness of the EFT for gravitating spinning objects is shown here once again, as was already demonstrated in \cite{Levi:2014gsa,Levi:2015msa,Levi:2015uxa}, which obtained all spin dependent sectors, required up to the 4PN accuracy, and was initiated at the NNLO level in \cite{Levi:2011eq,Levi:2014sba}. The EFT of spinning objects allows to directly obtain the EOM from the potential via a proper variation of the action \cite{Levi:2014sba,Levi:2015msa}, and to obtain corresponding Hamiltonians in a straightforward manner \cite{Levi:2015msa}. In a forthcoming paper we are going to derive these for the NNLO spin-squared potential obtained in this paper. 

The spin-squared sector is an intricate one, as it requires the consideration of the point particle action beyond minimal coupling, and involves mainly contributions with the spin-squared worldline couplings, which are quite complex, compared to the worldline couplings, that follow from the minimal coupling part of the action. Yet, there are also contributions to this sector, involving the linear in spin couplings, as we go up in the nonlinearity of the interaction, and in the loop order. Thus, there is an increase in the number  of Feynman diagrams and in complexity with respect to the NLO, which is even more excessive than that, which occurs in the spin1-spin2 sector \cite{Levi:2008nh, Levi:2011eq, Levi:2015msa}. We have here 64 diagrams, of which more are higher loop ones, where we have 11 two-loops, and tensor integrals up to order 6. The application of the ``NRG'' fields \cite{Kol:2007bc,Kol:2010ze}, and the gauge choices of the EFT for gravitating spinning objects \cite{Levi:2015msa} then help to render the computations more efficient. Lastly, we checked most of the computations, using Mathematica. 

Finally, it should also be noted, that the relation between the Wilson coefficients, appearing in this sector, and in \cite{Levi:2014gsa}, required for generic compact objects at the 4PN accuracy, and the multipole moments used in numerical codes, e.g.~\cite{Laarakkers:1997hb, Pappas:2012ns, Yagi:2014bxa}, as well as the relation with the Geroch-Hansen multipoles for black holes \cite{Hansen:1974zz,Ryan:1995wh}, should be worked
out via a formal EFT matching procedure. Clearly, this involves subtleties 
\cite{Pappas:2012ns}, and is left for future work.

\acknowledgments

ML is grateful for the kind hospitality at the Max Planck Institute for 
Gravitational Physics (AEI) during the final stages of this work. 
This work has been done within the Labex ILP (reference ANR-10-LABX-63) part 
of the Idex SUPER, and received financial French state aid managed by the 
Agence Nationale de la Recherche, as part of the programme Investissements 
d'Avenir under the reference ANR-11-IDEX-0004-02.

\bibliographystyle{jhep}
\bibliography{gwbibtex}

\end{document}